\documentclass[12pt]{article} 

\newcommand{\doctype}{TECH}

\usepackage{ifthen}

\long\def\comment#1{}

\usepackage{natbib}

\usepackage{amsfonts,graphicx}
\usepackage{amsfonts,amsmath,amssymb,amsthm}
\usepackage{epic,eepic,eepicemu}
\usepackage{epsf}
\usepackage{graphics}
\usepackage{psfrag}
\usepackage{times}
\usepackage{epsfig}
\usepackage{subfig}

\usepackage{fullpage}

\newtheorem{theorem}{Theorem}
\newtheorem{proposition}[theorem]{Proposition}

\makeatletter
\def\@cite#1#2{[\if@tempswa #2 \fi #1]}
\makeatother

\def\real{{\mathbb R}}

\def\E{{\mathbb E}}

\def\indicator{{\mathbb I}}

\def\var{{\rm Var}}

\def\cov{{\rm Cov}}
\def\corr{{\rm Corr}}

\renewcommand\vec[1]{\ensuremath\boldsymbol{#1}}

\newcommand{\myparagraph}[1]{\noindent {\bf{#1}}}

\makeatletter
\long\def\@makecaption#1#2{
        \vskip 0.8ex
        \setbox\@tempboxa\hbox{\small {\bf #1:} #2}
        \parindent 1.5em  
        \dimen0=\hsize
        \advance\dimen0 by -3em
        \ifdim \wd\@tempboxa >\dimen0
                \hbox to \hsize{
                        \parindent 0em
                        \hfil 
                        \parbox{\dimen0}{\def\baselinestretch{0.96}\small
                                {\bf #1.} #2
                                } 
                        \hfil}
        \else \hbox to \hsize{\hfil \box\@tempboxa \hfil}
        \fi
        }
\makeatother

\makeatletter
\long\def\barenote#1{
    \insert\footins{\footnotesize
    \interlinepenalty\interfootnotelinepenalty 
    \splittopskip\footnotesep
    \splitmaxdepth \dp\strutbox \floatingpenalty \@MM
    \hsize\columnwidth \@parboxrestore
    {\rule{\z@}{\footnotesep}\ignorespaces
      #1\strut}}}
\makeatother

\title{Inference of global clusters from locally distributed
data}
\date{}

\comment{
\author{XuanLong Nguyen \\
Department of Statistics \\
University of Michigan \\
\emph{xuanlong@umich.edu}}
}
\begin{document}

\comment{
\ifthenelse{\equal{\doctype}{IEEE}}
{\typeout{}}
{ \begin{center}
Technical Report 504 \\
Department of Statistics \\
University of Michigan \\
Ann Arbor MI 48109 \\
\bigskip
\end{center}
}
}

\begin{center}
{\LARGE Inference of global clusters from locally distributed
data} \\
\vspace{.5cm}
XuanLong Nguyen \\
Department of Statistics \\
University of Michigan
\end{center}

\begin{abstract}
We consider the problem of analyzing the heterogeneity of
clustering distributions for multiple groups of observed data, each
of which is indexed by a covariate value, and inferring
global clusters arising from observations aggregated over 
the covariate domain. 
We propose a novel Bayesian nonparametric method reposing on
the formalism of spatial modeling and a nested hierarchy of
Dirichlet processes.
We provide an analysis of the model properties, relating and
contrasting the notions of local and global clusters.
We also provide an efficient inference algorithm,
and demonstrate the utility of our method in several data examples,
including the problem of object tracking and a global
clustering analysis of functional data where the functional
identity information is not available.
\footnote{This work was partially supported by NSF-CDI grant No. 0940671.
The author wishes to thank the referees and the Associate Editor
for valuable comments that help improve the presentation of this work.}
\end{abstract}

{\small
\myparagraph{Keywords:} global clustering, local clustering,
nonparametric Bayes, hierarchical Dirichlet process,
Gaussian process, graphical model, spatial dependence,
Markov chain Monte Carlo, model identifiability
}

\section{Introduction}

In many applications it is common to separate observed data
into groups (populations) indexed by some covariate $u$.
A particularly fruitful characterization of grouped data is 
the use of mixture distributions to describe the populations in 
terms of clusters of similar behaviors. Viewing observations
associated with a group as local data, and the clusters 
associated with a group as local clusters, it is often of 
interest to assess how the local heterogeneity is described by 
the changing values of covariate $u$. Moreover, in some applications the 
primary interest is to extract some sort of global clustering 
patterns that arise out of the aggregated observations.

Consider, for instance, a problem of tracking multiple objects
moving in a geographical area. Using covariate $u$ to 
index the time point, at a given time point $u$ we are
provided with a snapshot of the locations of the objects,
which tend to be grouped into local clusters. Over time,
the objects may switch their local clusters. We are not really
interested in the movement of each individual object.
It is the paths over which the local clusters evolve that
are our primary interest. Such paths are the global clusters.
Note that the number of global and local clusters are unknown,
and are to be inferred directly from the locally observed groups
of data.

The problem of estimating global clustering patterns out of locally 
observed groups of data also arises in the context of functional data
analysis where the functional identity information is not available.
By the absence of functional identity information,
we mean the data are not actually given as a collection
of sampled functional curves (even if such functional curves
exist in reality or conceptually),
due to confidentiality constraints or the impracticality
of matching the identity of individual functional curves.
As another example, the progesterone hormone behaviors
recorded by a number of women on a given day in their monthly
menstrual cycle is associated with a local group, which
are clustered into typical behaviors. Such local clusters
and the number of clusters may evolve throughout the monthly cycle. 
Moreover, aggregating the data over days in the cycle, 
there might exist one or more typical monthly (``global'' trend) 
hormone behaviors due to contraception or medical treatments. 
These are the global clusters. Due to privacy concern, the subject
identity of the hormone levels are neither known nor matched across
the time points $u$. In other words, the data are given not as
a collection of hormone curves, but as a collection of hormone
levels observed over time.

In the foregoing examples, the covariate $u$ indexes the time. 
In other applications, the covariate might index geographical locations
where the observations are collected.
More generally, observations associated with different 
groups may also be of different data types. For instance, consider 
the assets of a number of individuals (or countries), where 
the observed data can be subdivided into holdings according to 
different currency types (e.g., USD, gold, bonds). Here, 
each $u$ is associated with a currency type, and a global 
cluster may be taken to represent a typical portforlio of 
currency holdings by a given individual. In view of a substantial
existing body of work drawing from the spatial statistics literature
that we shall describe in the sequel, throughout this paper a 
covariate value $u$ is sometimes referred to as a spatial 
location unless specified otherwise. Therefore, the dependence 
on varying covariate values $u$ of the local heterogeneity of data 
is also sometimes referred to as the spatial dependence among 
groups of data collected at varying local sites.

We propose in this paper a model-based approach to learning
global clusters from locally distributed data. 
Because the number of both global and local clusters are assumed to be unknown,
and because the local clusters may vary with the covariate $u$, a natural approach
to handling this uncertainty is based on Dirichlet process mixtures and
their variants. A Dirichlet process $\textrm{DP}(\alpha_0,G_0)$
defines a distribution on (random) probability measures, where $\alpha_0$
is called the concentration parameter, and parameter $G_0$ denotes the base
probability measure or centering distribution~\citep{Ferguson}.
A random draw $G$ from the Dirichlet
process (DP) is a discrete measure (with probability 1), which
admits the well-known  ``stick-breaking'' representation~\citep{Sethuraman}:
\begin{equation}
\label{Eqn-Sethu-1}
G = \sum_{k=1}^{\infty} \pi_k \delta_{\phi_k},
\end{equation}
where the $\phi_k$'s are independent random variables distributed
according to $G_0$, $\delta_{\phi_k}$ denotes an atomic distribution
concentrated at $\phi_k$, and the stick breaking weights $\pi_k$ are random
and depend only on parameter $\alpha_0$. Due to the discrete nature of the DP
realizations, Dirichlet processes and their variants have become an effective
tool in mixture modeling and learning of clustered data. The basic
idea is to use the DP as a prior on the mixture components in
a mixture model, where each mixture component is associated with an
atom in $G$. The posterior distribution of the atoms provides the probability
distribution on mixture components, and also yields a probability distribution of
partitions of the data. The resultant mixture model, generally known as the
Dirichlet process mixture, was pioneered by the work of~\cite{Antoniak}
and subsequentially developed by many others (e.g.,~\citep{Lo,Escobar-West,MacEachern-Muller}).

A Dirichlet process (DP) mixture can be utilized to model each group of
observations, so a key issue is how to model and assess the local heterogeneity
among a collection of DP mixtures. In fact,
there is an extensive literature in Bayesian nonparametrics 
that focuses on coupling multiple Dirichlet process mixture distributions
(e.g.,~\cite{MacEachern-99,Muller-Quintana-Rosner,DeIorio-etal,Ishwaran-James,Teh-etal}).
A common theme has been to utilize the Bayesian
hierarchical modeling framework, where the parameters are 
conditionally independent draws from a probability distribution. 
In particular, suppose that the $u$-indexed group is modeled using a mixing distribution
$G_u$. We highlight the hierarchical Dirichlet process (HDP)
introduced by~\cite{Teh-etal}, a framework that we shall subsequentially
generalize, which posits that
$G_u|\alpha_0,G_0 \sim \textrm{DP}(\alpha_0, G_0)$ for some
base measure $G_0$ and concentration parameter $\alpha_0$. 
Moreover, $G_0$ is also random, and is distributed 
according to another DP: $G_0|\gamma,H \sim \textrm{DP}(\gamma,H)$.
The HDP model and other aforementioned work
are inadequate for our problem, because we are interested in modeling the 
linkage among the groups \emph{not} through the exchangeability
assumption among the groups, but through the more explicit dependence 
on changing values of a covariate $u$. 

Coupling multiple DP-distributed mixture distributions
can be described under a general framework outlined by~\cite{MacEachern-99}.
In this framework, a DP-distributed random measure can be represented
by the random ``stick'' and ``atom'' random variables (see Eq.~\eqref{Eqn-Sethu-1}),
which are general stochastic processes indexed by 
$u\in V$. 
%
Starting from this representation, there are a number of proposals 
for co-varying infinite mixture models~\citep{Duan-etal-07,Petrone-etal-09,
Rodriguez-etal-09,Dunson-kpp-08,Nguyen-Gelfand-09}. These proposals
were designed for functional data only, i.e., where the data
are given as a collection of sampled functions of $u$, and thus not 
suitable for our problem, because functional identity information 
is assumed unknown in our setting.
In this regard, the work of~\cite{Griffin-Steel-06,Dunson-Park,Rodriguez-etal-09b} 
are somewhat closer to our setting. These authors introduced 
spatial dependency of the local DP mixtures through 
the stick variables in a number of interesting ways,
while \cite{Rodriguez-etal-09b} 
additionally considered spatially varying atom variables, resulting
in a flexible model. These work focused mostly on the problem
of interpolation and prediction, not clustering. In particular, 
they did not consider 
the problem of inferring global clusters from locally observed data groups, 
which is our primary goal.

To draw inferences about global clustering patterns from locally grouped
data, in this paper we will introduce an explicit notion of and model for 
global clusters,
through which the dependence among locally distributed groups of data
can be described.
This allows us to not only assess the dependence of local 
clusters associated with multiple groups of data indexed by $u$, 
but also to extract the global clusters that arise from 
the aggregated observations.
From the outset, we use a spatial stochastic process, and more generally
a graphical model $H$ indexed over $u\in V$ to characterize the centering
distribution of global clusters. Spatial stochastic process and graphical
models are versatile and customary choice for modeling of multivariate
data~\citep{Cressie-93,Lauritzen-96,Jordan-04}. To ``link'' global clusters to local clusters,
we appeal to a hierarchical and nonparametric Bayesian formalism:
The distribution $Q$ of global clusters is random and distributed
according to a DP: $Q|H\sim \textrm{DP}(\gamma,H)$. For each $u$, 
the distribution $G_u$ of local clusters is assumed random, and is distributed
according to a DP: $G_u|Q \stackrel{indep}{\sim} \textrm{DP}(\alpha_u, Q_u)$,
where $Q_u$ denotes the marginal distribution at $u$ induced by the
stochastic process $Q$. In other words, in the first stage, the Dirichlet process $Q$ 
provides support for \emph{global atoms}, which in turn provide support for 
the \emph{local atoms} of lower dimensions for multiple groups 
in the second stage. Due to the use of hierarchy and the 
discreteness property of the DP realizations,
there is sharing of global atoms across the groups. Because
different groups may share only \emph{disjoint}
components of the global atoms, the spatial dependency 
among the groups is induced by the spatial distribution of the global atoms.
We shall refer to the described hierarchical specification as the
nested Hierarchical Dirichlet process (nHDP) model.

\comment{Formally, or each group indexed 
by $u$, we assume that $G_u \sim \textrm{DP}(\alpha_u,Q_u)$,
where $Q_u$ is a base measure for the DP associated with $u$.
To link the $Q_u$'s together, we let $Q_u$ be the marginal
distribution at $u$ of a stochastic process $Q$ indexed by $u\in V$.
We force $Q$ to be discrete and yet have broad support, 
by assuming that $Q$ is a random draw from a Dirichlet 
process $\textrm{DP}(\gamma, H)$. To induce the dependence
on the covariate $u\in V$, $H$ is taken to be a spatial process
indexed by $u \in V$, or more generally a graphical model defined
on the collection of variables indexed by $V$ In summary,
we obtain the following hierarchical specification:
\begin{eqnarray}
Q | \gamma, H & \sim & \textrm{DP}(\gamma, H) \notag \\
G_u |\alpha_u, Q & \stackrel{indep}{\sim|} & \textrm{DP}(\alpha_u, Q_u) 
\;\; \textrm{for each}\; u\in V,
\label{Defn-gdp}
\end{eqnarray}
which we shall refer to as a nested Hierarchical Dirichlet process (nHDP).
}

The idea of incorporating spatial dependence in the base measure of Dirichlet 
processes goes back to~\cite{Cifarelli-Regazzini,Muliere-Petrone,Gelfand-etal-05}, 
although not in a fully nonparametric hierarchical framework as is considered
here.
The proposed nHDP is an instantiation of the 
nonparametric and hierarchical modeling 
philosophy eloquently advocated in~\cite{Teh-Jordan-10}, but there is
a crucial distinction: Whereas Teh and Jordan generally
advocated for a \emph{recursive} construction of Bayesian hierarchy, 
as exemplified by
the popular HDP~\citep{Teh-etal}, the nHDP features a richer
\emph{nested} hierarchy: instead of taking a joint distribution,
one can take marginal distributions of a random distribution
to be the base measure to a DP in the next stage of
the hierarchy. This feature is essential to bring about the relationship
between global clusters and local clusters in our model. 
In fact, the nHDP generalizes the HDP model in the following sense:
If $H$ places a prior with probability one
on constant functions (i.e., if $\vec{\phi} = (\phi_u)_{u\in V} \sim H$ 
then $\phi_u = \phi_v \forall u,v \in V$) then the nHDP is reduced to
the HDP. 

Most closely related to our work is the hybrid DP of~\cite{Petrone-etal-09}, 
which also considers global and local clustering, and which in fact
serves as an inspiration for this work. Because the hybrid DP is designed for
functional data, it cannot be applied to situations where functional (curve)
identity information is not available, i.e., when the data are not given as
a collection of curves. When such functional id information is indeed
available, it makes sense to model the behavior of individual curves 
directly, and this ability may provide an advantage
over the nHDP. On the other hand, the hybrid DP is a rather complex
model, and in our experiment (see Section~\ref{sec-examples}), it
tends to overfit the data due to the model complexity. In fact, we
show that the nHDP provides a more satisfactory clustering performance
for the global clusters despite not using any functional id information,
while the hybrid DP requires not only such information,
it also requires the number of global clusters (``pure species'') 
to be pre-specified. 
\comment{Also related is the latent stick-breaking process,
which was introduced by~\cite{Rodriguez-etal-09} as a model for functional data.
Both~\cite{Rodriguez-etal-09} and~\cite{Petrone-etal-09} utilized
a latent Gaussian process to regulate the (local) cluster switching
behavior of a random curve, with the former focusing primarily
on prediction and interpolation applications instead of clustering.}
It is worth noting that in the proposed nHDP,
by not directly modeling the local cluster switching behavior,
our model is significantly simpler from both viewpoints of model
complexity and computational efficiency of statistical inference.

\comment{There are several recent interesting proposals for combining the graphical 
modeling formalism with Dirichlet processes. \cite{Caron-Doucet-09} and~\cite{Blei-Frazier-09} 
proposed methods for exploiting the spatial/graphical topology
of the underlying probability space (not the covariates), starting from
the a P\'olya-urn characterization of the Dirichlet process.
Closer in spirit to our work is~\cite{Heinz-09},
who considered quite restrictive classes of base measure under which 
a collection of (marginal) Dirichlet processes admits conditional 
independence properties.  These are fundamentally different approaches
to ours and they do not share the same goals.}

The paper outline is as follows. Section~\ref{sec-gdp}
provides a brief background of Dirichlet processes, the HDP,
and we then proceed to define the nHDP mixture model.
Section~\ref{sec-properties} explores the model properties,
including a stick-breaking characterization,
an analysis of the underlying graphical and
spatial dependency, a P\'olya-urn
sampling characterization. We also offer a discussion of
a rather interesting issue intrinsic to our problem and 
the solution, namely, the conditions under which global clusters
can be identified based on only locally grouped data.
As with most nonparametric Bayesian methods, inference is 
an important issue. We demonstrate in Section~\ref{sec-inference}
that the confluence of graphical/spatial with
hierarchical modeling allows for efficient computations
of the relevant posterior distributions.
Section~\ref{sec-examples} presents 
several experimental results, including a comparison to a recent
approach in the literature. Section~\ref{sec-discussions} concludes 
the paper. 
%

\section{Model formalization}
\label{sec-gdp}

\subsection{Background}
We start with a brief background on Dirichlet processes~\citep{Ferguson}, 
and then proceed to hierarchical Dirichlet processes~\citep{Teh-etal}. 
Let $(\Theta_0,\mathcal{B},G_0)$ be a probability space,
and $\alpha_0 > 0$. A Dirichlet process $\textrm{DP}(\alpha_0,G_0)$
is defined to be the distribution of a random probability measure $G$
over $(\Theta_0,\mathcal{B})$ such that, for any finite measurable
partition $(A_1,\ldots,A_r)$ of $\Theta_0$, the random vector
$(G(A_1),\ldots,G(A_r))$ is distributed as a finite dimensional
Dirichlet distribution with parameters $(\alpha_0G_0(A_1),\ldots,
\alpha_0G_0(A_r))$. $\alpha_0$ is referred to as the concentration parameter, 
which governs the amount of variability of $G$ around the centering
distribution $G_0$.
A DP-distributed probability measure $G$ is discrete with probability one.
Moreover, it has a constructive representation due to~\cite{Sethuraman}:
$G = \sum_{k=1}^{\infty}\pi_k \delta_{\phi_k}$,
where $(\phi_k)_{k=1}^{\infty}$ are iid draws from $G_0$, and $\delta_{\phi_k}$ denotes
an atomic probability measure concentrated at atom $\phi_k$.  The elements of the sequence
$\vec{\pi} = (\pi_k)_{k=1}^{\infty}$ are referred to as ``stick-breaking''
weights, and can be expressed in terms of independent beta variables:
$\pi_k = \pi'_k \prod_{l=1}^{k-1}(1-\pi'_l)$, where $(\pi'_l)_{l=1}^{\infty}$
are iid draws from $\textrm{Beta}(1,\alpha_0)$. 
Note that $\vec{\pi}$ satisfies $\sum_{k=1}^{\infty} \pi_k = 1$ with
probability one, and can be viewed as a random probabity measure
on the positive integers. For notational convenience, we write
$\vec{\pi}\sim \textrm{GEM}(\alpha_0)$, following~\cite{Pittman-02}.

A useful viewpoint for the Dirichlet process is given by
the P\'olya urn scheme, which shows that draws from the Dirichlet
process are both discrete and exhibit a clustering property. From
a computational perspective, the P\'olya urn scheme provides a
method for sampling from the random distribution $G$, by integrating
out $G$. More concretely, let atoms $\theta_1,\theta_2,\ldots$ are iid 
random variables distributed according to $G$. Because $G$ is random, 
$\theta_1,\theta_2,\ldots$ are exchangeable. \cite{Blackwell-MacQueen} showed that the
conditional distribution of $\theta_i$ given $\theta_1,\ldots,\theta_{i-1}$
has the following form:
\[[\theta_i|\theta_1,\ldots,\theta_{i-1},\alpha_0,G_0] \sim
\sum_{l=1}^{i-1}\frac{1}{i-1+\alpha_0}\delta_{\theta_l} + \frac{\alpha_0}{i-1+\alpha_0}G_0.\]
This expression shows that $\theta_i$ has a positive probability of being equal to one
of the previous draws $\theta_1,\ldots,\theta_{i-1}$. Moreover, the more
often an atom is drawn, the more likely it is to be drawn in the
future, suggesting a clustering property induced by the random
measure $G$. The induced distribution over random partitions
of $\{\theta_i\}$ is also known as the Chinese restaurant 
process~\citep{Aldous-85}.

A Dirichlet process mixture model utilizes $G$ as the prior on
the mixture component $\theta$. Combining with a likelihood
function $P(y|\theta) = F(y|\theta)$, the DP mixture model is given as:
$\theta_i | G \sim G$; 
$y_i|\theta_i \stackrel{ind}{\sim} F(\cdot|\theta_i)$.
Such mixture models have been studied in the pioneering work of
~\cite{Antoniak} and subsequentially by a number
of authors~\citep{Lo,Escobar-West,MacEachern-Muller}, 
For more recent and elegant accounts on the theories and 
wide-ranging applications of DP mixture modeling, 
see~\cite{Hjort-etal-10}.

\paragraph{Hierarchical Dirichlet Processes.}
Next, we proceed giving a brief background on the HDP formalism
of~\cite{Teh-etal}, which is typically motivated from the setting of
grouped data. Under this setting, the observations are organized into groups indexed
by a covariate $u\in V$, where $V$ is the index set. Let $y_{u1}$,
$y_{u2},\ldots, y_{u n_u}$  be the observations associated with group $u$.
For each $u$, the $\{y_{ui}\}_{i}$ are assumed to be exchangeable. 
This suggests the use of mixture modeling: The $y_{ui}$ are  
assumed identically and independently drawn from
a mixture distribution. Specifically, let $\theta_{ui} \in \Theta_u$ denote 
the parameter specifying the mixture component associated with $y_{ui}$.
Under the HDP formalism, $\Theta_u$ is the same space for all $u\in V$, i.e.,
$\Theta_u \equiv \Theta_0$ for all $u$, and $\Theta_0$ is endowed with
the Borel $\sigma$-algebra of subsets of $\Theta_0$. 
$\theta_{ui}$ is referred to as \emph{local factors} indexed by 
covariate $u$. Let $F(\cdot|\theta_{ui})$ denote the distribution of 
observation $y_{ui}$ given the local factor $\theta_{ui}$. Let $G_u$ 
denote a prior distribution for the local factors $(\theta_{ui})_{i=1}^{n_u}$. 
We assume that the local factors $\theta_{ui}$'s are conditionally
independent given $G_u$. As a result we have the following
specification:
\begin{equation}
\theta_{ui} | G_u \stackrel{iid}{\sim}  G_u; \;\;
y_{ui} | \theta_{ui}  \stackrel{iid}{\sim}  
F(\cdot|\theta_{ui}),\;\mbox{for any}\; u\in V;i=1,\ldots,n_u.
\label{Eqn-mixture}
\end{equation}

Under the HDP formalism, to statistically couple the collection of mixing distributions $G_u$, we posit 
that random probability measures $G_u$ are conditionally independent, with distributions given by
a Dirichlet process with base probability measure $G_0$:
\[G_u | \alpha_0, G_0 \stackrel{iid}{\sim} \textrm{DP}(\alpha_0, G_0).\]
Moreover, the HDP framework takes a fully nonparametric and hierarchical 
specification, by positing that $G_0$ is also a random probability
measure, which is distributed according to another Dirichlet process
with concentration parameter $\gamma$ and base probability measure $H$:
\[G_0 |\gamma, H \sim \textrm{DP}(\gamma,H).\]
An interesting property of the HDP is that because $G_u$'s are discrete random 
probability measures (with probability one) whose support are given
by the support of $G_0$. Moreover, $G_0$ is also a discrete measure,
thus the collection of $G_u$ are random discrete measures sharing the
same countable support.
In addition, because the random partitions induced by the collection of
$\theta_{ui}$ within each group $u$ are distributed according to a 
Chinese restaurant process, the collection of these Chinese restaurant
processes are statistically coupled. In fact, they are exchangeable,
and the distribution for the collection of such stoschastic
processes is known as the Chinese restaurant franchise~\citep{Teh-etal}.

\subsection{Nested hierarchy of DPs for global clustering analysis}
\myparagraph{Setting and notations.} In this paper we are interested in the
same setting of grouped data as that of the HDP that is described
by Eq.~\eqref{Eqn-mixture}. Specifically, the observations $y_{u1}, y_{u2},\ldots, y_{un_u}$ 
within each group $u$ are iid draws from a mixture distribution. 
The local factor $\theta_{ui} \in \Theta_u$ denotes the parameter specifying the mixture 
component associated with $y_{ui}$. The $(\theta_{ui})_{i=1}^{n_u}$
are iid draws from the mixing distribution $G_u$.

Implicit in the HDP model is the assumptions that the spaces $\Theta_u$ 
all coincide, and that random distributions $G_u$ are 
exchangeble. Both assumptions will be relaxed. 
Moreover, our goal here is the inference of global clusters, which are associated
with global factors that lie in the product space $\Theta := \prod_{u\in V}
\Theta_u$. To this end, $\Theta$ is endowed with a $\sigma$-algebra $\mathcal{B}$
to yield a measurable
space $(\Theta, \mathcal{B})$. Within this paper 
and in the data illustrations, $\Theta = \real^{V}$, and $\mathcal{B}$
corresponds to the Borel $\sigma$-algebra of subsets of $\real^{V}$,
%
Formally, a \emph{global factor}, which are denoted by $\vec{\psi}$ 
or $\vec{\phi}$ in the sequel, is a high dimensional vector (or function)
in $\Theta$ whose components are indexed by covariate $u$. That is, 
$\vec{\psi} = 
(\psi_u)_{u\in V} \in \Theta$, and $\vec{\phi} = (\phi_u)_{u\in V} \in \Theta$.
As a matter of notations,
we always use $i$ to denote the numbering index for $\theta_u$
(so we have $\theta_{ui}$). We always use $t$ and $k$ to denote the
number index for instances of $\vec{\psi}$'s and $\vec{\phi}$'s, 
respectively (e.g., $\vec{\psi}_t$ and $\vec{\phi}_k$). The components
of a vector $\vec{\psi}_t$ ($\vec{\phi}_k$) are denoted
by $\vec{\psi}_{ut}$ ($\vec{\phi}_{uk}$). We may also use
letters $v$ and $w$ beside $u$ to denote the group indices.

\paragraph{Model description.}
Our modeling goal is to specify a distribution $Q$ on
the global factors $\vec{\psi}$, and to relate $Q$ to the
collection of mixing distributions $G_u$ associated with
the groups of data. Such resultant model shall
enable us to infer about the \emph{global} clusters associated
with a global factor $\vec{\psi}$ on the basis of data collected
\emph{locally} by the collection of groups indexed by $u$.
At a high level, the random probability measures $Q$ and
the $G_u$'s are ``glued'' together under the nonparametric and
hierarchical framework, while the probabilistic linkage 
among the groups are governed by a stochastic process $\vec{\phi}
= (\phi_u)_{u\in V}$ 
indexed by $u\in V$ and distributed according to $H$. 
Customary choices of such stochastic processes include either 
a spatial process, or a graphical model $H$. 



Specifically, let $Q_u$ denote the induced marginal distribution of $\psi_u$.
Our model posits that for each $u \in V$, $G_u$ is a random
measure distributed as a DP with concentration
parameter $\alpha_u$, and base probability measure $Q_u$:
$G_u | \alpha_u, Q \sim \textrm{DP} (\alpha_u, Q_u)$.
Conditioning on $Q$, the distributions $G_u$ are independent,
and $G_u$ varies around the centering distribution 
$Q_u$, with the amount
of variability given by $\alpha_u$. 
The probability measure $Q$ is random, and distributed 
as a DP with concentration parameter $\gamma$ and
base probability measure $H$:
$Q | \gamma, H \sim \textrm{DP} (\gamma, H)$, where
$H$ is taken to be a spatial process indexed by $u \in V$, or 
more generally a graphical model defined on the collection 
of variables indexed by $V$.
In summary, collecting the described specifications gives
the \emph{nested Hierarchical Dirichlet process} (nHDP) mixture model:
\begin{eqnarray*}
Q | \gamma, H & \sim & \textrm{DP} (\gamma, H), \\
G_u | \alpha_u, Q & \stackrel{indep}{\sim} & \textrm{DP} (\alpha_u, Q_u),\; \textrm{for all}\; u\in V \\
\theta_{ui} | G_u & \stackrel{iid}{\sim} & G_u, \;\; 
y_{ui} | \theta_{ui} \stackrel{iid}{\sim}  F(\cdot|\theta_{ui}) \; \textrm{for all}\; u,i,
\end{eqnarray*}
%
%
%

As we shall see in the next section, the $\vec{\phi}_k$'s, which
are draws from $H$, provide the support for
global factors $\vec{\psi}_t \sim Q$, which in turn provide the support for
the local factors $\theta_{ui}\sim G_u$. The global and local factors 
provide distinct representations for both global clusters and local clusters 
that we envision being present in data. Local factors $\theta_{ui}$'s provide the support for 
local cluster centers at each $u$. 
The global factors $\vec{\psi}$ in turn provide the support for the local clusters, but
they also provide the support for global cluster centers in the
data, when observations are aggregated across different groups.

\paragraph{Relations to the HDP.} Both the HDP and nHDP are instances
of the nonparametric and hierarchical modeling framework involving
hierarchy of Dirichlet processes~\citep{Teh-Jordan-10}. At a high-level, the distinction 
here is that while the HDP is a recursive hierarchy of random probability 
measures generally operating on the same probability space, the nHDP features 
a nested hierarchy, in which the probability spaces associated with 
different levels in the hierarchy are distinct but related in the
following way: the probability distribution associated with a particular level,
say $G_u$, has support in the support of the marginal distribution
of a probability distribution (i.e., $Q$) in the upper level in the 
hierarchy. Accordingly, for $u \neq v$, $G_u$ and $G_v$ have
support in distinct components of vectors $\vec{\psi}$.
For a more explicit comparison, it is simple to see that if $H$ places distribution 
for \emph{constant} global factors $\vec{\phi}$
with probability one (e.g., for any $\vec{\phi} \sim H$ there holds
$\phi_{u} = \phi_{v} \forall u,v \in V$), then we obtain the HDP 
of~\cite{Teh-etal}.
\section{Model properties}
\label{sec-properties}
\subsection{Stick-breaking representation and graphical or spatial dependency}
\label{sec-stick}
Given that the multivariate base measure $Q$ is distributed
as a Dirichlet process, it can be expressed using Sethuraman's 
stick-breaking representation:
$Q = \sum_{k=1}^{\infty} \beta_k \delta_{\vec{\phi}_k}$.
Each atom $\vec{\phi}_k$ is multivariate and denoted by
$\vec{\phi}_k = (\phi_{uk}: u \in V)$. The $\vec{\phi}_k$'s are independent draws
from $H$, and $\vec{\beta} = (\beta_k)_{k=1}^{\infty} \sim \textrm{GEM}(\gamma)$.
The $\vec{\phi}_k$'s and $\vec{\beta}$ are mutually independent.
The marginal induced by $Q$ at each location $u \in V$ is:
%
$Q_u = \sum_{k=1}^{\infty} \beta_k \delta_{\phi_{uk}}$.
%
Since each $Q_u$ has support at the points $(\phi_{uk})_{k=1}^{\infty}$,
each $G_u$ necessarily has support at these points as well, and can
be written as:
\begin{equation}
\label{Eqn-Gu}
G_u = \sum_{k=1}^{\infty} \pi_{uk} \delta_{\phi_{uk}};
\;\;
Q_u = \sum_{k=1}^{\infty} \beta_k \delta_{\phi_{uk}}.
\end{equation}

Let $\vec{\pi}_u = (\pi_{uk})_{k=1}^{\infty}$. Since $G_u$'s are 
independent given $Q$, the weights $\vec{\pi}_u$'s are independent
given $\beta$. Moreover, because $G_u|\alpha_u,Q \sim \textrm{DP}(\alpha_u,Q_u)$
it is possible
to derive the relationship between weights $\vec{\pi}_u$'s and 
$\vec{\beta}$. Following~\cite{Teh-etal}, if $H$ is non-atomic, it is
necessary and sufficient for $G_u$ defined by Eq.~\eqref{Eqn-Gu} to
satisfy $G_u \sim \textrm{DP}(\alpha_u Q_u)$ that the following holds:
$\vec{\pi}_u \sim \textrm{DP}(\alpha_u, \beta)$,
where $\vec{\pi}_u$ and $\vec{\beta}$ are interpreted as probability
measures on the set of positive integers. 

The connection between the nHDP and the HDP of~\cite{Teh-etal} 
can be observed clearly here: The stick-breaking weights of the
nHDP-distributed $G_u$ have the same distributions as those 
of the HDP, while the atoms $\phi_{uk}$ are linked by a graphical model
distribution, or more generally a stochastic process indexed by $u$.

The spatial/graphical dependency given by base measure $H$ induces 
the dependency between the DP-distributed $G_u$'s. We shall
explore this in details by considering specific examples of $H$.


\begin{figure}[t]
\psfrag{H}{$H$}
\psfrag{Hu}{$H_u$}
\psfrag{Hv}{$H_v$}
\psfrag{Hw}{$H_w$}
\psfrag{Q}{$Q$}
\psfrag{Qu}{$Q_u$}
\psfrag{Qv}{$Q_v$}
\psfrag{Qw}{$Q_w$}
\psfrag{Gu}{$G_u$}
\psfrag{Gv}{$G_v$}
\psfrag{Gw}{$G_w$}
\psfrag{theu}{$\theta_{ui}$}
\psfrag{thev}{$\theta_{vi}$}
\psfrag{thew}{$\theta_{wi}$}
\psfrag{gam}{$\gamma$}
\psfrag{beta}{$\vec{\beta}$}
\psfrag{phiu}{$\phi_{uk}$}
\psfrag{phiv}{$\phi_{vk}$}
\psfrag{phiw}{$\phi_{wk}$}
\psfrag{piu}{$\vec{\pi}_u$}
\psfrag{piv}{$\vec{\pi}_v$}
\psfrag{piw}{$\vec{\pi}_w$}
\psfrag{inf}{$\infty$}
\psfrag{phi}{$\{\vec{\phi}_k\}$}
\psfrag{psi}{$\{\vec{\psi}_t\}$}
\psfrag{psiu}{$\psi_{ut}$}
\psfrag{psiv}{$\psi_{vt}$}
\psfrag{thetau}{$\{\theta_{ui}\}$}
\psfrag{thetav}{$\{\theta_{vi}\}$}
\begin{center}
\begin{tabular}{ccc}
\includegraphics[keepaspectratio,width = 0.20\textwidth]{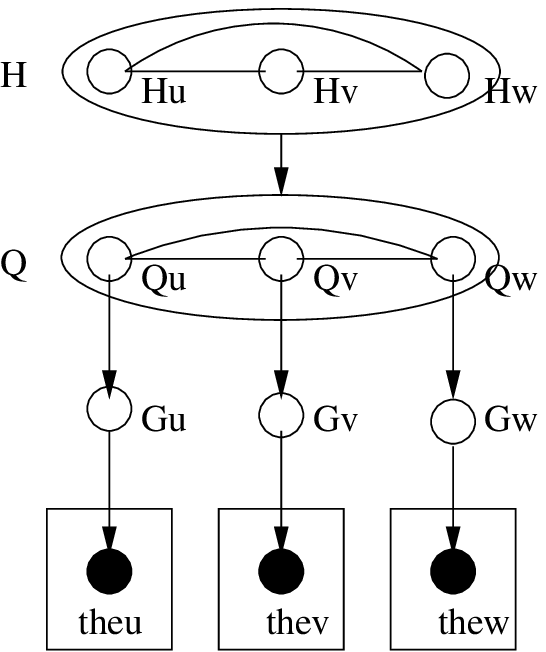} &
\; &
\includegraphics[keepaspectratio,width = 0.30\textwidth]{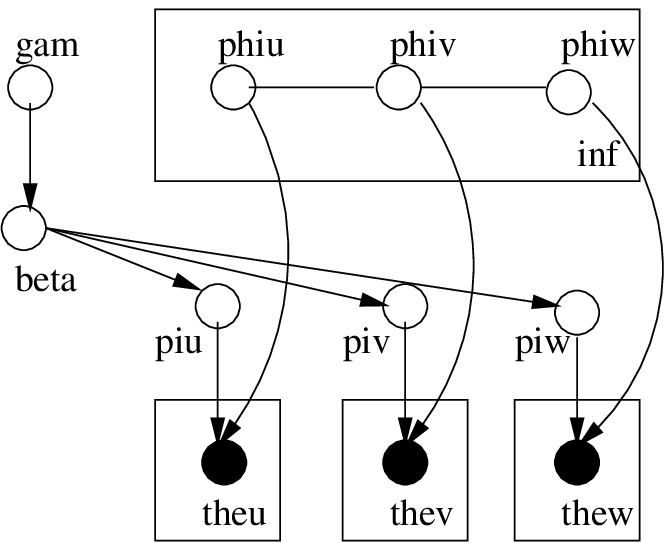} 
\end{tabular}
\end{center}
\caption{Left: The nHDP is depicted as a graphical model, where each
unshaded node represents a random distribution. Right: A graphical model
representation of the nHDP using the stick-breaking parameterisation.}
\label{Fig-gm}
\end{figure}

\comment{
\begin{figure}[h]
\psfrag{phi}{$\{\vec{\phi}_k\}$}
\psfrag{psi}{$\{\vec{\psi}_t\}$}
\psfrag{phiu}{$\phi_{uk}$}
\psfrag{phiv}{$\phi_{vk}$}
\psfrag{psiu}{$\psi_{ut}$}
\psfrag{psiv}{$\psi_{vt}$}
\psfrag{thetau}{$\{\theta_{ui}\}$}
\psfrag{thetav}{$\{\theta_{vi}\}$}
\begin{center}
\begin{tabular}{c}
\includegraphics[keepaspectratio,width = 0.35\textwidth]{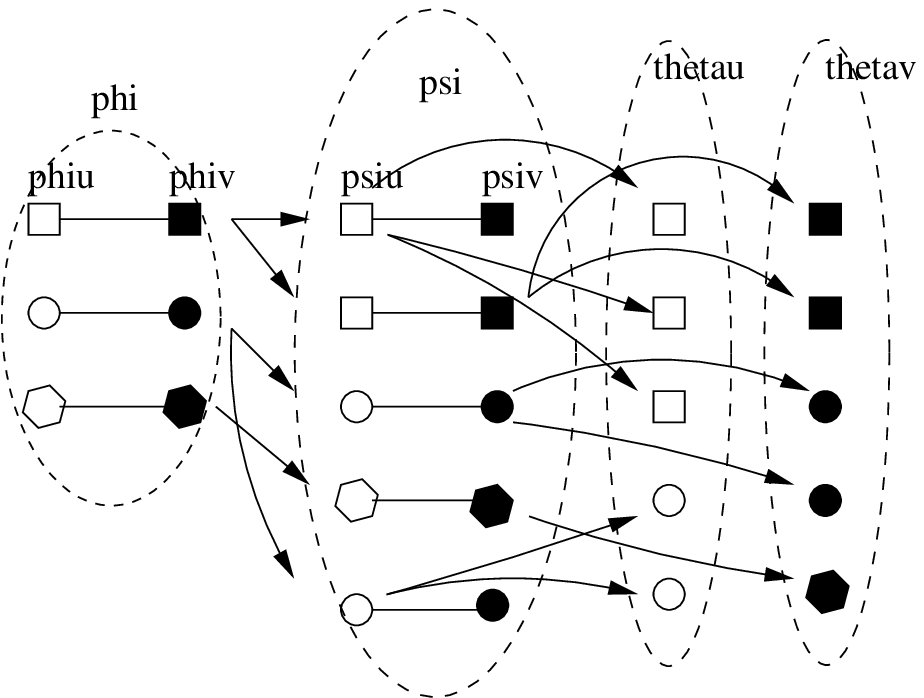} 
\end{tabular}
\end{center}
\caption{Illustration of the assignments of mixture component membership via
global and local factor variables for two groups indexed by $u$ and $v$.}
\label{Fig-assignment}
\end{figure}
}

\myparagraph{Example 1 (Graphical model $H$).}
For concreteness, we consider a graphical model $H$ of three
variables $\phi_u,\phi_v, \phi_w$ which are associated with three
locations $u,v,w \in V$. Moreover, assume the conditional
independence relation:
$\phi_u \perp \phi_w | \phi_v$. 
Let $\vec{\psi} = (\psi_u,\psi_v,\psi_w)$ be a random draw from $Q$.
Because $Q \sim \textrm{DP}(\gamma,H)$, $\vec{\psi}$ also has 
distribution $H$ once $Q$ is integrated out.
Thus, $\psi_u \perp \psi_w | \psi_v$.

At each location $u\in V$, the marginal distribution $Q_u$ of variable
$\psi_u$ is random and $Q_u|\gamma,H $ $\sim \textrm{DP}(\gamma,H_u)$. Moreover,
in general the $Q_u$'s are mutually dependent regardless of any
(conditional) independence relations that $H$ might confer.
This fact can be easily seen from Eq.~\eqref{Eqn-Gu}. With probability 1, 
all $Q_u$'s share the same $\vec{\beta}$. It follows that 
$Q_u \perp Q_w | Q_v, \vec{\beta}$.
Because $\vec{\beta}$ is random, the conditional independence relation
no longer holds among $Q_u, Q_w, Q_v$ in general.
From a modeling standpoint, the dependency among the $Q_u$'s is
natural for our purpose, as $Q$ provides the distribution for the 
global factors associated with the global clusters that we are also interested
in inferring.

Turning now to distributions $G_u$ for local factors $\theta_{ui}$, we note
that $G_u,G_v, G_w$ are independent given $Q$. Moreover, for each $u \in V$,
the support of $G_u$ is the same as that of $Q_u$ (i.e., $\theta_{ui}$ for
$i=1,2,\ldots$ take value among $(\psi_{ut})_{t=1}^{\infty}$). 
Integrating over the random $Q$, for any measurable partition $A \subset \Theta_u$, 
there holds:
$\E [G_u(A)| H]
= \E[\E[G_u(A)|Q]|H] = \E[Q_u(A)|H] = H_u(A)$.
%
In sum, the global factors $\vec{\psi}$'s take values in
the set of $(\vec{\phi}_k)_{k=1}^{\infty} \sim H$, and provide the support 
set for the local 
factors $\theta_{ui}$'s at each $u \in V$. The prior means of the local 
factors $\theta_{ui}$'s are also derived from the prior mean of the global factors. 

\myparagraph{Example 2 (Spatial model $H$).} 
To quantify more detailed dependency among DP-distributed $G_u$'s,
let $V$ be a finite subset of $\real^r$ and $H$ be a
second-order stochastic process 
indexed by $v\in V$. A customary choice for $H$ is a Gaussian process.
In effect, $\vec{\phi} = (\phi_u: u\in V) \sim N(\vec{\mu},\Sigma)$,
where the covariance $\vec{\Sigma}$ has entries of the exponential form:
$\rho(u,v) = \sigma^2 \exp-\{\omega \|u - v\|\}$.


For any measurable partitions $A\subset\Theta_u$, and $B\subset \Theta_v$, 
we are interested in expressions for variation and correlation measures
under $Q$ and $G_u$'s.
Let $H_{uv}(A,B) := p(\phi_u \in A, \phi_v \in B|H)$.
%
Define $g(\gamma) = 1/(\gamma+1)$. 
Applying stick-breaking representation for $Q_u$, it is simple to derive that:

\begin{proposition} For any pair of distinct locations $u,v)$, there holds:
\begin{eqnarray}
\cov(Q_u(A),Q_v(B)|H) & = & g(\gamma) (H_{uv}(A,B) - H_u(A)H_v(B)), \\
\var(Q_u(A)|H) & = & g(\gamma)(H_u(A) - H_u(A)^2), \\
\corr(Q_u(A),Q_v(B)) & := &
\frac{\cov(Q_u(A),Q_v(B)|H)}{\var(Q_u(A)|H)^{1/2}\var(Q_v(B)|H)^{1/2}} \notag \\
& = & \frac{(H_{uv}(A,B) - H_u(A)H_v(B))}
{(H_u(A) - H_u(A)^2)^{1/2}(H_v(B) - H_v(B)^2)^{1/2}}.
\end{eqnarray}
\end{proposition}

For any pair of locations $u,v \in V$, if $\|u-v\| \rightarrow \infty$,
it follows that $\rho(u,v) = \cov(\phi_u,\phi_v|H) \\ \rightarrow 0$.
Due to standard properties of Gaussian variables, $\phi_u$ and $\phi_v$ 
become less dependent of each other, and $H_{uv}(A,B) - H_u(A)H_v(B) 
\rightarrow 0$, so that $\corr(Q_u(A),Q_v(B)) 
\rightarrow 0$. On the other hand, if $u-v\rightarrow 0$, we obtain
that $\corr(Q_u(A),Q_v(A)) \rightarrow 1$, as desired. 

Turning to distributions $G_u$'s for the local factors, the following 
result can be shown:
%
%
\begin{proposition} For any pair of $u,v \in V$, there holds:
\begin{eqnarray}
\var(G_u(A)|H)  & = & \E[\var(G_u(A)|Q)|H] + \var(\E[G_u(A)|Q]|H) \notag \\
& = & (g(\gamma) + g(\alpha_u) - g(\gamma)g(\alpha_u)) (H_u(A) - H_u(A)^2), \\
\label{Eqn-anova-local}
\corr(G_u(A),G_v(B)) & = & \frac{g(\gamma)\corr(Q_u(A),Q_v(B)|H)}
{(g(\gamma) + g(\alpha_u) - g(\gamma)g(\alpha_u))^{1/2}
(g(\gamma) + g(\alpha_v) - g(\gamma)g(\alpha_v))^{1/2}}. \notag
\end{eqnarray}
where $g(\alpha_u) = 1/(\alpha_u+1)$.
\end{proposition}

Eq.~\eqref{Eqn-anova-local} exhibits an interesting decomposition of variance.
Note that $\var(G_u(A)|H) \geq \var(Q_u(A) |H)$.
That is, the variation of a local factor is greater than that of the global factor
evaluated at the same location, where the extra variation is governed by concentration
parameter $\alpha_u$. If $\alpha_u \rightarrow \infty$ so that
$g(\alpha_u) \rightarrow 0$, the local variation at $u$ disappears,
with the remaining variation contributed by the global factors only.
If $\alpha_u \rightarrow 0$ so that
$g(\alpha_u) \rightarrow 1$, the local variation contributed by $G_u$
completely dominates the global variation contributed by $Q_u$.

Finally, turning to correlation measures in the two 
stages in our hierachical model, we note that
\comment{
\begin{equation}
\corr(G_u(A),G_v(B)) = \frac{g(\gamma)\corr(Q_u(A),Q_v(B)|H)}
{(g(\gamma) + g(\alpha_u) - g(\gamma)g(\alpha_u))^{1/2}
(g(\gamma) + g(\alpha_v) - g(\gamma)g(\alpha_v))^{1/2}}.
\end{equation}
}
$\corr(G_u(A),G_v(B)|H)  \leq\corr(Q_u(A),Q_v(B)|H)$.
That is, the correlation across the locations in $V$ 
among the distributions $G_u$'s of the local 
factors is bounded from above by the correlation among 
the distribution $Q_u$'s for the global factors.
Note that $\corr(G_u(A),G_v(B))$ vanishes
as $\|u - v\| \rightarrow \infty$. 
The correlation measure increases as either $\alpha_u$
or $\alpha_v$ increases. 
The dependence on $\gamma$ is 
quite interesting. As $\gamma$ ranges from $0$ to $\infty$ so 
that $g(\gamma)$ decreases from $1$ to $0$, and as a 
result the correlation measure ratio $\corr(G_u(A),G_v(B))/
\corr(Q_u(A),Q_v(B))$ decreases from $1$ to $0$. 
\comment{
The described properties demonstrate the useful role of 
the Bayesian hierarchy in the nHDP model, as they bring about
the needed and rich distinctions in behaviors for the 
distributions of the global and the local factors.
}

\subsection{P\'olya-urn characterization}
\label{sec-polya}
The P\'olya-urn characterization of the canonical Dirichlet process
is fully retained by the nHDP.
It is also useful in highlighting both local
clustering and global clustering aspects that are described
by the nHDP mixture. In the sequel,
the P\'olya-urn characterization is given as a sampling scheme
for both the global and local factors.
Recall that the global factors $\vec{\phi}_1,\vec{\phi}_2,\ldots$ 
are i.i.d. random variables distributed 
according to $H$. We also introduced random vectors $\vec{\psi}_t$ 
which are i.i.d. draws from $Q$. Both $\vec{\phi}_k$ and $\vec{\psi}_t$
are multivariate, denoted by $\vec{\phi}_k = (\phi_{uk})_{u\in V}$
and $\vec{\psi}_t = (\psi_{ut})_{u\in V}$. Finally, for each location $u \in V$,
the local factor variables $\theta_{ui}$ are distributed according
to $G_u$. 

Note that each $\vec{\psi}_t$ is associated with one $\vec{\phi}_{k}$,
and each $\theta_{ui}$ is associated with one $\psi_{ut}$.
Let $t_{ui}$ be the index of the $\psi_{ut}$ associated with the
local factor $\theta_{ui}$,
and $k_{t}$ be the index of the $\vec{\phi}_{k}$ associated with
the global factor $\vec{\psi}_t$.
Let $K$ be the present number of distinct global factors
$\vec{\phi}_k$. The sampling process starts with $K=0$ and increases
$K$ as needed.
We also need notations for counts. We use notation $n_{ut}$ to denote
the present number of local factors $\theta_{ul}$ taking value 
$\psi_{ut}$.  $n_{u}$ denotes the number of local factors at group
$u$ (which is also the number of observations at group $u$).
$n_{u\cdot k}$ is the number of local factors at $u$ taking value $\phi_{uk}$.
Let $m_{u}$ denote the number of factors $\vec{\psi}_t$ 
that provide supports for group $u$.
The notation $q_{k}$ denotes the number of global factors 
$\vec{\psi}_{t}$'s taking value $\vec{\phi}_k$, while $q_{\cdot}$
denotes the total number of global factors $\vec{\psi}_t$'s.
To be precise:
%
\begin{eqnarray*}
&& n_{ut}  = \sum_{i} \indicator(t_{ui} = t); \;\;\;
n_{u\cdot k}  = \sum_{t} n_{ut}\indicator(k_t = k); \;\;\;
n_{u} = \sum_{t} n_{ut}; \\
&& m_{u} = \sum_{t} \indicator(n_{ut} > 0); \;\;\;
q_k = \sum_{t} \indicator(k_t = k); \;\;\;
q_{\cdot}  =  \sum_{k} q_k.
\end{eqnarray*}

First, consider the conditional distribution for $\theta_{ui}$
given $\theta_{u1},\theta_{u2},\ldots,\theta_{u,i-1}$, and $Q$,
where the $G_u$ is integrated out:
\begin{equation}
\label{Eqn-sample-theta-1}
\theta_{ui}|\theta_{u1},\ldots, \theta_{u,i-1}, \alpha_u, Q
\sim \sum_{t=1}^{m_u}\frac{n_{ut}}{i-1+\alpha_u} \delta_{\psi_{ut}}
+ \frac{\alpha_u}{i-1+\alpha_u}Q_u.
\end{equation}
This is a mixture, and a realization from this mixture can be obtained
by drawing from the terms on the right-hand side with probabilities
given by the corresponding mixing proportions. If a term in the first
summation is chosen, then we set $\theta_{ui} = \psi_{ut}$ for
the chosen $t$, and let $t_{ui} = t$, and increment $n_{ut}$. 
If the second term is chosen, then we increment $m_{u}$ by one, 
draw $\vec{\psi}_{u m_{u}} \sim Q_u$. In addition, we set 
$\theta_{ui} = \psi_{u m_{u}}$, and $t_{ui} = m_{u}$.

Now we proceed to integrate out $Q$. Since $Q$ appears only in its
role as the distribution of the variable $\vec{\psi}_t$, we only
need to draw sample $\vec{\psi}_t$ from $Q$. 
The samples from $Q$ can be obtained
via the conditional distribution of $\vec{\psi}_t$ as follows:
\begin{equation}
\label{Eqn-sample-psi-1}
\vec{\psi}_t | \{\vec{\psi}_{l}\}_{l\neq t}, \gamma, H
\sim \sum_{k=1}^{K} \frac{q_k}{q_{\cdot}+ \gamma} \delta_{\vec{\phi}_k} 
+ \frac{\gamma}{q_{\cdot} + \gamma} H.
\end{equation}
If we draw $\vec{\psi}_t$ via choosing a term in the summation on the 
right-hand side of this equation, we set $\vec{\psi}_t = \phi_k$, and
let $k_t = k$ for the chosen $k$, and increment $q_k$. If the second term 
is chosen then we increment $K$ by one, draw $\vec{\phi}_K \sim H$ and set 
$\vec{\psi}_{t} = \vec{\phi}_K$, $k_{jt} = K$, and $q_K = 1$.

The P\'olya-urn characterization of the nHDP can be illustrated
by the following culinary metaphor. Suppose that there are three groups of dishes (e.g.,
appetizer, main course and dessert) indexed by $u$, $v$ and
$w$. View a global factor $\vec{\phi}_k$'s as a typical meal box
where each $\phi_{uk}$, $\phi_{vk}$ and $\phi_{wk}$ is associated with 
a dish group. In an electic eatery, the dishes are sold in meal boxes,
while customers come in, buy dishes and share among one another
according to the following process. A new customer
can join either one of the three groups of dishes. Upon
joining the group, she orders a dish to contribute to the group,
i.e., a local factor $\theta_{ui}$,
based on its popularity within the group. She can also
choose to order a new dish, but to do so, she needs to order
the entire meal box, i.e. a global factor $\vec{\psi}_t$. 
A meal box is chosen based on its popularity as a whole,
across all eating groups.

The ``sharing'' of global factors (meal box) across indices
$u$ can be seen by noting that the ``pool'' of present global factors 
$\{\vec{\psi_l}\}$ has support in the discrete set of global factor values
$\vec{\phi}_1,\vec{\phi}_2,\ldots$. Moreover, the spatial (graphical)
distribution of the global factors induces the spatial dependence
among local factors associated with each group indexed by $u$.
See Fig.~\ref{Fig-assignment} for an illustration.

\begin{figure}
\psfrag{phi}{$\{\vec{\phi}_k\}$}
\psfrag{psi}{$\{\vec{\psi}_t\}$}
\psfrag{phiu}{$\phi_{uk}$}
\psfrag{phiv}{$\phi_{vk}$}
\psfrag{psiu}{$\psi_{ut}$}
\psfrag{psiv}{$\psi_{vt}$}
\psfrag{thetau}{$\{\theta_{ui}\}$}
\psfrag{thetav}{$\{\theta_{vi}\}$}
\begin{center}
\begin{tabular}{c}
\includegraphics[keepaspectratio,width = 0.45\textwidth]{figs/memassign2.eps} 
\end{tabular}
\end{center}
\caption{Illustration of the assignments of mixture component membership via
global and local factor variables for two groups indexed by $u$ and $v$.}
\label{Fig-assignment}
\end{figure}

\subsection{Model identifiability and complexity}
\label{sec-identify}
This section investigates
the nHDP mixture's inferential behavior, including issues 
related to the model identifiability.
It is useful to recall that a DP mixture model can be viewed as the 
infinite limit of finite mixture models~\citep{Neal-92,Ishwaran-Zarepour-02b}.
The nHDP can also be viewed as the limit of a finite mixture counterpart. 
Indeed, consider the following finite mixture model:
\begin{align}
\vec{\beta} | \gamma \sim \textrm{Dir}(\gamma/L, \ldots \gamma/L)
\;\;\;\;
\vec{\pi}_u | \alpha_u, \vec{\beta} \sim \textrm{Dir}(\alpha_u\vec{\beta}) 
\;\;\;\;
\vec{\phi}_k \sim H \notag \\
Q^L = \sum_{k=1}^{L} \beta_k \delta_{\vec{\phi}_k} 
\;\;\;\;\;\;\;\;
G_u^{L} = \sum_{k=1}^{L} \pi_{uk} \delta_{\phi_{uk}}.
\label{Eqn-QL}
\end{align}
It is a known fact that as $L\rightarrow 0$,
$Q^L \Rightarrow Q$ weakly, in the sense that for any real-valued
bounded and continuous function $g$, there holds 
$\int g \; dQ^L \rightarrow \int g \;dQ$ in distribution~\citep{Muliere-Secchi-95}.
\footnote{A stronger result was obtained by~\cite{Ishwaran-Zarepour-02b}, Theorem 2, 
in which convergence holds for any integrable function $g$ with respect to $H$.}
Because for each $u\in V$, there holds $G_u^L \sim \textrm{DP}(\alpha_u Q^L)$, it 
also follows that $G_u^L \Rightarrow G_u$ weakly.
The above characterization provides a convenient means of understanding
the behavior of the nHDP mixture by studying the behavior of its finite mixture
counterpart with $L$ global mixture components, as $L \rightarrow 
\infty$.

\noindent {\bf Information denseness of nHDP prior.}
For concreteness in this section we shall assume that
for any $u\in V$ the likelihood $F(y_{u}|\phi_{u})$
is specified by the normal distribution  whose parameters
such as mean and variance are represented by $\phi_{u}$. 
Write $\phi_{u} = (\mu_{u},\sigma_u^2) \in (\real \times \real_+)$.
Recall that conditionally on $Q$, $G_u$'s are independent 
across $u\in V$. Given $G_u$, the marginal distribution on 
observation $y_u$ has the following density:
\begin{equation}
\label{Eqn-fu}
f_u(y_u | G_u) = \int F(y_u|\phi_{u}) d G_u(\phi_u).
\end{equation}
Thus, each $f_u$ is the density of a location-scale mixture of normal
distribution. The $f_u$'s are random due to the randomness of $G_u$'s.
In other words, the nHDP places a prior distribution, which we 
denote by $\Pi$, over the collection of random measures 
$(G_u)_{u\in V}$. This in turn induces a prior over the 
joint density of $\vec{y} := (y_u)_{u\in V}$, which we call $\Pi$ as well.
Replacing the mixing distributions $Q$ and $G_u$ by the finite mixture
$Q^L$ and $G_u^L$'s (as specified by Eq.~\eqref{Eqn-QL}), we obtain
the corresponding marginal density:
\begin{equation}
\label{Eqn-fu}
f_u^L(y_u | G_u) = \int F(y_u|\phi_{u}) d G_u^L(\phi_u).
\end{equation}
Let $\Pi_L$ to denote the induced prior distribution for $\{f_u^L\}_{u\in V}$.
From the above, $\Pi_L \Rightarrow \Pi$ weakly. 

We shall show that for each $u\in V$ the prior $\Pi_L$ is 
information dense in the space of finite mixtures as $L\rightarrow \infty$. 
Indeed, for any group index $u$, consider
any finite mixture of normals $f_{u,0}$ associated with
mixing distributions $Q_0$ and $G_{u,0}$ of the form:
\begin{align}
Q_0 = \sum_{k=1}^{d} \beta_{k,0} \delta_{\vec{\phi}_{k,0}},
\;\;\;\;\;\;\;\;
G_{u,0} = \sum_{k=1}^{d} \pi_{uk,0} \delta_{\phi_{uk,0}},
\label{Eqn-Q0}
\end{align}

\begin{proposition} Suppose that the base measure 
$H$ places positive probability on a rectangle containing the 
support of $Q_0$, then the prior $\Pi_L$ places a 
positive probability in arbitrarily small Kullback-Leibler 
neighborhood of $f_{u,0}$ for $L$ sufficiently large. That is, for 
any $\epsilon > 0$,  there holds:
$\Pi_L(f_{u}: D(f_{u,0}||f_{u}) < \epsilon) > 0$
for any sufficiently large $L$. 
\end{proposition}

At a high level, this result
implies that the nHDP provides a prior over the space of
mixture distributions that is ``well spread'' in the 
Kullback-Leibler topology. A proof of this result
can be obtained using the same proof techniques 
of~\cite{Ishwaran-Zarepour-02} for a similar result
applied to (non-hierarchical) finite-dimensional Dirichlet distributions, and is 
therefore omitted.
An immediate consequence of the information denseness property is the weak 
consistency of the posterior distribution of 
$y_u$ for any $u\in V$, thanks to the asymptotic theory of~\cite{Schwartz-65}. 

\noindent {\bf Identifiability of factors $\vec{\phi}$.}
The above results are relevant from the viewpoint of density estimation 
(for the joint vector $\vec{y}$). From a clustering
viewpoint, we are also interested in the ability 
of the nHDP prior in recovering the underlying local
factors $\phi_{uk}$'s, as well as the global factors 
$\vec{\phi}_k$'s for the global clusters.  
This is done by studying the identifiability of the
finite mixtures that lie in the union of the support
of $\Pi_L$ for all $L < \infty$. This is the set of 
all densities $(f_u^L)_{u\in V; L < \infty}$ whose corresponding
mixing distributions are given by  Eq.~\eqref{Eqn-QL}.

Recall that each marginal $f_u^L$ is a normal mixture, and
the $L$ mixture components are parameterised by 
$\phi_{uk} = (\mu_{uk},\sigma_{uk}^2)$ for $k=1,\ldots, L$.
Again, let $f_{u,0}$ be the ``true'' marginal density of a mixture
distribution for group $u$ that has $d$ mixture components,
and the associated mixing distributions $Q_{0}$ and $G_{u,0}$
are given by Eq.~\eqref{Eqn-Q0}.
The parameter for the $k$-th component for each $k=1,\ldots,d$ 
is denoted by $\phi_{uk,0} = (\mu_{uk,0},\sigma_{uk,0}^2)$. The following
is a direct consequence of Theorem 2 of~\cite{Ishwaran-Zarepour-02}:

\begin{proposition} 
Suppose that for any $u\in V$,
$f_{u}(y_u) = f_{u,0}(y_u) \;\;\textrm{for almost all}\; y_u$.
In addition, the mixing distributions $G_u^L$ satisfy the following condition:
\[\int_{\real\times \real_+} \exp \biggr (\frac{\mu_u^2}{2(\sigma_u^*-\sigma_u)}
\biggr ) G_u^L(d \phi_u) < \infty,\]
for any $u\in V$, where $\sigma_u^* = \min\{\sigma_{u1,0},\ldots,\sigma_{uk,0}\}$.
Then,  $G_u = G_{u,0} \;\textrm{for all}\; u\in V$.
\end{proposition}
In other words, this result claims that it is possible to 
identify all local clusters
specified by $\phi_{uk}$ and $\pi_{uk}$ for $k=1,\ldots, d$, 
up to the ordering of the mixture component index $k$.
A more substantial issue is the identifiability
of global factors. Under additional conditions of 
``true'' global factors $\vec{\phi}_{k,0}$'s, and the distribution of 
global factors $Q^L$, the identification
of global factors $\vec{\phi}_{k,0}$'s is possible.
Viewing a global factor $\vec{\phi}_k = (\phi_{uk})_{u\in V}$ 
(likewise, $\vec{\phi}_{k,0}$)
as a function of $u\in v$, a trivial example is that when
$\vec{\phi}_{k,0}$ are constant functions, and that base measure
$H$ (and consequentially $Q^L$) places probability 1 on such 
set of functions, then the identifiability of local factors
implies the identifiability of global factors. 
A nontrivial condition is that the ``true'' global
factors $\vec{\phi}_{k,0}$ as a function of $u$ can be
parameterised by a small number of parameters (e.g. a linear
function, or an appropriately defined smooth function in $u\in V$). 
Then, it is possible that the identifiability of
local factors also implies the identifiability of global factors.
An in-depth theoretical treatment of this important issue is beyond 
the scope of the present paper.

The above observations suggest several prudent guidelines for
prior specifications (via the base measure $H$). To ensure
good inferential behavior for the local factors $\phi_u$'s, it
is essential that the base measure $H_u$ places sufficiently small 
tail probabilities on both $\mu_u$ and $\sigma_u$. In addition, 
if it is believed the underlying global factors are smooth
function in the domain $V$, placing
a very vague prior $H$ over the global factors (such as a factorial
distribution $H = \prod_{u\in V} H_u$ by assuming the $\phi_u$
are independent across $u\in V$) may not do the job.
Instead, an appropriate base measure $H$ that puts most of its
mass on smooth functions is needed. Indeed, these
observations are also confirmed by our empirical experiments
in Section 5.


\section{Inference}
\label{sec-inference}
In this section we shall describe posterior inference methods for the nested
Hierarchical
Dirichlet process mixture. 
We describe two different sampling approaches: The ``marginal approach''
proceeds by integrating out the DP-distributed random measures, while the ``conditional approach''
exploits the stick-breaking representation. The former approach
arises directly from the P\'olya-urn characterization of the nHDP.
However its implementation is more involved due to book-keeping
of the indices. Within this section we shall describe the conditional
approach, leaving the details of the marginal approach to
the supplemental material.
Both sampling methods draw from the basic
features of the sampling methods developed for the Hierarchical 
Dirichlet Process of~\cite{Teh-etal}, in addition to the computational
issues that arise when high-dimensional global factors are sampled.

\comment{Due to the hierarchical aspects of our model, we borrow the
basic features of the sampling methods developed by~\cite{Teh-etal}
for their hierarchical Dirichlet process 
mixtures. The unique aspect for our model is that 
in addition to sampling about the mixture membership variables, 
we also need to integrate over or sample the global factors 
$\vec{\phi} \sim H$ which typically have very high dimension.}

For the reader's convenience, we recall key notations and 
introduce a few more for the sampling algorithms.
$t_{ui}$ is the index of the $\psi_{ut}$ 
associated with the local factor $\theta_{ui}$, i.e., 
$\theta_{ui} = \psi_{u t_{ui}}$; and $k_t$ is the index of the 
$\phi_{k}$ associated with the global factor $\vec{\psi}_t$, i.e., 
$\vec{\psi}_t = \vec{\phi}_{k_t}$. The local
and global atoms are related by $\theta_{ui} = \psi_{ut_{ui}}
= \phi_{uk_{t_{ui}}}$.  Let $z_{ui} = k_{t_{ui}}$ denote 
the mixture component associated with observation $y_{ui}$.
Turning to count variables, $n_{ut}^{-ui}$ denotes the number of local
atoms $\theta_{ul}$'s that are associated with $\vec{\psi}_t$,
excluding atom $\theta_{ui}$. $n_{u\cdot k}^{-ui}$ denotes the number
of local atoms $\theta_{ul}$ that such that $z_{ul} = k$,
leaving out $\theta_{ui}$.
$\vec{t}^{-ui}$ denotes the vector of 
all $t_{ul}$'s leaving out element $t_{ui}$. Likewise, $\vec{k}^{-t}$
denotes the vector of all $k_r$'s leaving out element $k_t$.
In the sequel, the concentration parameters $\gamma,\alpha_u$,
and parameters for $H$ are assumed fixed. In practice,
we also place standard prior distributions on these parameters,
following the approaches of~\cite{Escobar-West,Teh-etal} for
$\gamma,\alpha_u$, and, e.g.,~\cite{Gelfand-etal-05} for $H$'s.

The main idea of the conditional sampling approach is to exploit
the stick-breaking representation of DP-distributed $Q$
instead of integrating it out. Likewise,
we also consider not integrating over the base measure $H$.
%
Recall that a priori $Q \sim \textrm{DP}(\gamma, H)$. Due to a standard
property of the posterior of a Dirichlet process, conditioning on the
global factors $\vec{\phi}_k$'s and the index vector $\vec{k}$, 
$Q$ is distributed as $\textrm{DP}(\gamma + q_{\cdot}, \frac{\gamma H + 
\sum_{k=1}^{K}q_k \delta{\vec{\phi}_k}} {\gamma + q_{\cdot}})$. 
Note that vector $\vec{q}$ can be computed directly from $\vec{k}$.
Thus, an explicit representation of $Q$ is 
$Q  =  \sum_{k=1}^{K} \beta_k \delta_{\vec{\phi}_k} + \beta_{\textrm{new}} Q^{\textrm{new}}$,
where $Q^{\textrm{new}}  \sim \textrm{DP}(\gamma, H)$, and
\begin{eqnarray*}
\vec{\beta} & = & (\beta_1,\ldots, \beta_K, \beta_{\textrm{new}}) 
\sim \textrm{Dir}(q_1,\ldots,q_k, \gamma).
\end{eqnarray*}
Conditioning on $Q$, or equivalently conditioning on $\vec{\beta}, \vec{\phi}_k$'s
in the stick breaking representation, the distributions $G_u$'s 
associated with different locations $u \in V$ are decoupled (independent).
In particular, the posterior of $G_u$ given $Q$ and $\vec{k}, \vec{t}$ 
and the $\vec{\phi}_k$'s is distributed as $\textrm{DP}(\alpha_u + n_{u}, 
\frac{\alpha_u Q_u + \sum_{k=1}^{K}n_{u\cdot k} \delta_{\phi_{uk}}}
{\alpha_u + n_{u}})$. Thus, an explicit representation of the
conditional distribution of $G_u$ is given as 
$G_u  =  \sum_{k=1}^{K}\pi_{uk}\delta_{\phi_{uk}} +  \pi_{u \textrm{new}} G_{u}^{\textrm{new}}$,
where 
$G_{u}^{\textrm{new}} \sim  \textrm{DP}(\alpha_u \beta_{\textrm{new}}, Q_u^{\textrm{new}})$
and
\begin{eqnarray*}
\vec{\pi}_u & = & (\pi_{u1},\ldots, \pi_{uK}, \pi_{u \textrm{new}}) 
\sim \textrm{Dir}(\alpha_u \beta_1 + n_{u\cdot 1},\ldots,\alpha_u \beta_k
+ n_{u\cdot K}, \alpha_u \beta_{\textrm{new}}).
\end{eqnarray*}
In contrast to the marginal approach, we consider sampling directly 
in the mixture component variable
$z_{ui} = k_{t_{ui}}$, and in doing so we bypass the sampling steps
involving $\vec{k}$ and $\vec{t}$. Note that the likelihood 
of the data involves only the $z_{ui}$ variables and the global atoms
$\vec{\phi}_k$'s. The mixture proportion vector $\vec{\beta}$
involves only count vectors $\vec{q} = (q_1,\ldots,q_K)$. It suffices
to construct a Markov chain on the space of $(\vec{z}, \vec{q},
\vec{\beta}, \vec{\phi})$.

\myparagraph{Sampling $\vec{\beta}$.} As mentioned above,
$\vec{\beta} | \vec{q} \sim \textrm{Dir}(q_1,\ldots, q_K,\gamma)$.

\myparagraph{Sampling $\vec{z}$.} Recall that a priori 
$z_{ui}|\vec{\pi}_u,\vec{\beta} \sim \vec{\pi}_u$ where
$\vec{\pi}_u|\vec{\beta}, \alpha_u \sim \textrm{DP}(\alpha_u,\vec{\beta})$.
Let $n_{u\cdot k}^{-ui}$ denote the number of data items in the group $u$,
except $y_{ui}$, associated with the mixture component $k$. This can
be readily computed from the vector $\vec{z}$.
\begin{equation}
p(z_{ui} = k|\vec{z}^{-ui},\vec{q},\vec{\beta}, \vec{\phi}_k, \textrm{Data}) =
\begin{cases}
(n_{u\cdot k}^{-ui} + \alpha_u \beta_k) F(y_{ui}|\phi_{uk}) & \;\; \textrm{if} \; 
k \;\textrm{previously used} \\
\alpha_u \beta_{\textrm{new}} f_{uk^{\textrm{new}}}^{-y_{ui}}(y_{ui}) & \;\; \textrm{if}\;
k = k^{\textrm{new}}.
\end{cases}
\end{equation}
where 
%
\begin{equation}
\label{Eqn-integrate-phi-2}
f_{uk}^{-y_{ui}}(y_{ui}) = \frac{\int F(y_{ui}|\phi_{uk}) \prod_{u'i'\neq ui;z_{u'i'} = k}
F(y_{u'i'}|\phi_{u'k})H(\vec{\phi}_k) d\vec{\phi_k}}
{\int \prod_{u'i'\neq ui;z_{u'i'} = k}
F(y_{u'i'}|\phi_{u'k})H(\vec{\phi}_k) d\vec{\phi}_k}.
\end{equation}
Note that if $z_{ui}$ is taken to be $k^{\textrm{new}}$, then we update
$K = K+1$. (Obviously, $k^{\textrm{new}}$ takes the value of
the updated $K$).

\myparagraph{Sampling $\vec{q}$.}
To clarify the distribution for vector $\vec{q}$, we recall an
observation at the end of Section~\ref{sec-polya} 
that the set of global factors 
$\vec{\psi}_t$'s can be organized into disjoint subsets 
$\Psi_{u}$, each of which is associated with a location $u$. 
More precisely, $\vec{\psi}_t \in \Psi_u$ if and only if $n_{ut} > 0$. 
Within each group $u$, let $m_{uk}$ denote the number of 
$\vec{\psi}_t$'s taking value $\vec{\phi}_k$. Then, 
$q_k = \sum_{u\in V} m_{uk}$.

Conditioning on $\vec{z}$ we can collect all data items in
group $u$ that are associated with mixture component $\vec{\phi}_k$,
i.e., item indices $ui$ such that $z_{ui} = k$.  There are $n_{u\cdot k}$ 
such items, which are distributed according to a Dirichlet
process with concentration parameter $\alpha_u \beta_k$. The count
variable $m_{uk}$ corresponds to the number of mixture components
formed by the $n_{u\cdot k}$ items. It was shown by Antoniak (1974) that
the distribution of $m_{uk}$ has the form:
\[p(m_{uk} = m | \vec{z}, \vec{m}^{-uk}, \vec{\beta}) =
\frac{\Gamma(\alpha_u \beta_k)}{\Gamma(\alpha_u \beta_k + n_{u\cdot k})}
s(n_{u\cdot k}, m)(\alpha_u \beta_k)^m,\]
where $s(n,m)$ are unsigned Stirling number of the first kind. By
definition, $s(0,0) = s(1,1) = 1, s(n,0) = 0$ for $n>0$, and $s(n,m) = 0$
for $m> n$. For other entries, there holds $s(n+1,m) = s(n,m-1) + ns(n,m)$.

\myparagraph{Sampling $\vec{\phi}$.} The sampling of $\vec{\phi}_1,\ldots,\vec{\phi}_k$ 
follows from the following conditional probabilities:
\[p(\vec{\phi}_k| \vec{z},\textrm{Data})
\propto H(\vec{\phi}_k)\prod_{ui: z_{ui} = k} F(y_{ui}|\phi_{uk})
\;\textrm{for each}\; k=1,\ldots, K.\]
Let us index the set $V$ by $1,2,\ldots, M$, where $|V| = M$. We
return to our two examples.

As the first example, suppose that $\vec{\phi}_k$ is normally distributed,
i.e., under $H$, $\vec{\phi}_k \sim N(\vec{\mu}_k, \vec{\Sigma}_k)$, 
and that the likelihood $F(y_{ui}|\theta_{ui})$ is given as well
by $N(\theta_{ui},\sigma_\epsilon^2)$,
then the posterior distribution of $\vec{\phi}_k$ is also Gaussian with 
mean $\tilde{\vec{\mu}}_k$ and variance $\tilde{\vec{\Sigma}}_k$, where:
\begin{align}
\tilde{\vec{\Sigma}}_k^{-1} = \vec{\Sigma}_k^{-1} + \frac{1}{\sigma_\epsilon^2}\textrm{diag}(n_{1\cdot k},\ldots,n_{M\cdot k}), \notag \\
\tilde{\vec{\mu}}_k = \tilde{\vec{\Sigma}}_k \biggr (\vec{\Sigma}_k^{-1} \vec{\mu}_k
+ \frac{1}{\sigma_\epsilon^2} \biggr 
[\sum_{i}y_{1i}\indicator(z_{1i}=k) \ldots \sum_{i}y_{Mi}\indicator(z_{Mi}=k) \biggr ]^T \biggr ).
\end{align}
For the second example, we assume that $\vec{\phi}_k$ is very high dimensional, 
and the prior distribution $H$ is not tractable (e.g., a Markov random field).
Direct computation is no longer possible. A simple solution is to Gibbs sample 
each component of vector $\vec{\phi}_k$. Suppose that under a Markov
random field model $H$, the conditional probability $H(\phi_{uk}| \vec{\phi}_k^{-u})$
is simple to compute. Then, for any $u\in V$,
\[p(\phi_{uk}| \vec{\phi}_k^{-u}, \vec{z}, \textrm{Data}) \propto H(\phi_{uk}|\vec{\phi}_k^{-u})
\prod_{i:z_{ui} = k} F(y_{ui}|\phi_{uk}).\]

\paragraph{Computation of conditional density of data}
A major computational bottleneck in sampling methods for the nHDP
is the computation of conditional densities given by
Eq.~\eqref{Eqn-integrate-phi-2} and~\eqref{Eqn-integrate-phi}. 
In general, $\vec{\phi}$ is very high dimensional, and integrating over $\vec{\phi} \sim H$
is intractable. However it is possible to exploit the structure
of $H$ to alleviate this situation. As an example, if
$H$ is conjugate to $F$, the computation of these conditionals
can be achieved in closed form. Alternatively, if $H$ is specified 
as a graphical model where conditional independence
assumptions can be exploited, efficient inference methods
in graphical models can be brought to bear on our computational
problem.

\myparagraph{Example 1.} Suppose that the likelihood function $F$
is given by a Gaussian distribution, i.e., $y_{ui}|\theta_{ui} \sim N(\theta_{ui},
\sigma_\epsilon^2)$ for all $u,i$, and that the prior $H$ is conjugate,
i.e., $H$ is also a Gaussian distribution: $\vec{\phi}_k \sim 
N(\vec{\mu}_k,\vec{\Sigma}_k)$. Due to conjugacy, the computations
in Eq.~\eqref{Eqn-integrate-phi} 
are readily available in closed forms. Specifically, the density
in Eq.~\eqref{Eqn-integrate-phi} takes the following expression:
\begin{equation*}
f_{uk}^{-y_{ui}}(y_{ui}) = \frac{1}{(2\pi)^{1/2}\sigma_\epsilon}
\frac{|\vec{C}_{k+}|}{|\vec{C}_{k}|} \exp\biggr (-\frac{1}{2\sigma_\epsilon^2}y_{ui}^2
+ \frac{1}{2}{\vec{\mu}_{k+}^{-ui}}^T\vec{C}_{k+}^{-1}\vec{\mu}_{k+}^{-ui}-
\frac{1}{2}{\vec{\mu}_{k}^{-ui}}^T\vec{C}_{k}^{-1}\vec{\mu}_{k}^{-ui}\biggr ), 
\end{equation*}
where
\[\vec{C}_{k+}^{-1} = \vec{\Sigma}_k^{-1} + \frac{1}{\sigma_\epsilon^2}\textrm{diag}
(n_{1\cdot k}^{-ui},\ldots, 1+n_{u\cdot k}^{-ui}, \ldots,n_{M\cdot k}^{-ui}), \notag\]
\begin{align}
\vec{\mu}_{k+}^{-ui} = \vec{C}_{k+}\biggr (\vec{\Sigma}_k^{-1}\vec{\mu}_k 
+ \frac{1}{\sigma_\epsilon^2} \biggr [\cdots \sum_{i': z_{u'i'} = k} y_{u'i'} + y_{ui}\indicator(ui=u'i') \cdots \biggr ]^T \biggr ),
\notag \\
\vec{C}_k^{-1} = \vec{\Sigma}_k^{-1} + \frac{1}{\sigma_\epsilon^2}\textrm{diag}
(n_{1\cdot k}^{-ui},\ldots,n_{u\cdot k}^{-ui}, \ldots, n_{M\cdot k}^{-ui}), \notag \\
\vec{\mu}_{k}^{-ui} = \vec{C}_{k}\biggr (\vec{\Sigma}_k^{-1}\vec{\mu}_k 
+ \frac{1}{\sigma_\epsilon^2} \biggr [\cdots \sum_{i': z_{u'i'} = k; u'i'\neq ui} 
y_{u'i'} \cdots \biggr ]^T \biggr ).
\end{align}
It is straightforward to obtain required expressions for 
$f_{k}^{-\vec{y}_{t}}(\vec{y}_{t})$,
$f_{uk^{\textrm{new}}}^{-y_{ui}}(y_{ui})$, and
$f_{k^{\textrm{new}}}^{-\vec{y}_{t}}(\vec{y}_{t})$ --
the latter two quantities are given in the Appendix.

\myparagraph{Example 2.} If $H$ is a chain-structured model,
the conditional densities defined by
Eq.~\eqref{Eqn-integrate-phi}
are not available in closed forms, but we can still obtain
exact computation using an algorithm that is akin to the well-known
alpha-beta algorithm in the Hidden Markov model~\citep{Rabiner}.
The running time of such algorithm is proportional to the size
of the graph (i.e., $|V|$). For general graphical models, one can apply
a sum-product algorithm or approximate variational inference
methods~\citep{Wainwright-Jordan-08}.

\begin{figure}[t]
\begin{center}
\begin{tabular}{cc}
\includegraphics[keepaspectratio,width = 0.45\textwidth]{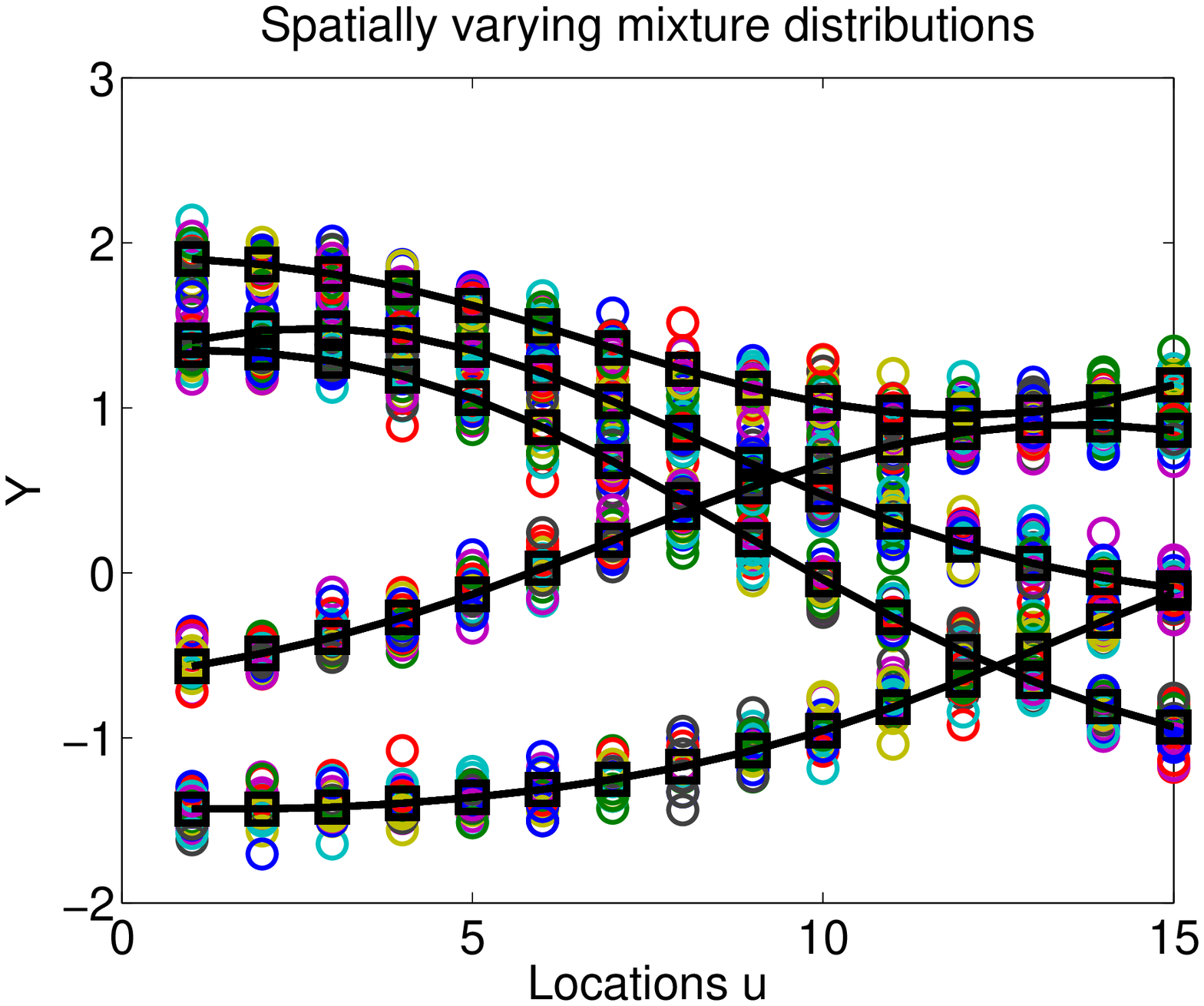}  &
\includegraphics[keepaspectratio,width = 0.45\textwidth]{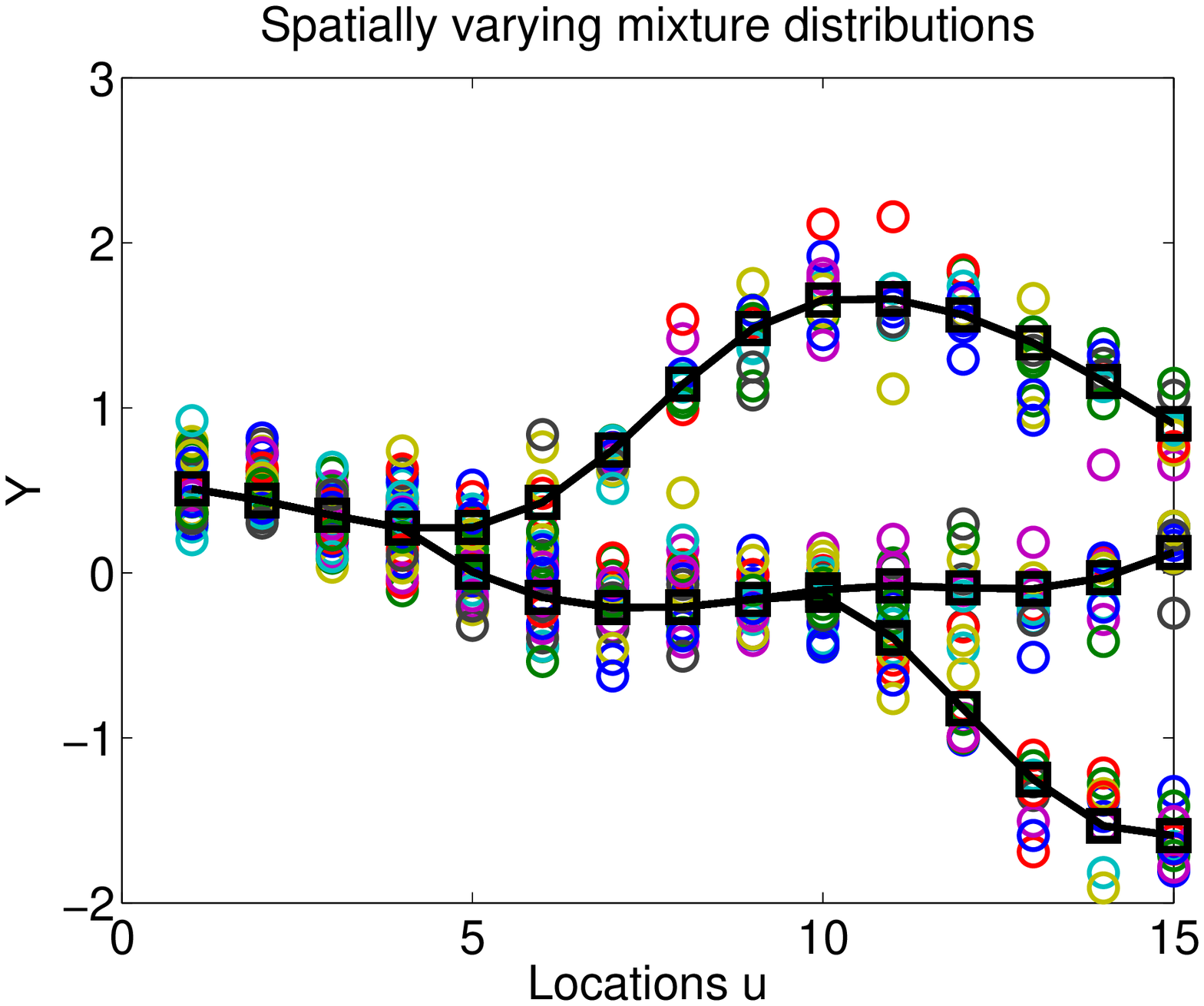} 
\end{tabular}
\end{center}
\caption{Left: Data set A illustrates a simulated problem of tracking
particles organized into clusters, which move in smooth paths. Right: Data set B 
illustrates bifurcating trajectories. In both cases, data are given
not as trajectories, but only as individual points denoted by circles at each $u$.}
\label{Fig-Data-AB}
\end{figure}

\begin{figure}[t]
\begin{center}
\begin{tabular}{cc}
\includegraphics[keepaspectratio,width = 0.25\textwidth]{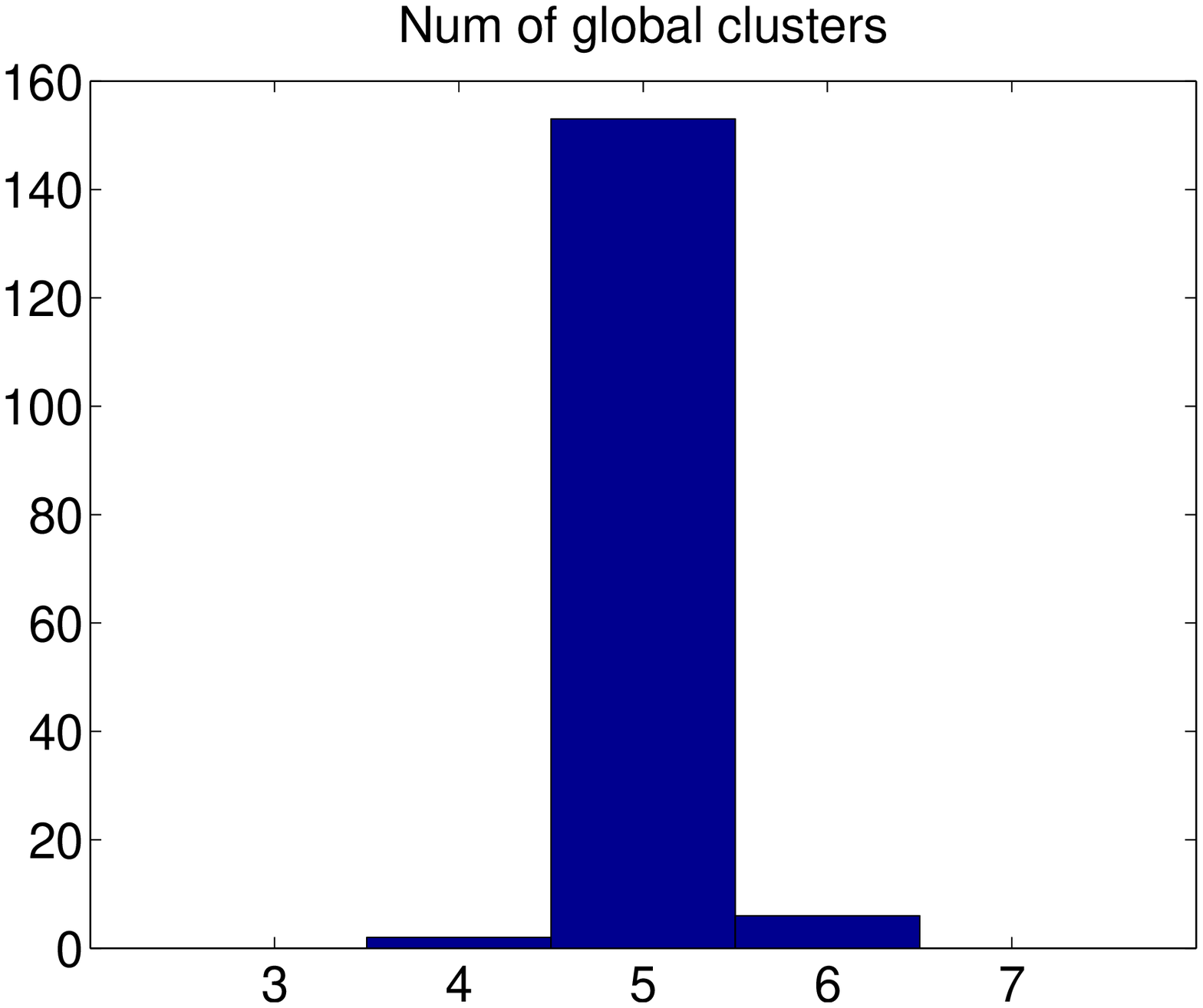} &
\includegraphics[keepaspectratio,width = 0.40\textwidth]{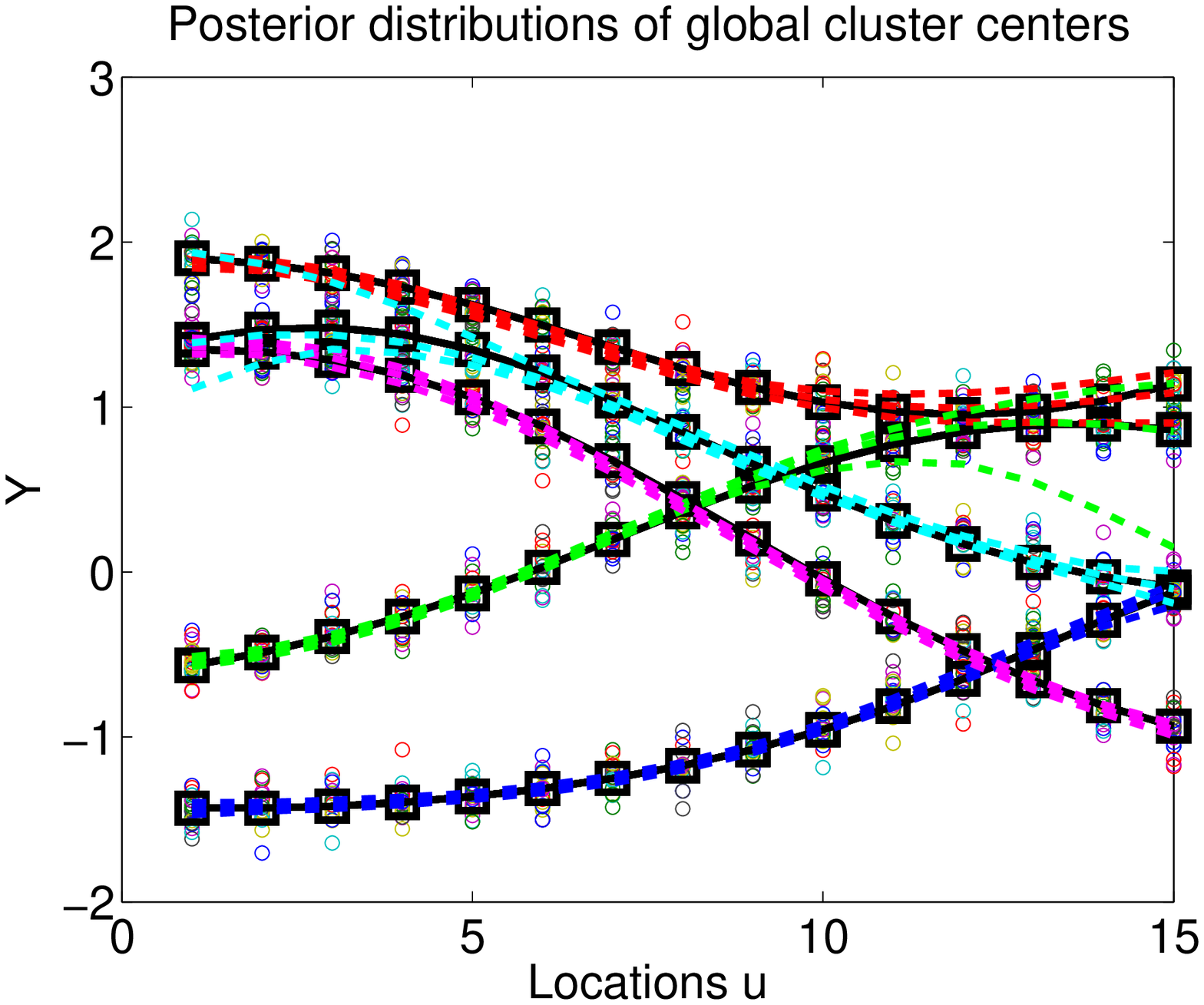} 
\end{tabular}
\end{center}
\caption{Data set A. Left: Posterior distribution of the number of global clusters.
Right: Posterior distributions of the global atoms. Dashed lines denote the mean curve
and (.05,.95) credible intervals.}

\label{Fig-global-A}
\end{figure}

\begin{figure}[t]
\begin{center}
\begin{tabular}{ccc}
\includegraphics[keepaspectratio,width = 0.2\textwidth]{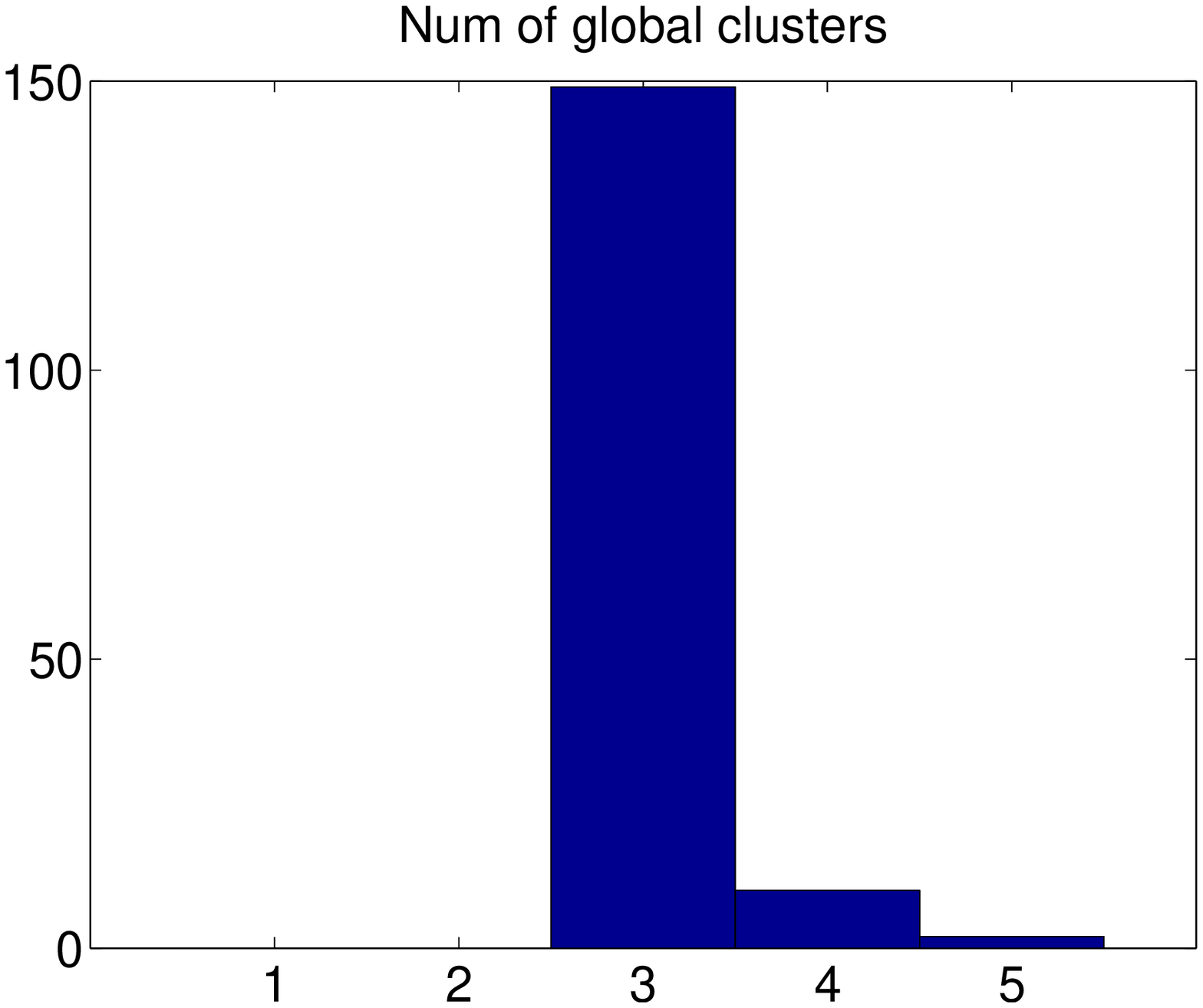} &
\;\;\; &
\includegraphics[keepaspectratio,width = 0.35\textwidth]{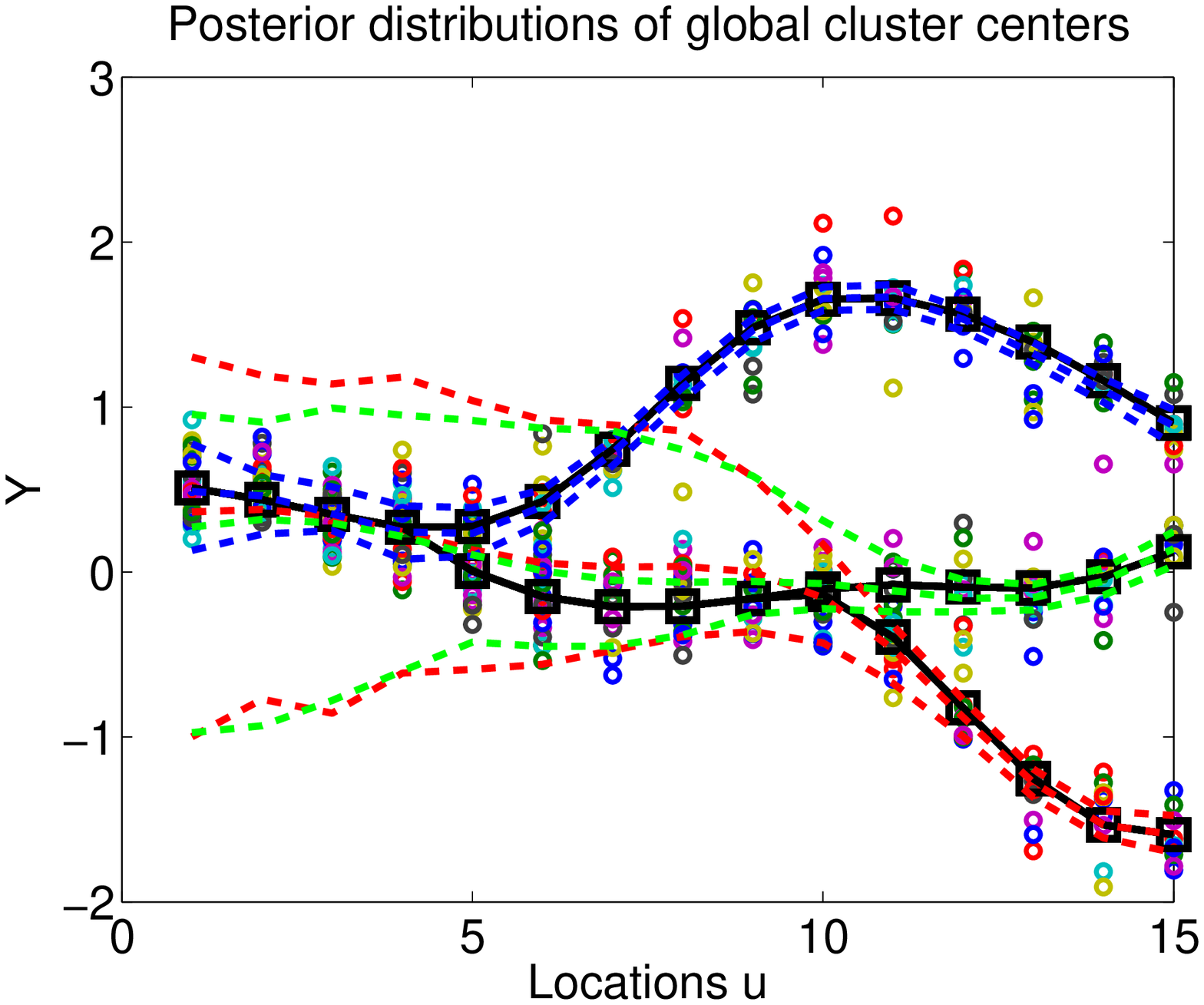}
\end{tabular}
\end{center}
\caption{Data set B. Left: Posterior distribution of the number of global clusters (atoms).
Right: Posterior distributions of the global atoms. Dashed lines denote the mean curve
and the (.05,.95) credible intervals.}
\label{Fig-global-B}
\end{figure}

\begin{figure}[h]
\begin{center}
\begin{tabular}{cc}
\includegraphics[keepaspectratio,width = 0.25\textwidth]{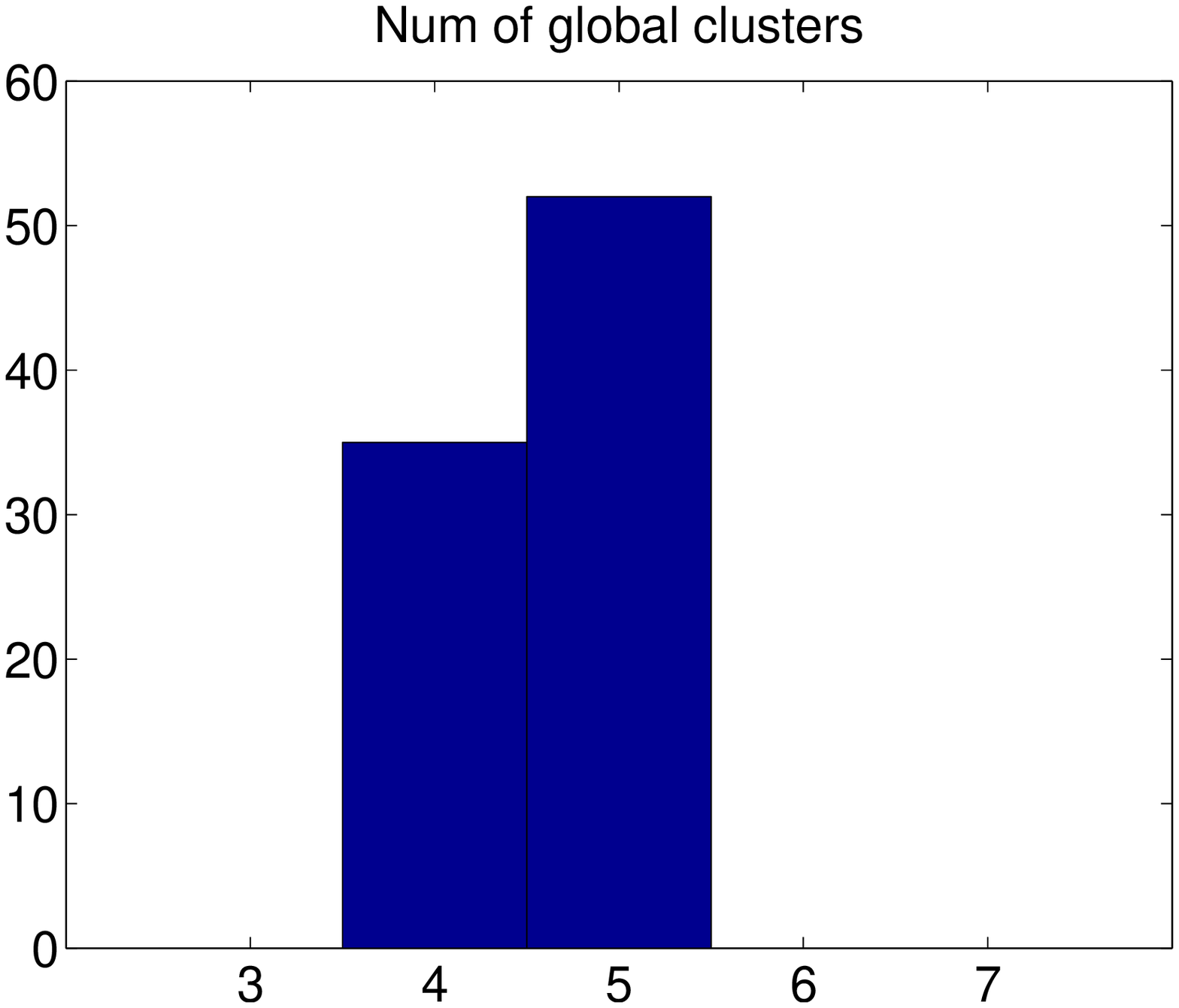} &
\includegraphics[keepaspectratio,width = 0.40\textwidth]{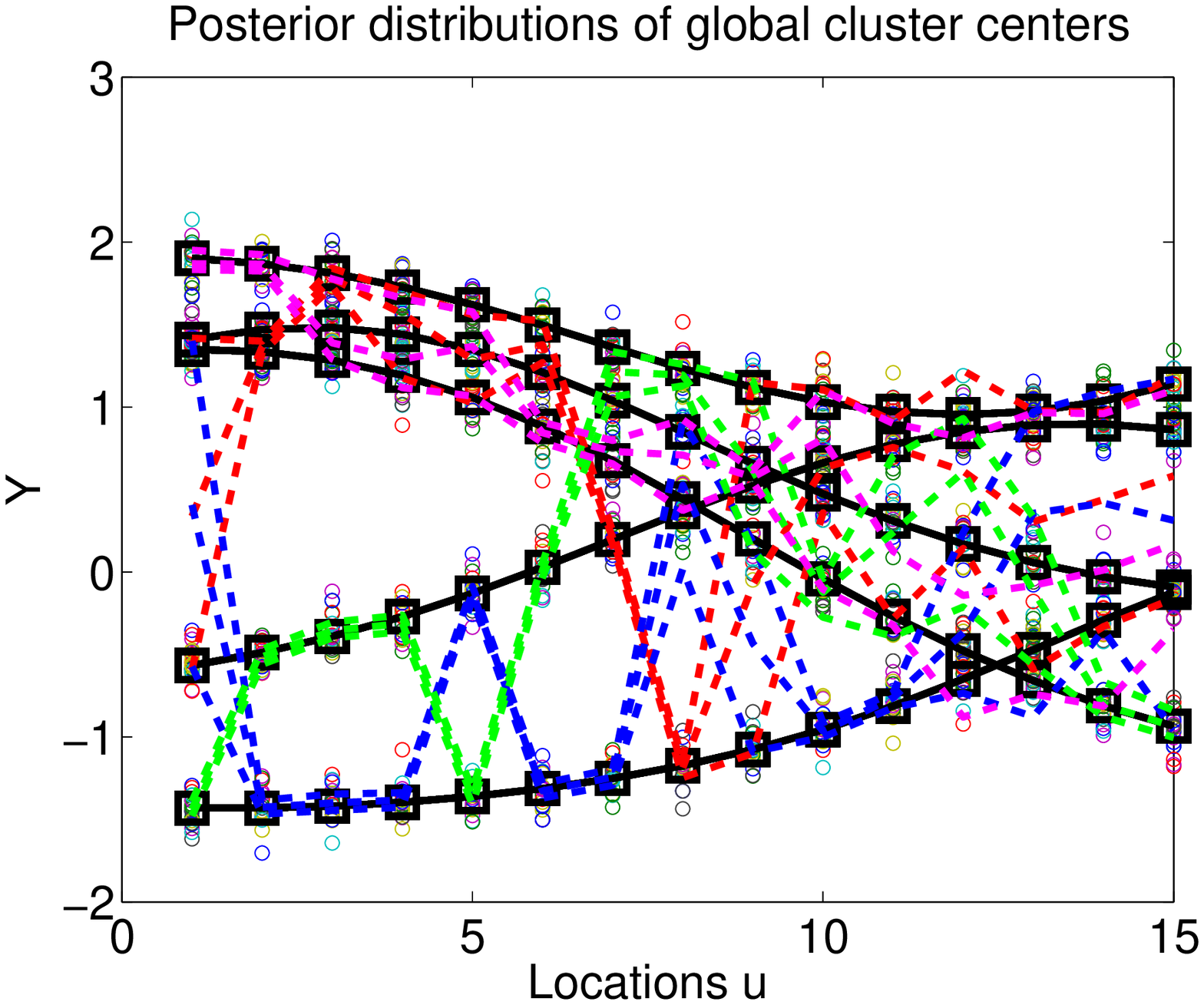} 
\end{tabular}
\end{center}
\caption{Effects of vague prior for $H$ results in weak identifiability of global clusters,
even as the local clusters are identified reasonably well.}
\label{Fig-identify}
\end{figure}

\begin{figure}[h]
\begin{center}
\begin{tabular}{cccc}
\includegraphics[keepaspectratio,width = .20\textwidth]{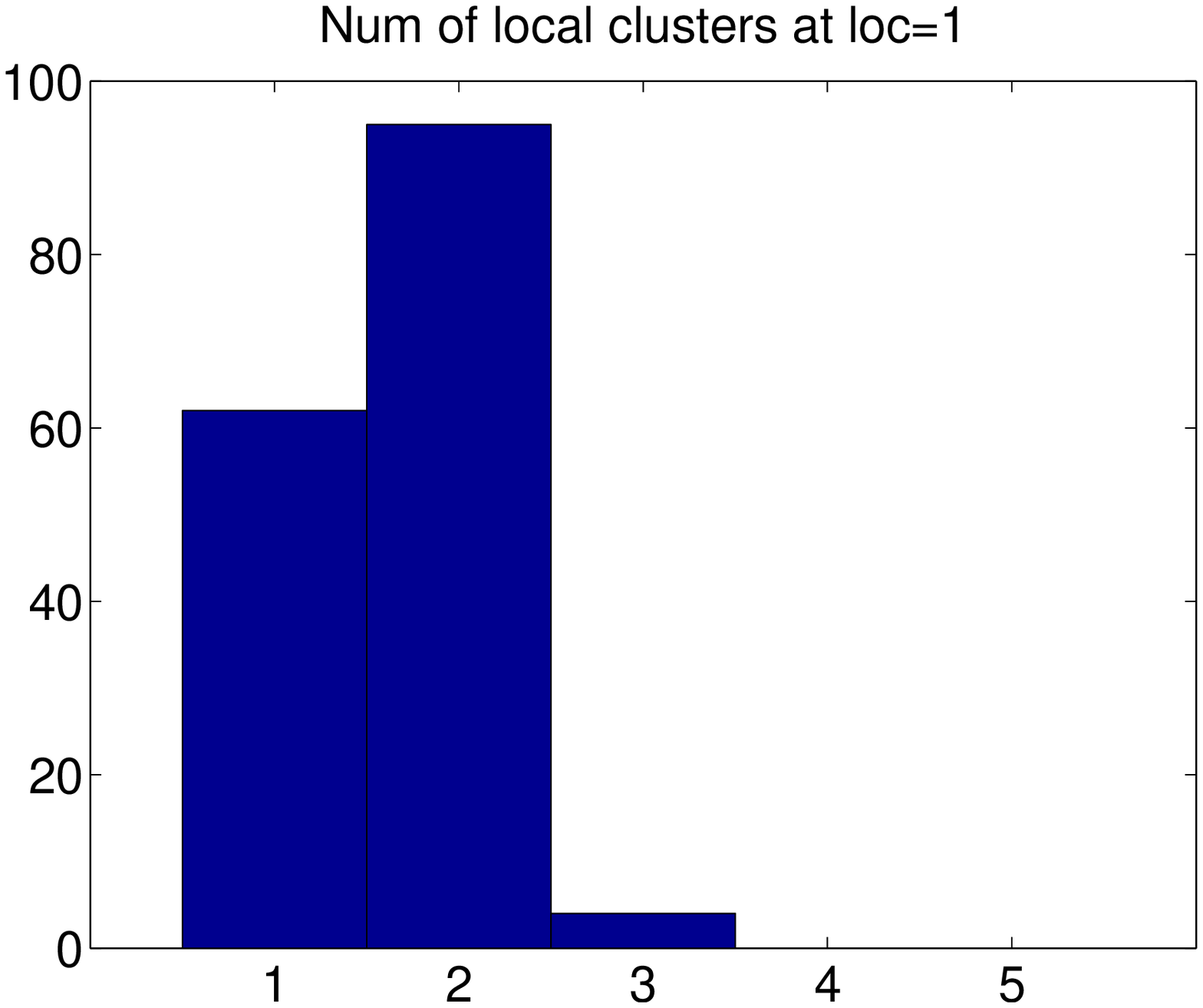} &
\includegraphics[keepaspectratio,width = .20\textwidth]{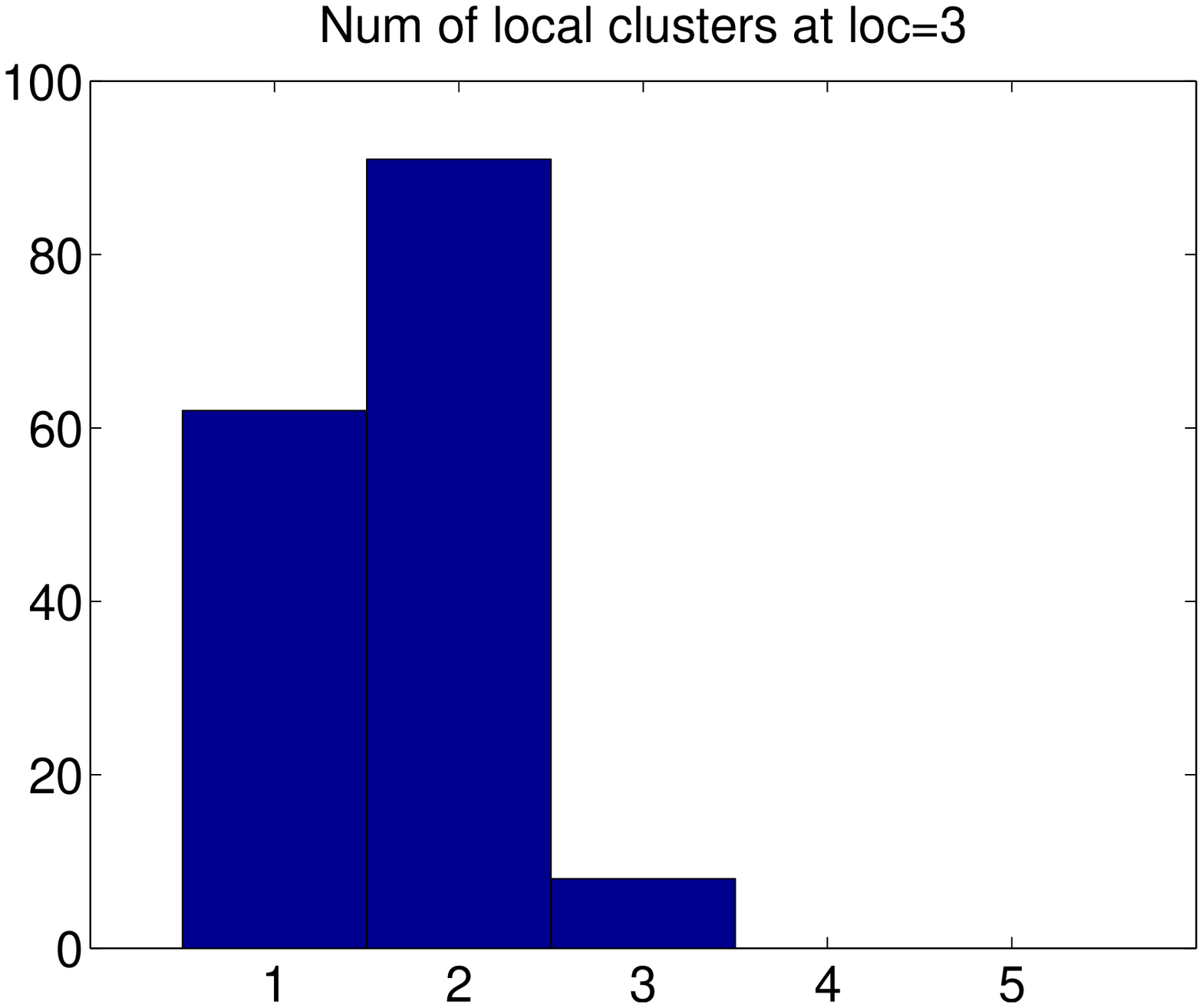} &
\includegraphics[keepaspectratio,width = .20\textwidth]{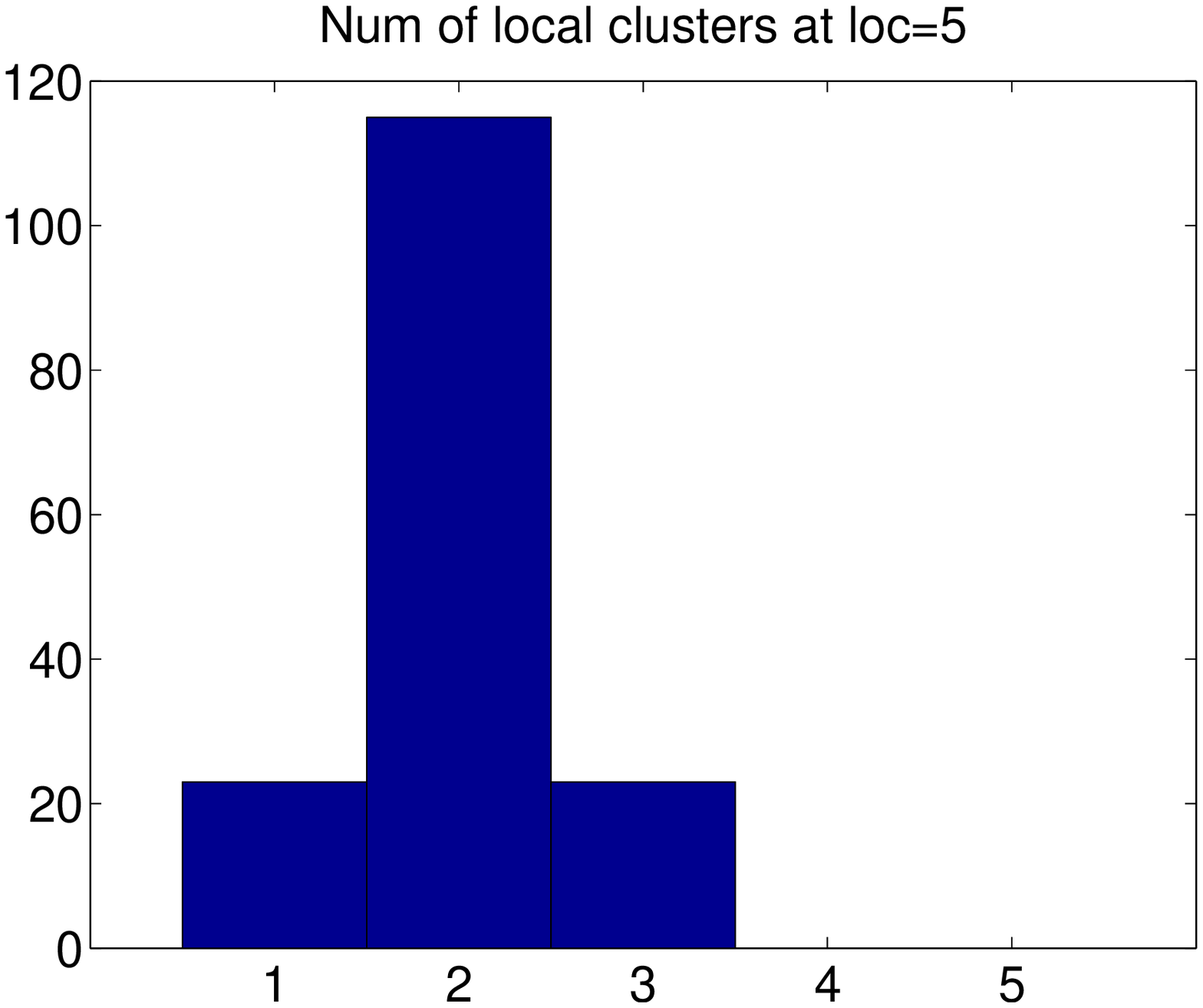} &
\includegraphics[keepaspectratio,width = .20\textwidth]{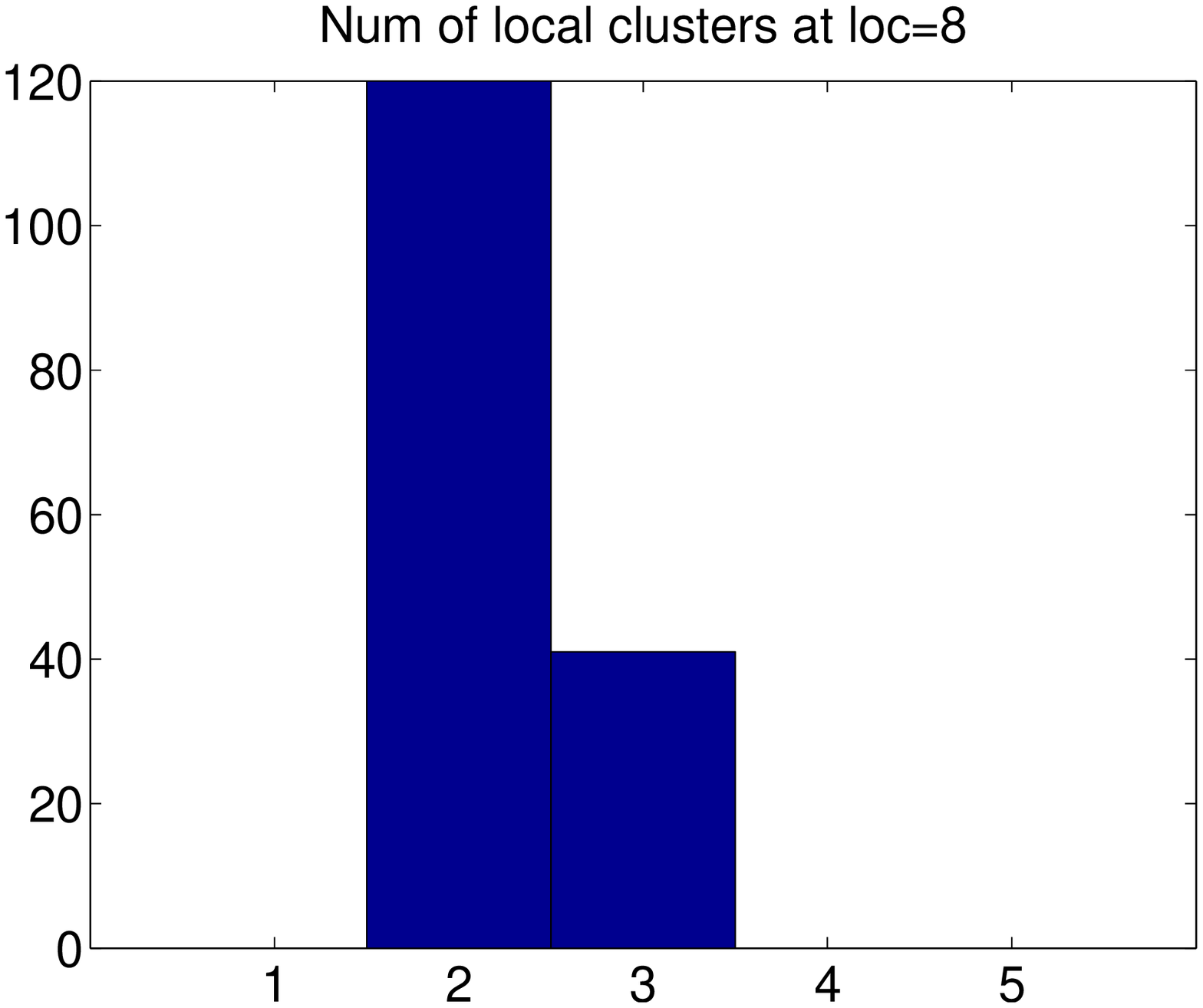} \\
\includegraphics[keepaspectratio,width = .20\textwidth]{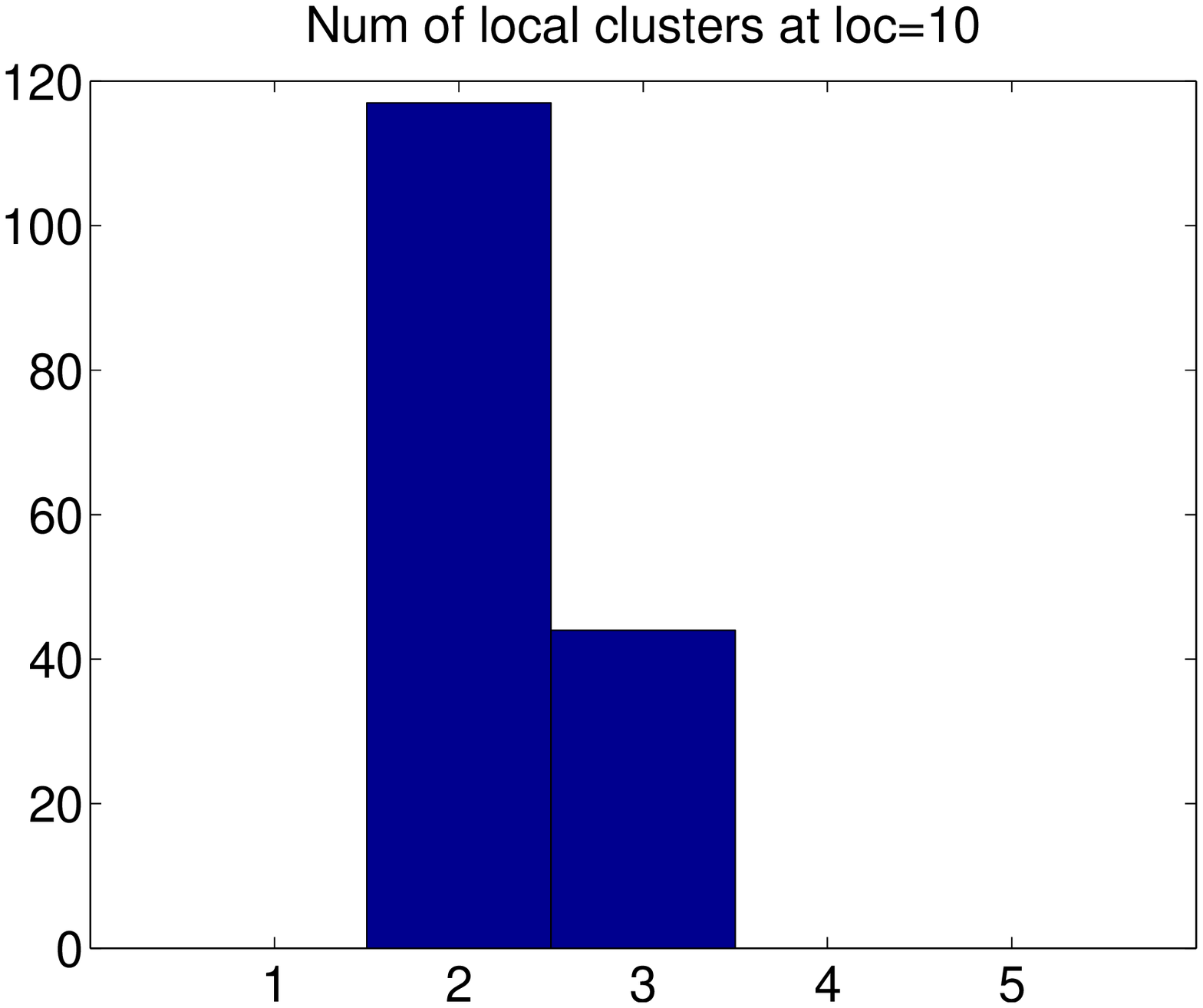} &
\includegraphics[keepaspectratio,width = .20\textwidth]{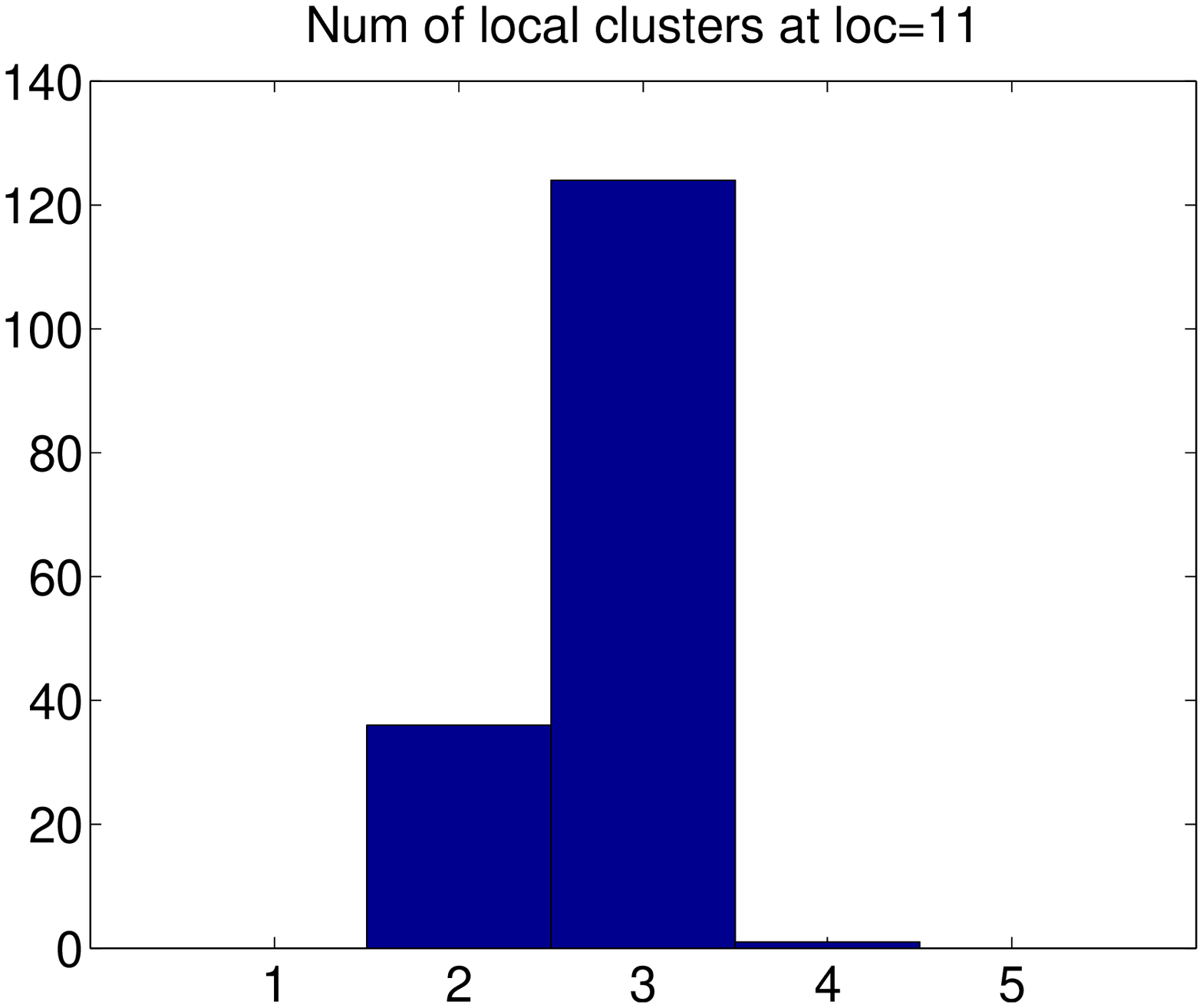} &
\includegraphics[keepaspectratio,width = .20\textwidth]{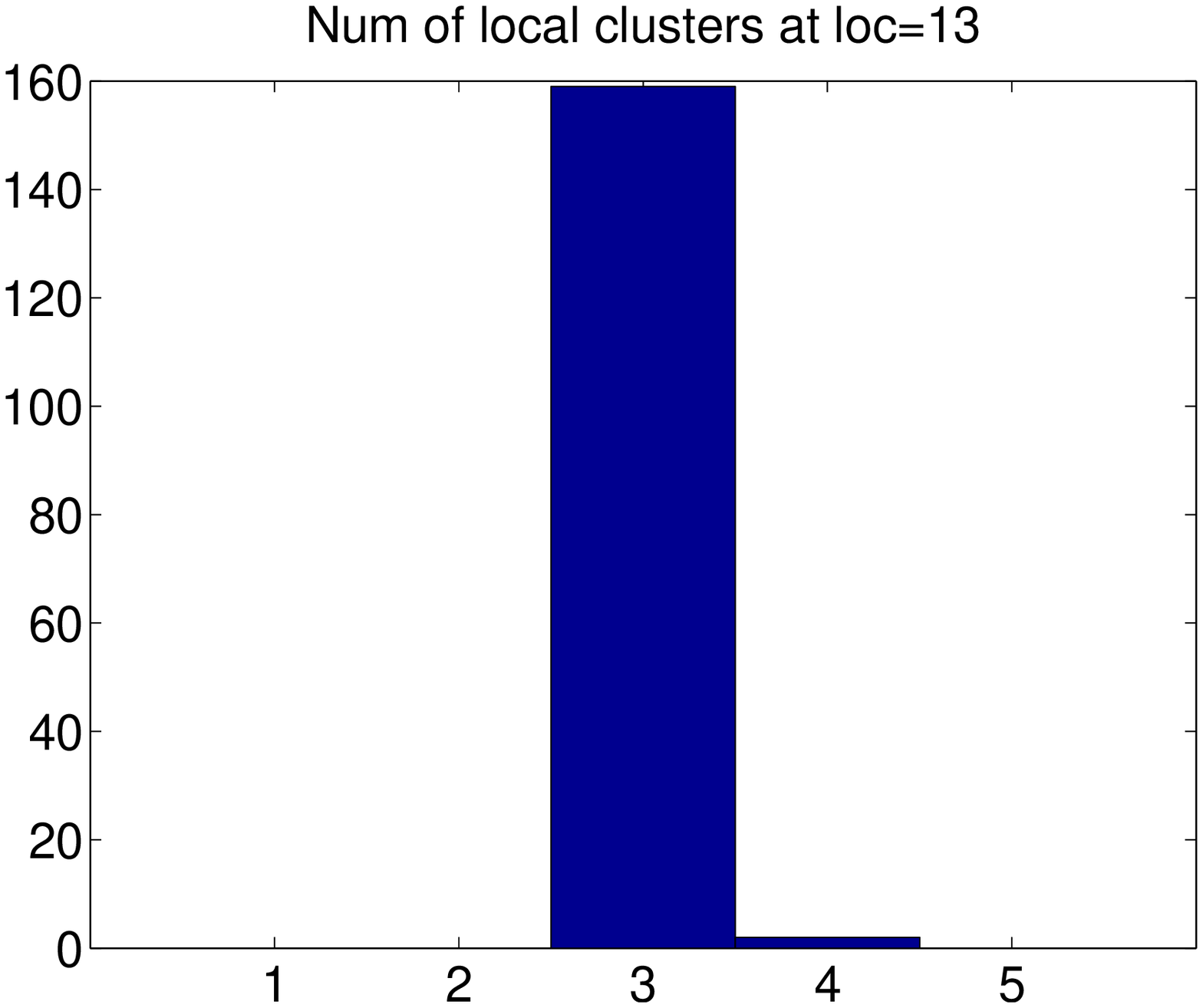} &
\includegraphics[keepaspectratio,width = .20\textwidth]{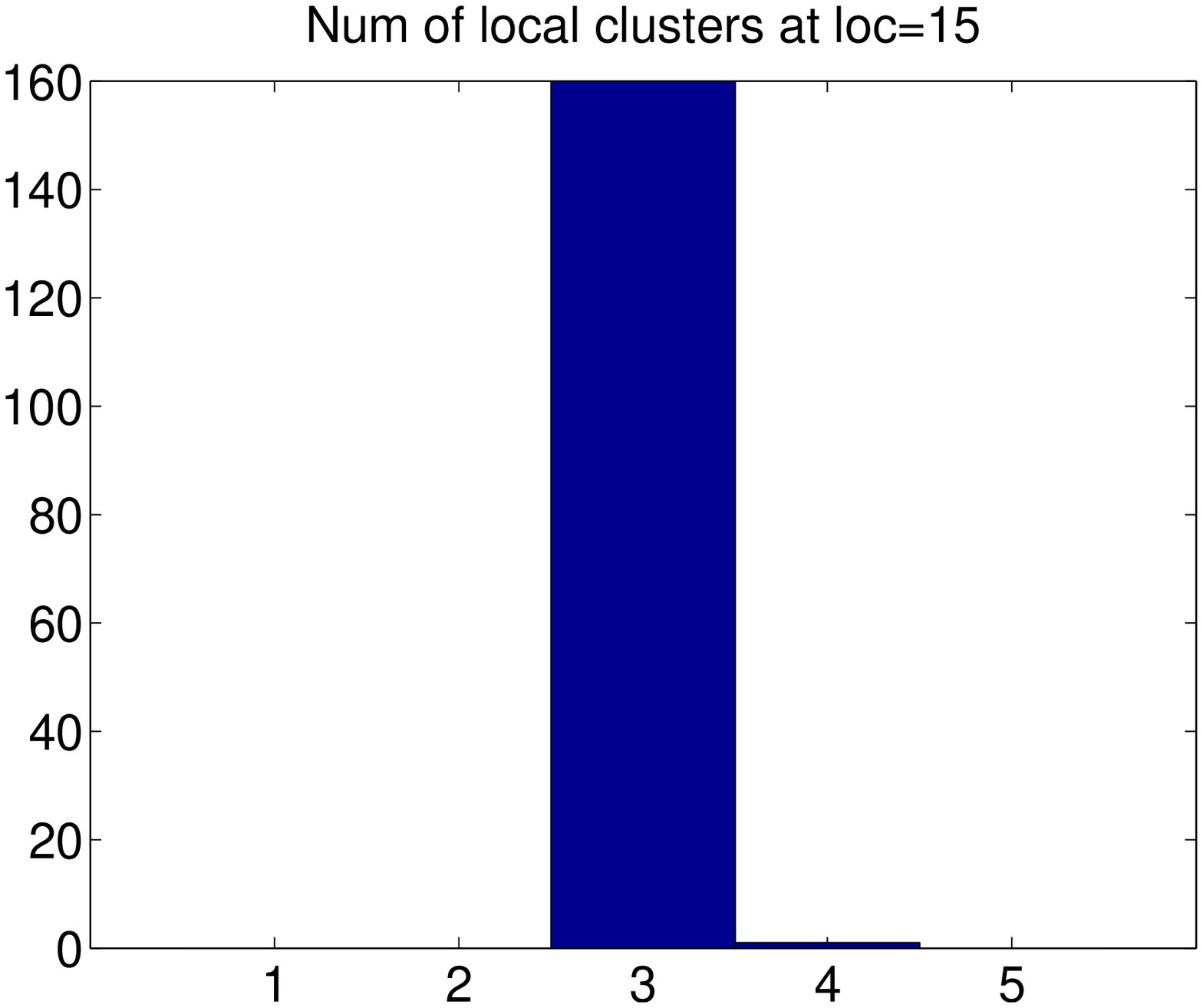}
\end{tabular}
\end{center}
\caption{Data set B: Posterior distribution of the number of local clusters associating with
different group index (location) $u$.}
\label{Fig-local-B}
\end{figure}

\comment{
\begin{figure}
\begin{center}
\begin{tabular}{ccc}
\includegraphics[keepaspectratio,width = 0.33\textwidth]{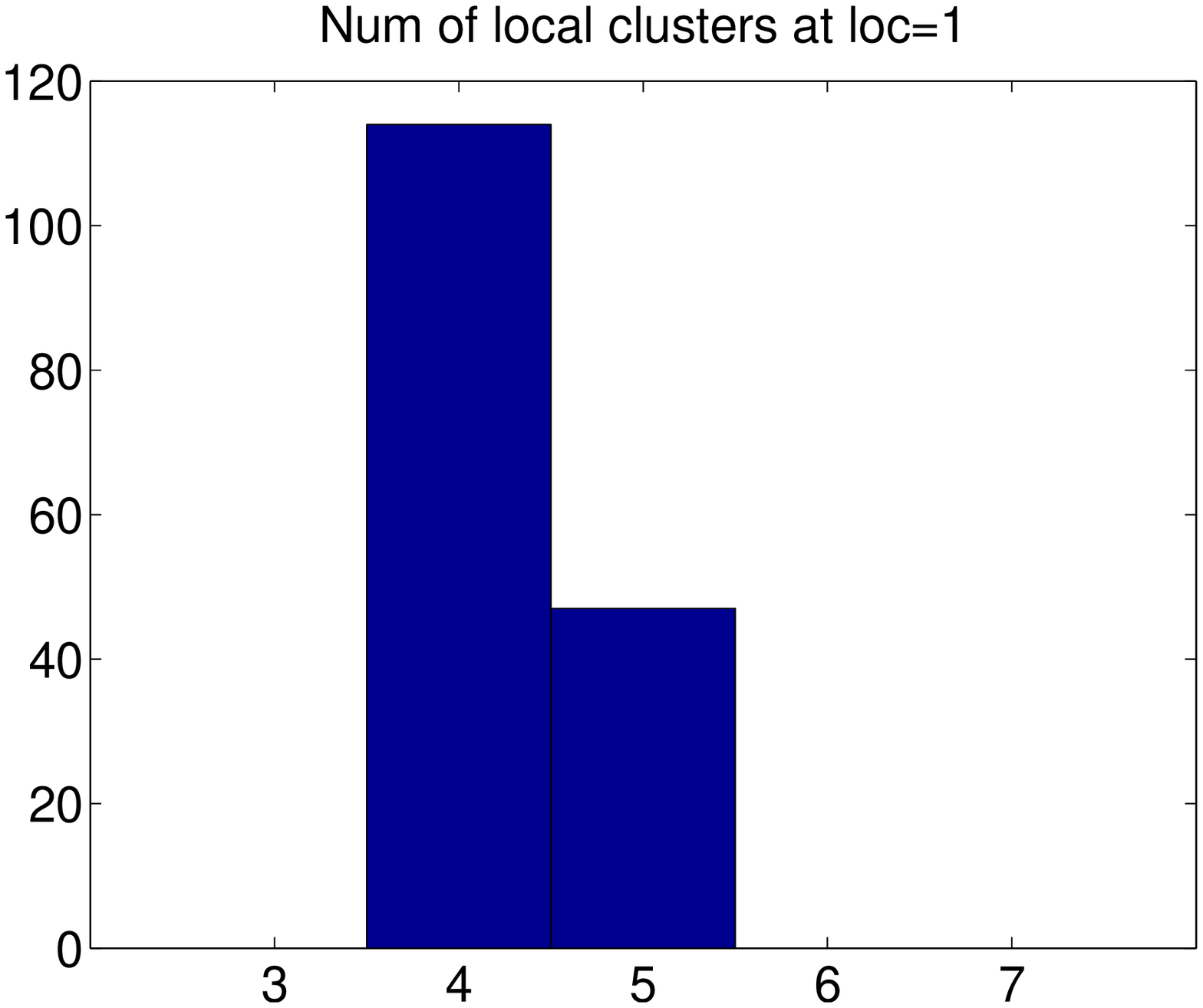} &
\includegraphics[keepaspectratio,width = 0.33\textwidth]{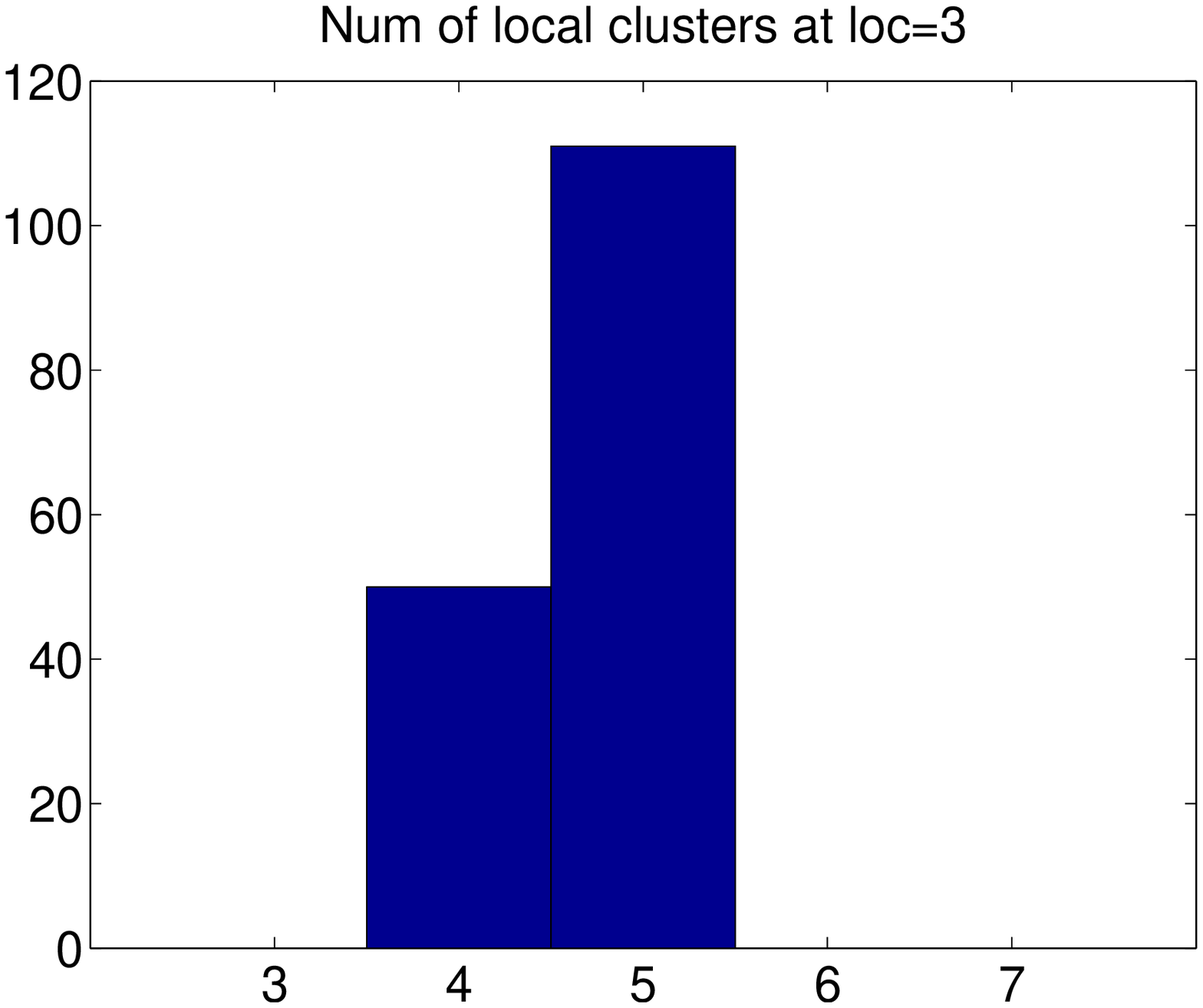} &
\includegraphics[keepaspectratio,width = 0.33\textwidth]{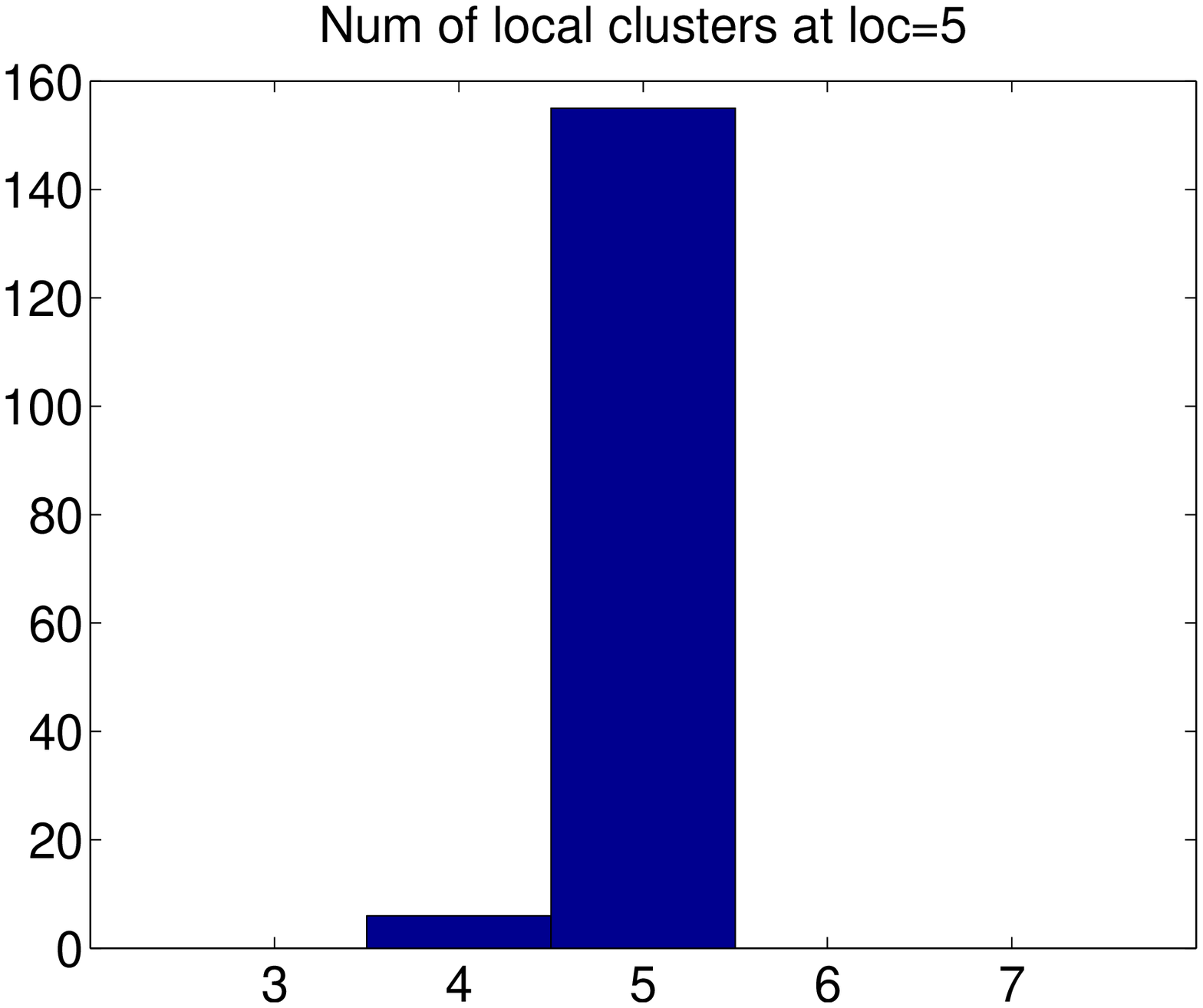} \\
\includegraphics[keepaspectratio,width = 0.33\textwidth]{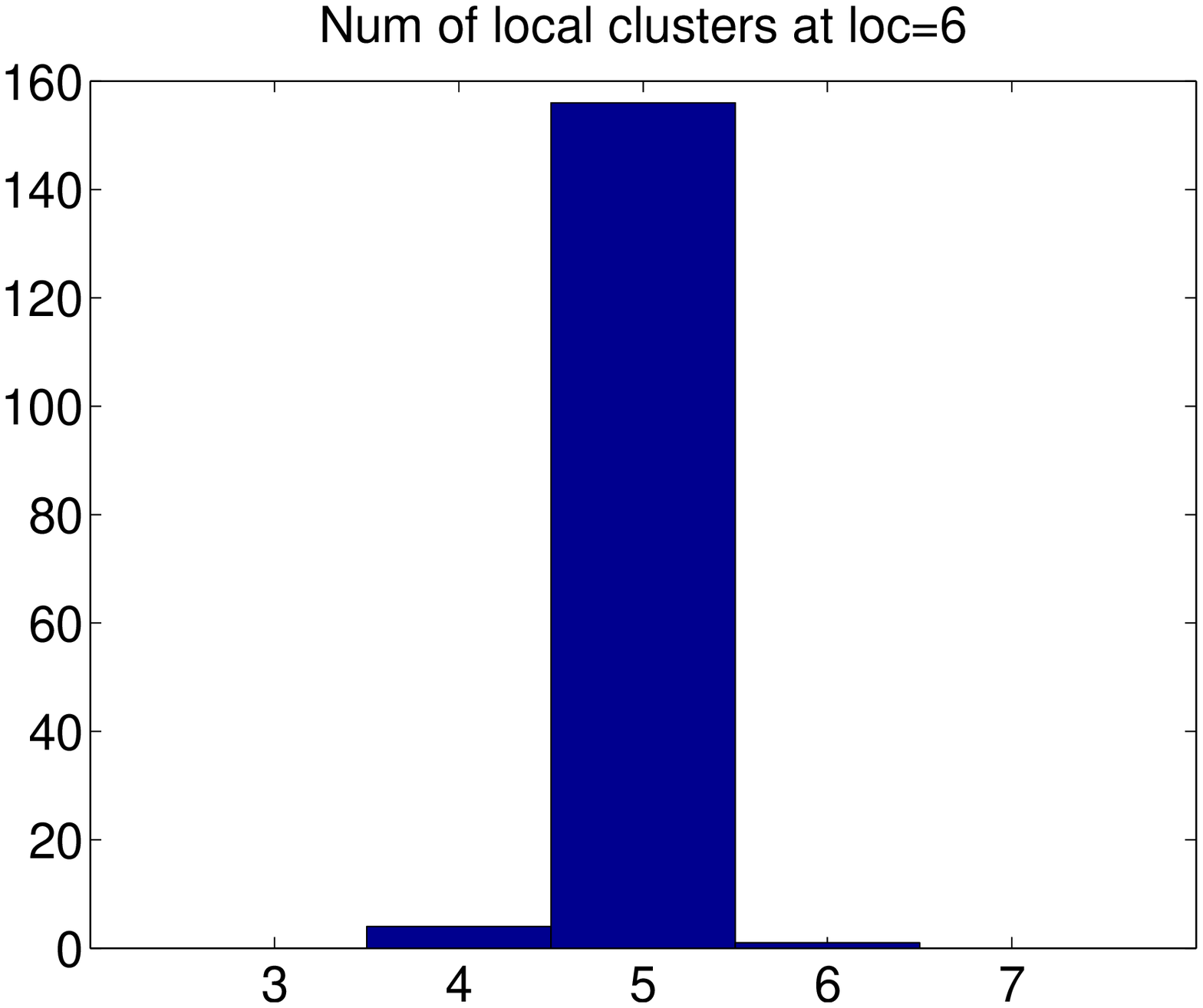} &
\includegraphics[keepaspectratio,width = 0.33\textwidth]{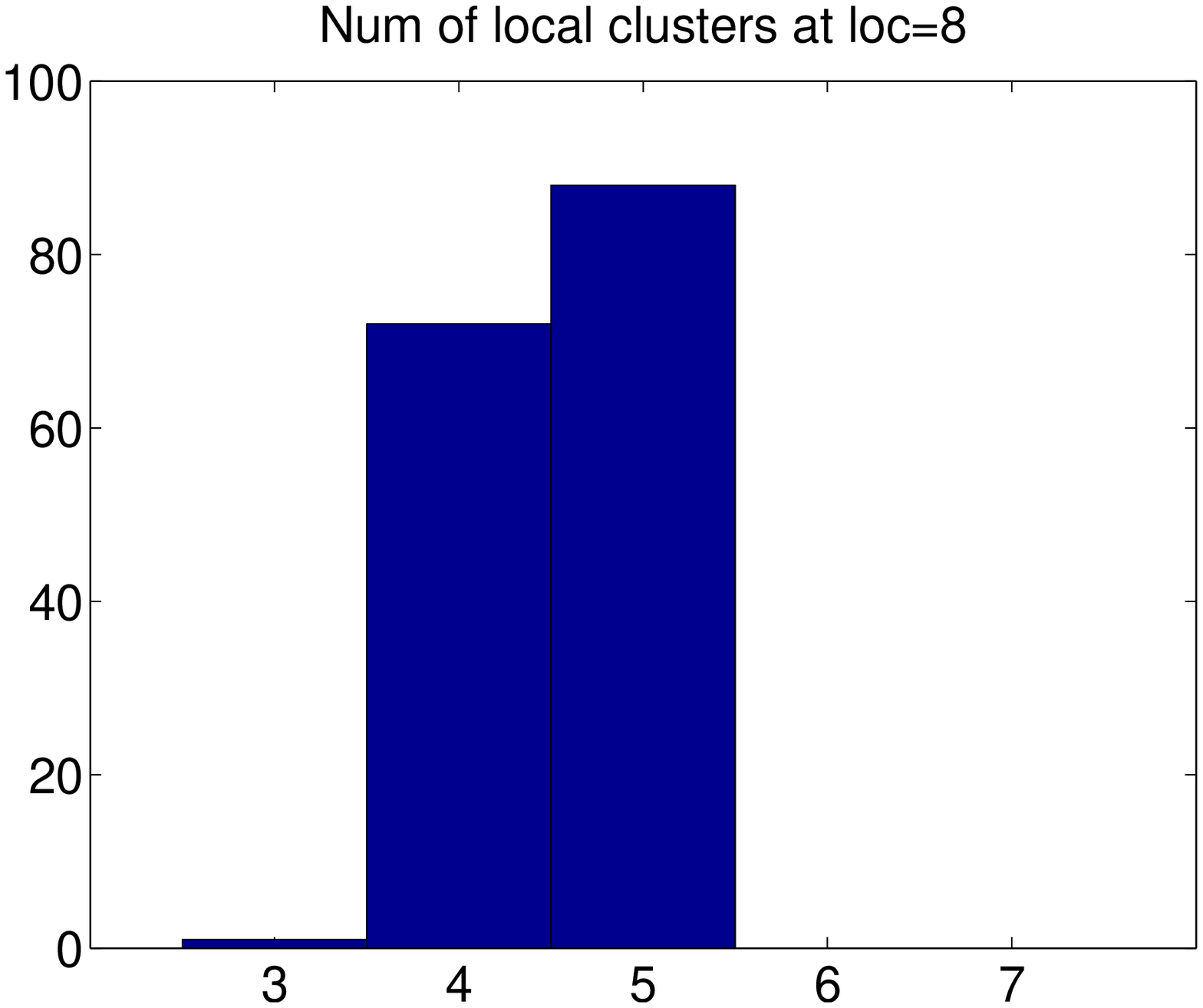} &
\includegraphics[keepaspectratio,width = 0.33\textwidth]{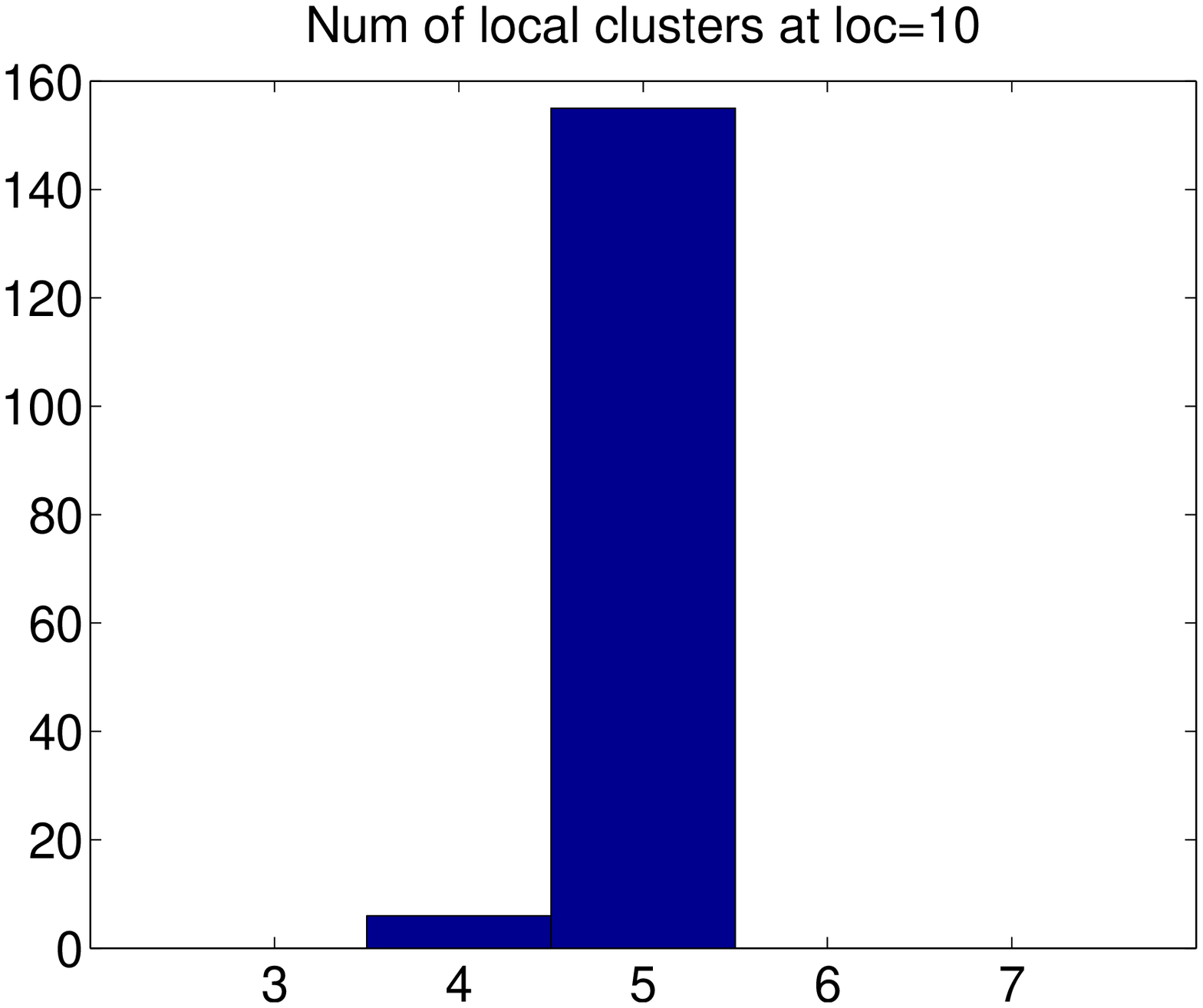} \\
\includegraphics[keepaspectratio,width = 0.33\textwidth]{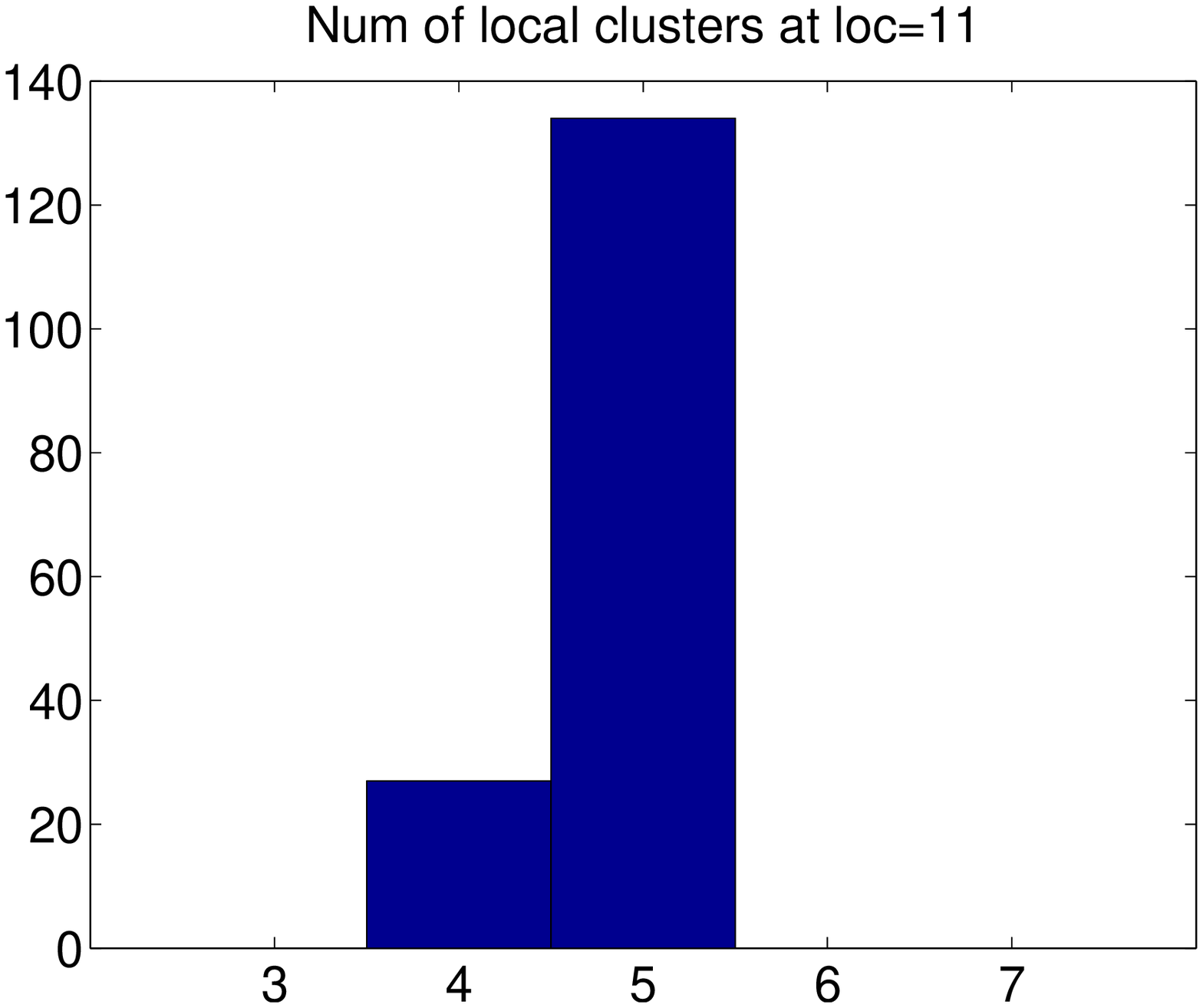} &
\includegraphics[keepaspectratio,width = 0.33\textwidth]{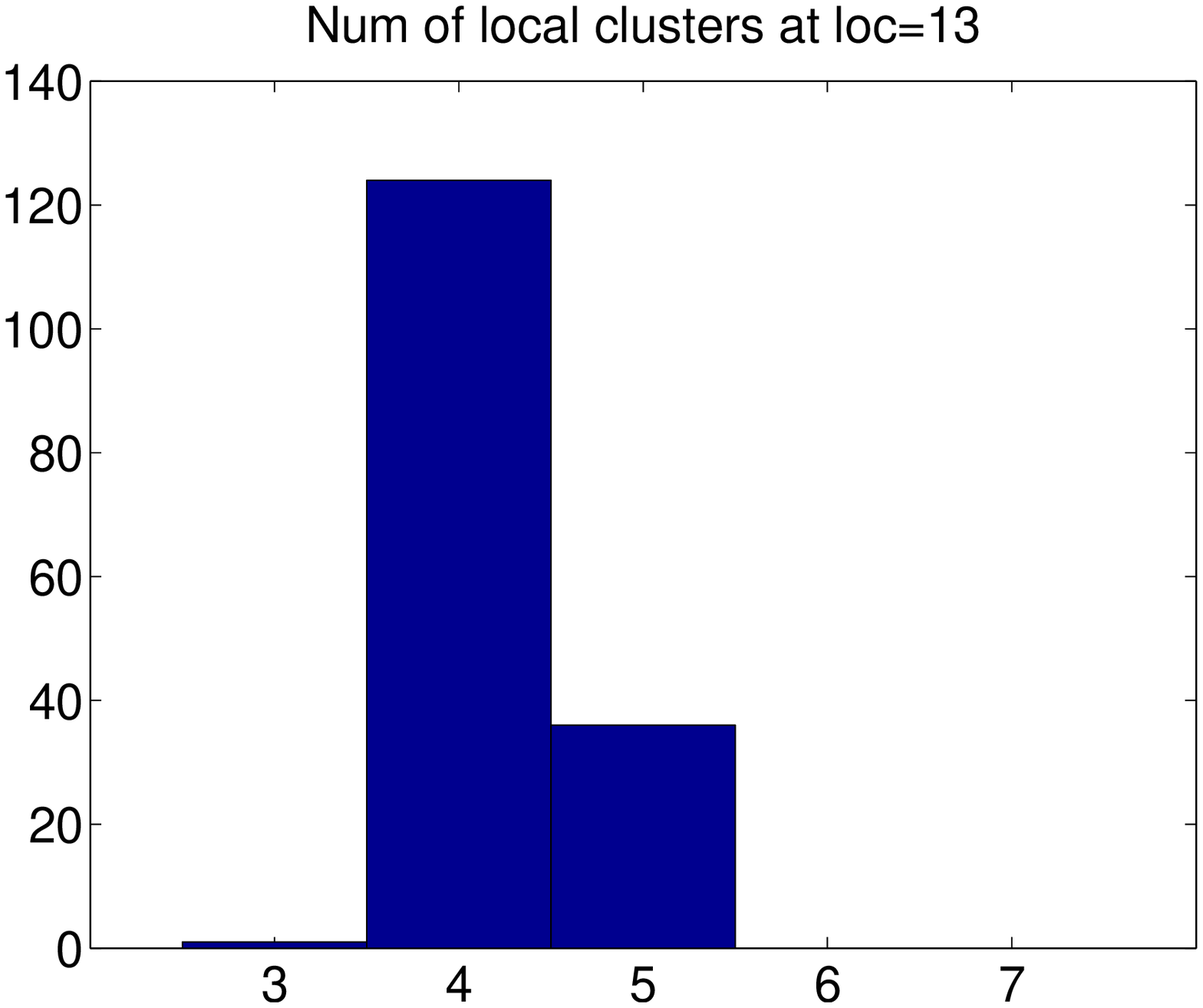} &
\includegraphics[keepaspectratio,width = 0.33\textwidth]{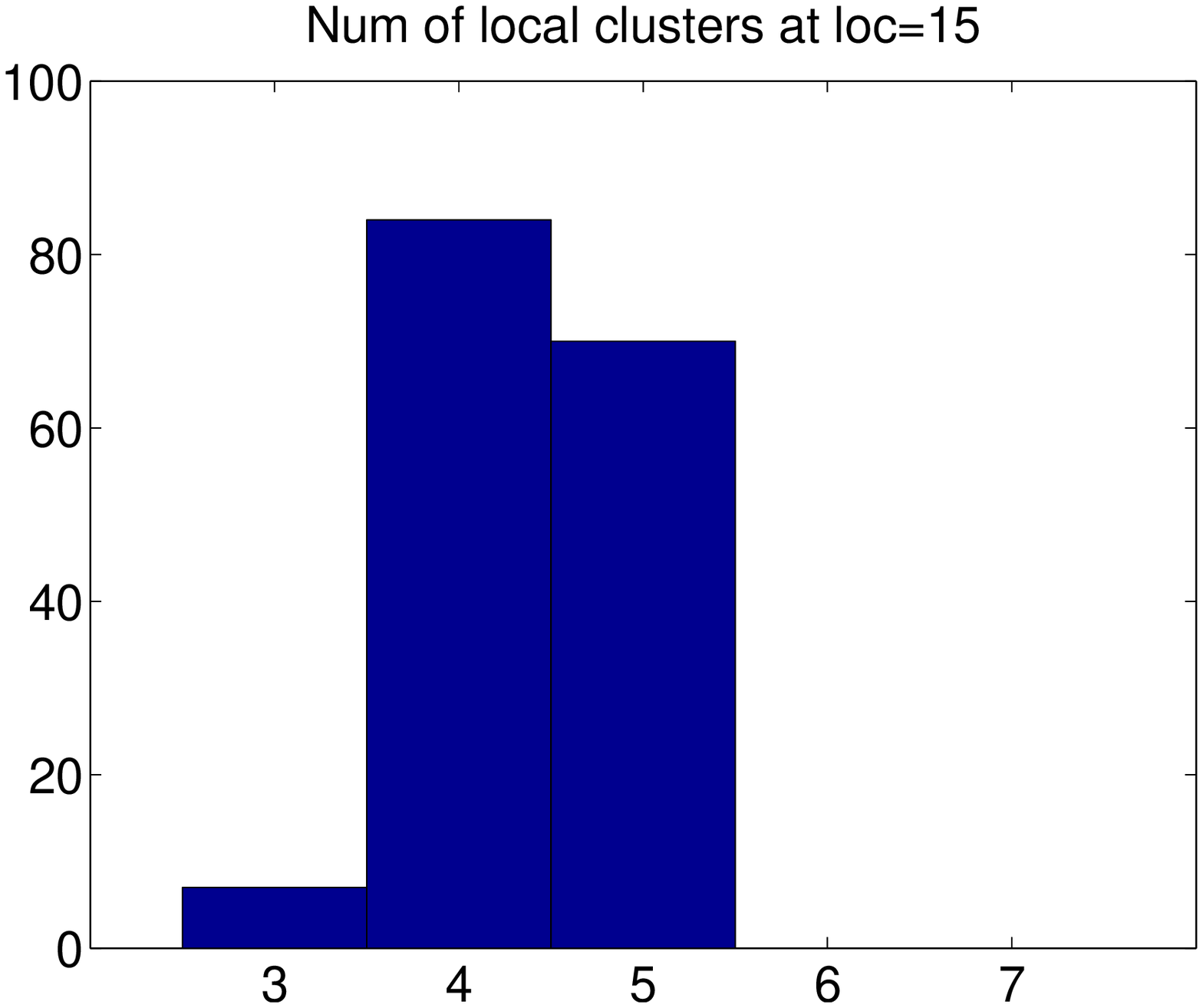}
\end{tabular}
\end{center}
\caption{Data set A: Posterior distribution of the number of local clusters associating with
different group index (location) $u$.}
\label{Fig-local-A}
\end{figure}
}

\begin{figure}[t]
\begin{center}
\begin{tabular}{c}
\includegraphics[keepaspectratio,width = 0.40\textwidth]{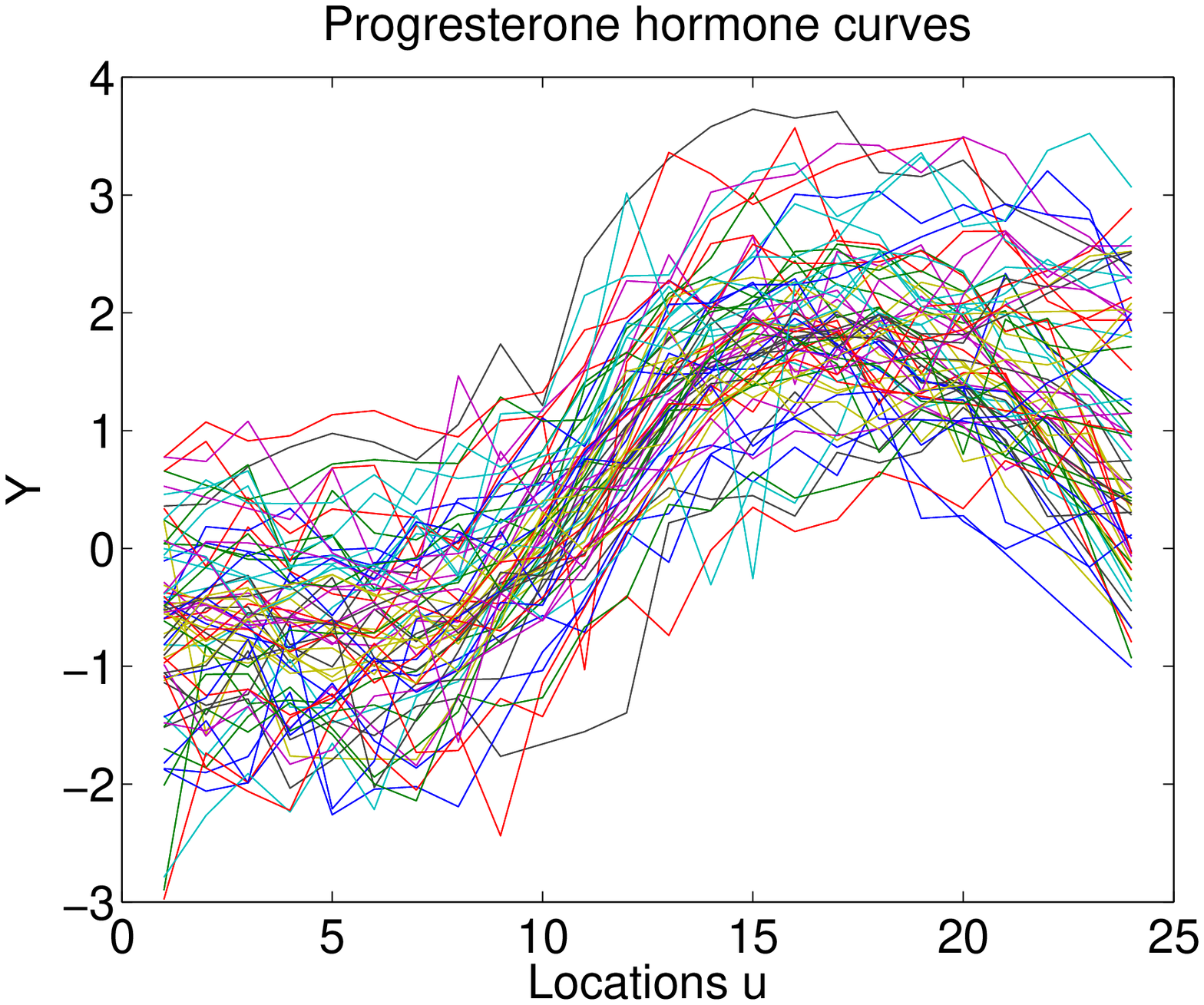} 
\end{tabular}
\end{center}
\caption{Progeresterone hormone curves.}
\label{Fig-Data-PGD}
\end{figure}

\begin{figure}[t]
\begin{center}
\begin{tabular}{cc}
\includegraphics[keepaspectratio,width = 0.40\textwidth]{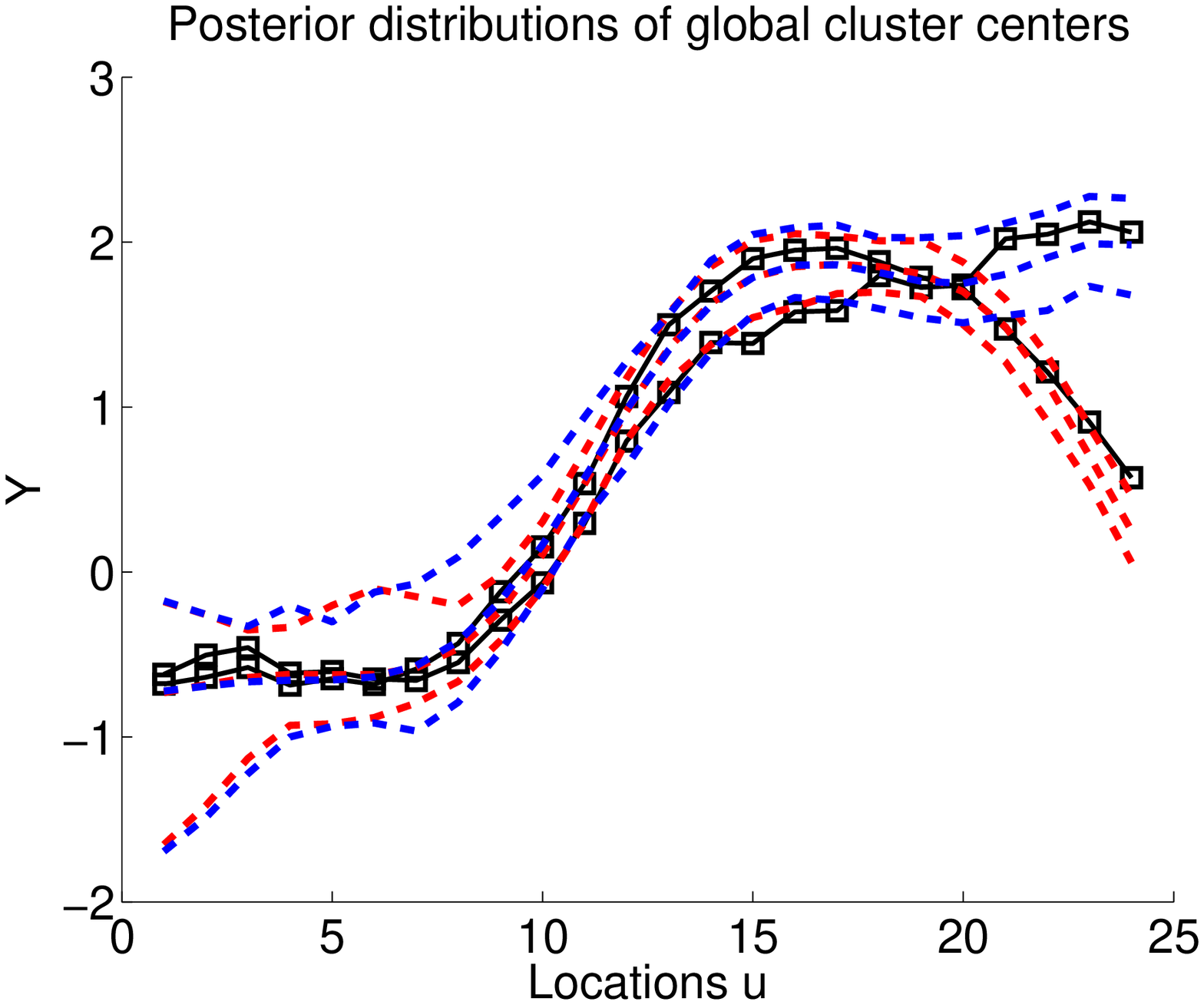}  &
\includegraphics[keepaspectratio,width = 0.40\textwidth]{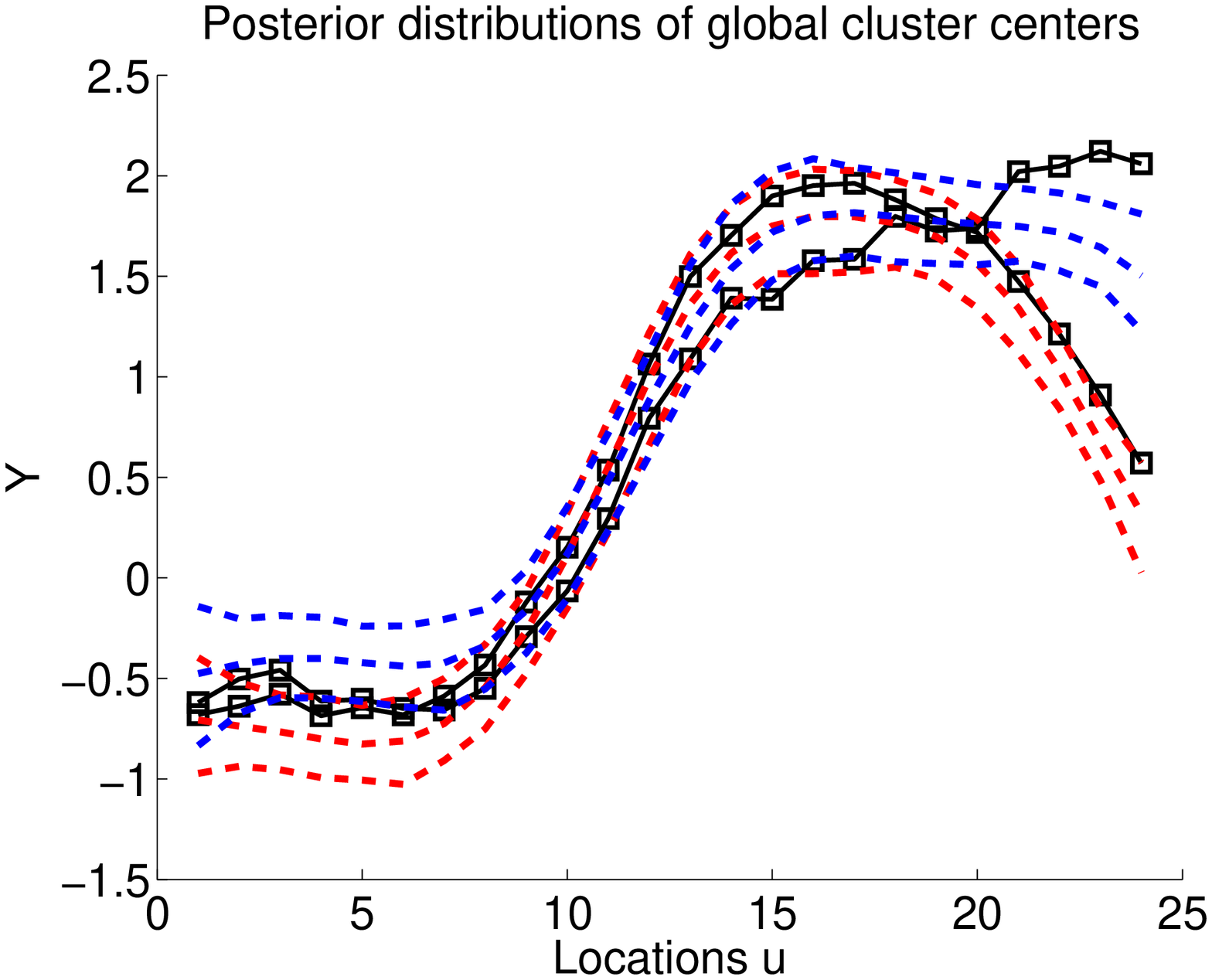}
\end{tabular}
\end{center}
\caption{Clustering results using the nHDP mixture model (Left), and the hybrid-DP of~\cite{Petrone-etal-09}
(Right). Mean and credible intervals
of global clusters (in dashed lines) are compared to sample mean curves of the contraceptive
group and no contraceptive group in black solid with square markers.}
\label{Fig-Data-PGD-dlp}
\end{figure}

\begin{figure}[t]
\begin{center}
\begin{tabular}{cc}
\includegraphics[keepaspectratio,width = 0.45\textwidth]{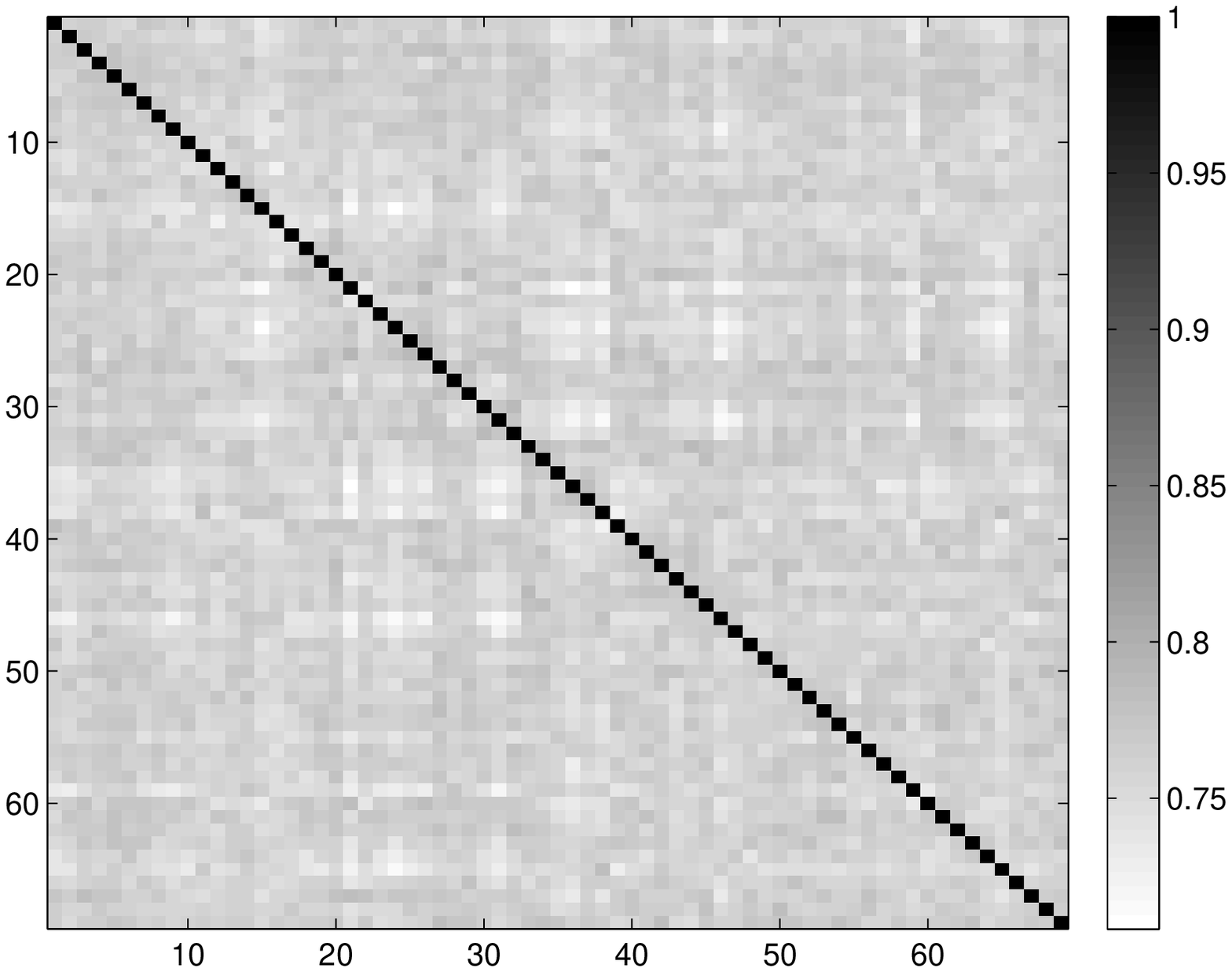}  &
\includegraphics[keepaspectratio,width = 0.45\textwidth]{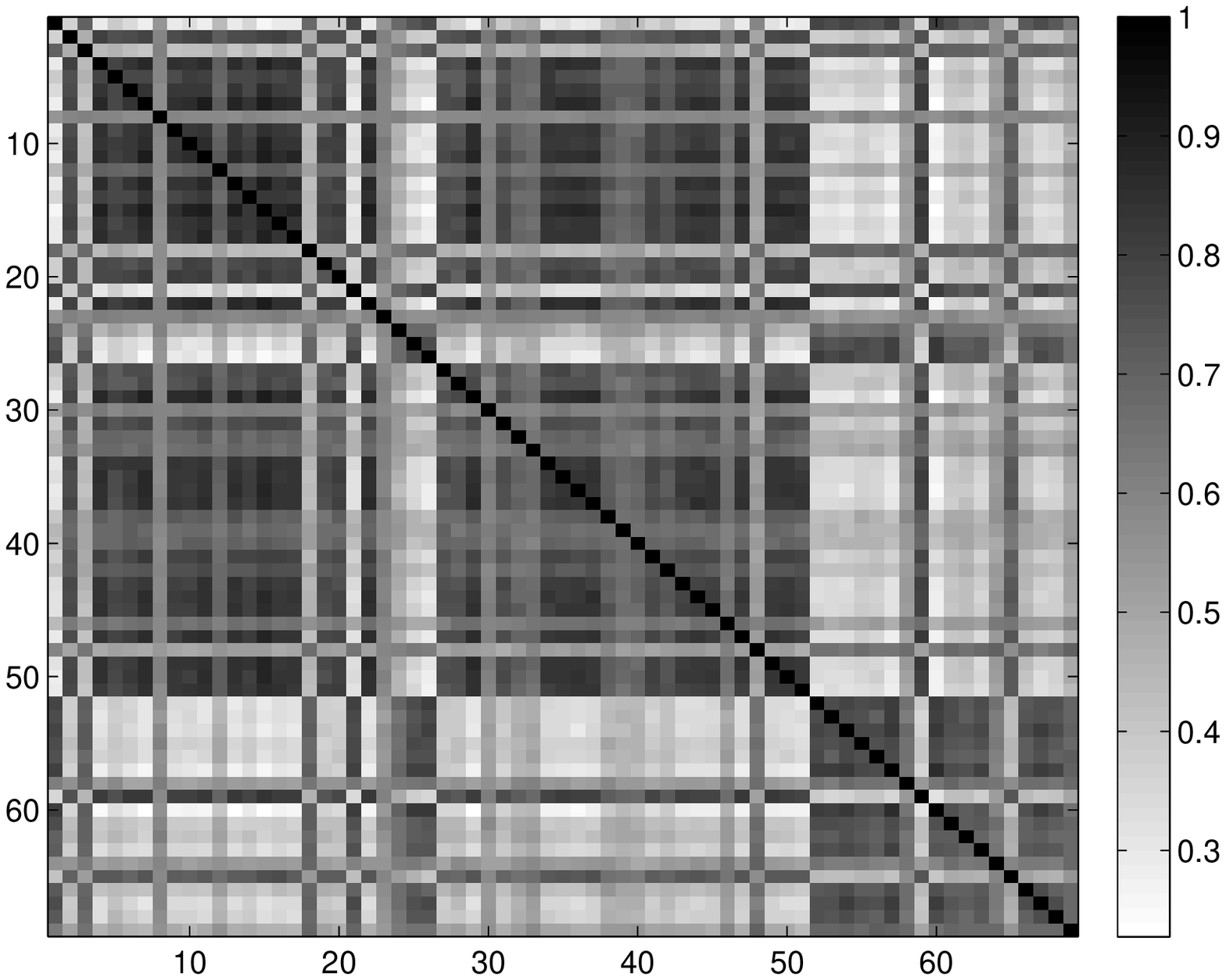} 
\end{tabular}
\end{center}
\caption{Pairwise comparison of individual hormone curves. Each entry
in the heatmap depicts the posterior probability that the two curves
share the same \emph{local} clusters, averaged over a fixed interval
([1,20] in the left, and [21,24] in the right figure) in the
menstrual cycle.}
\label{Fig-Heat}
\end{figure}

\begin{figure}[h]
\begin{center}
\begin{tabular}{ccccc}
\includegraphics[keepaspectratio,width = 0.18\textwidth]{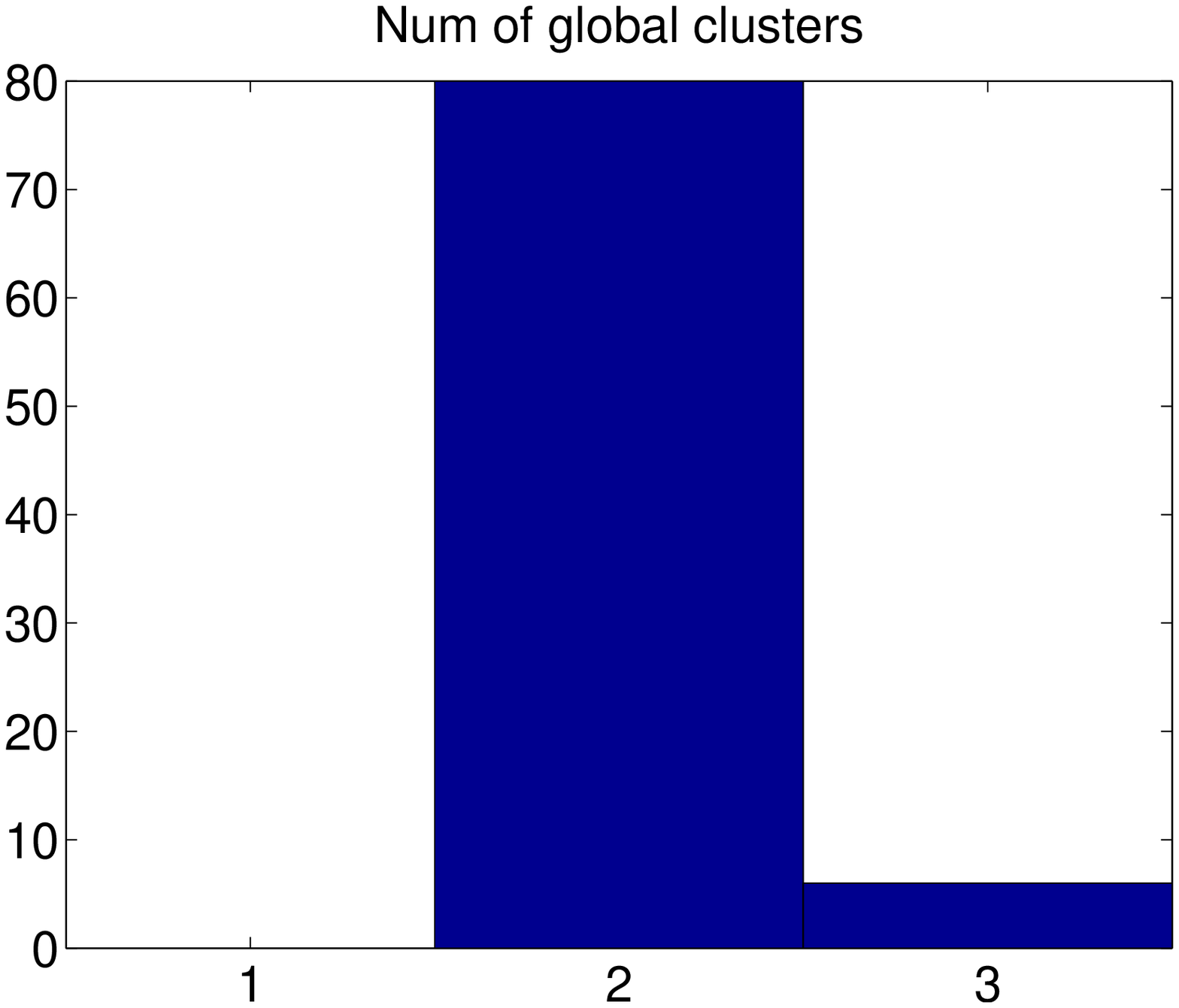} &
\includegraphics[keepaspectratio,width = 0.18\textwidth]{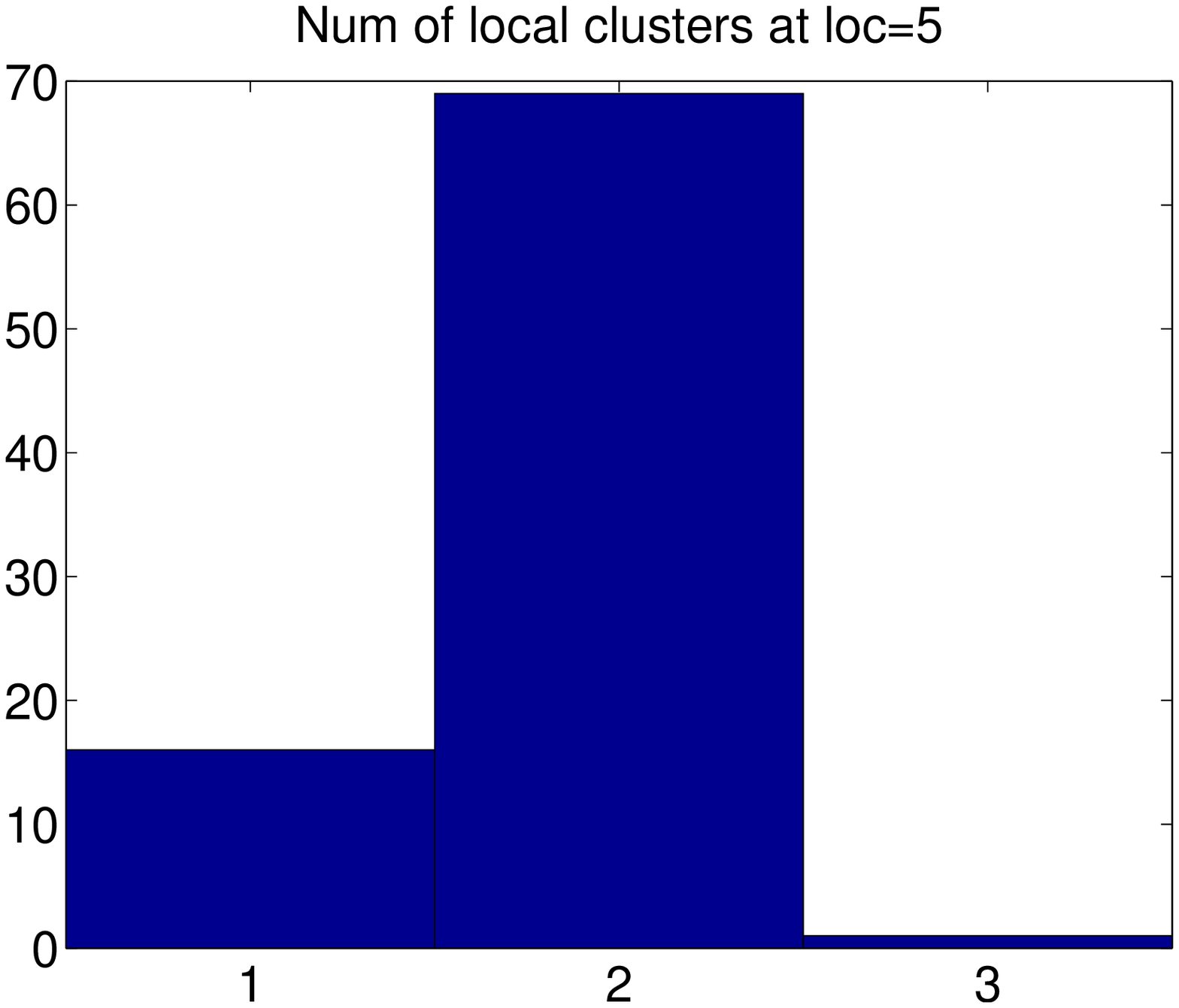} &
\includegraphics[keepaspectratio,width = 0.18\textwidth]{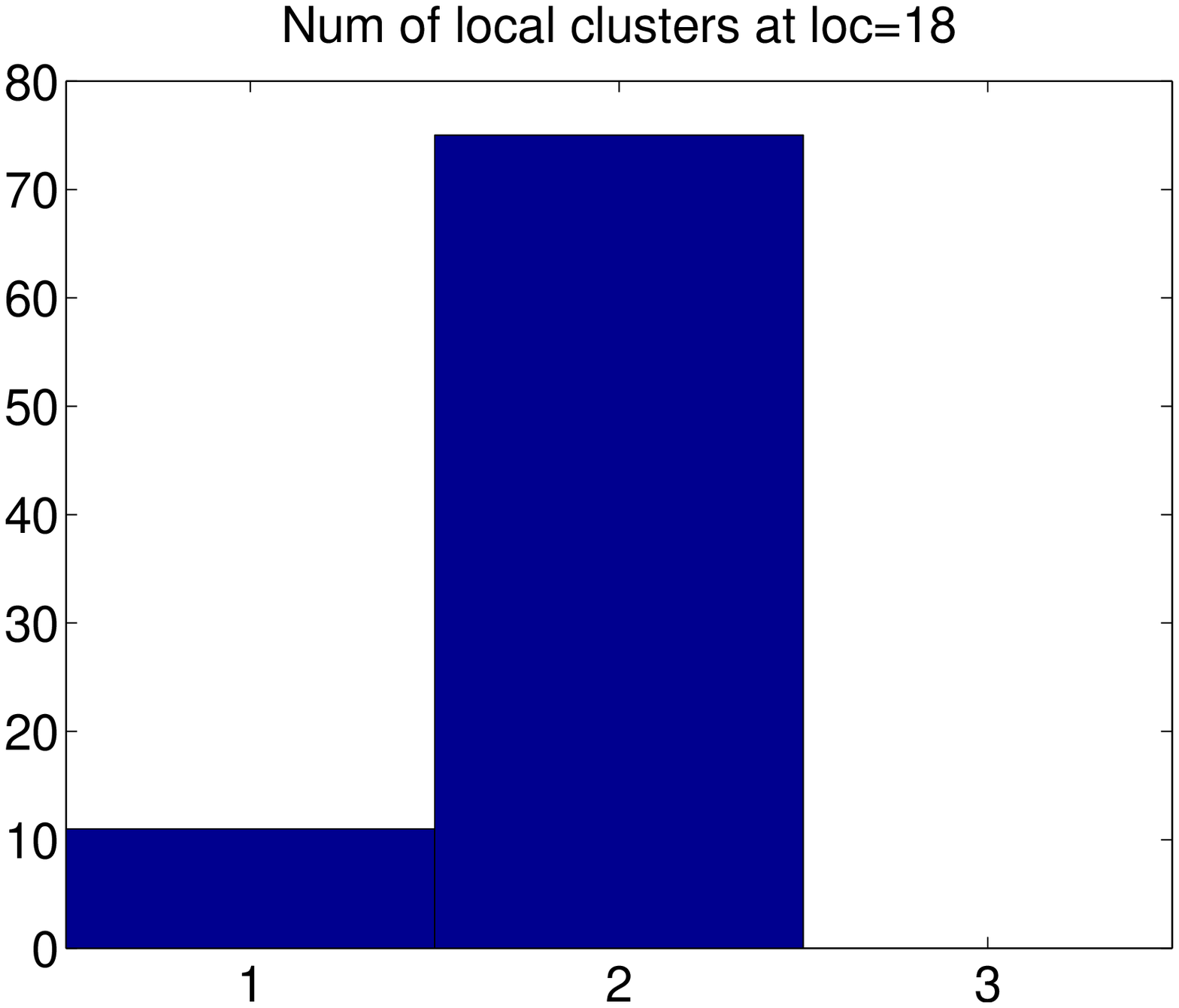} &
\includegraphics[keepaspectratio,width = 0.18\textwidth]{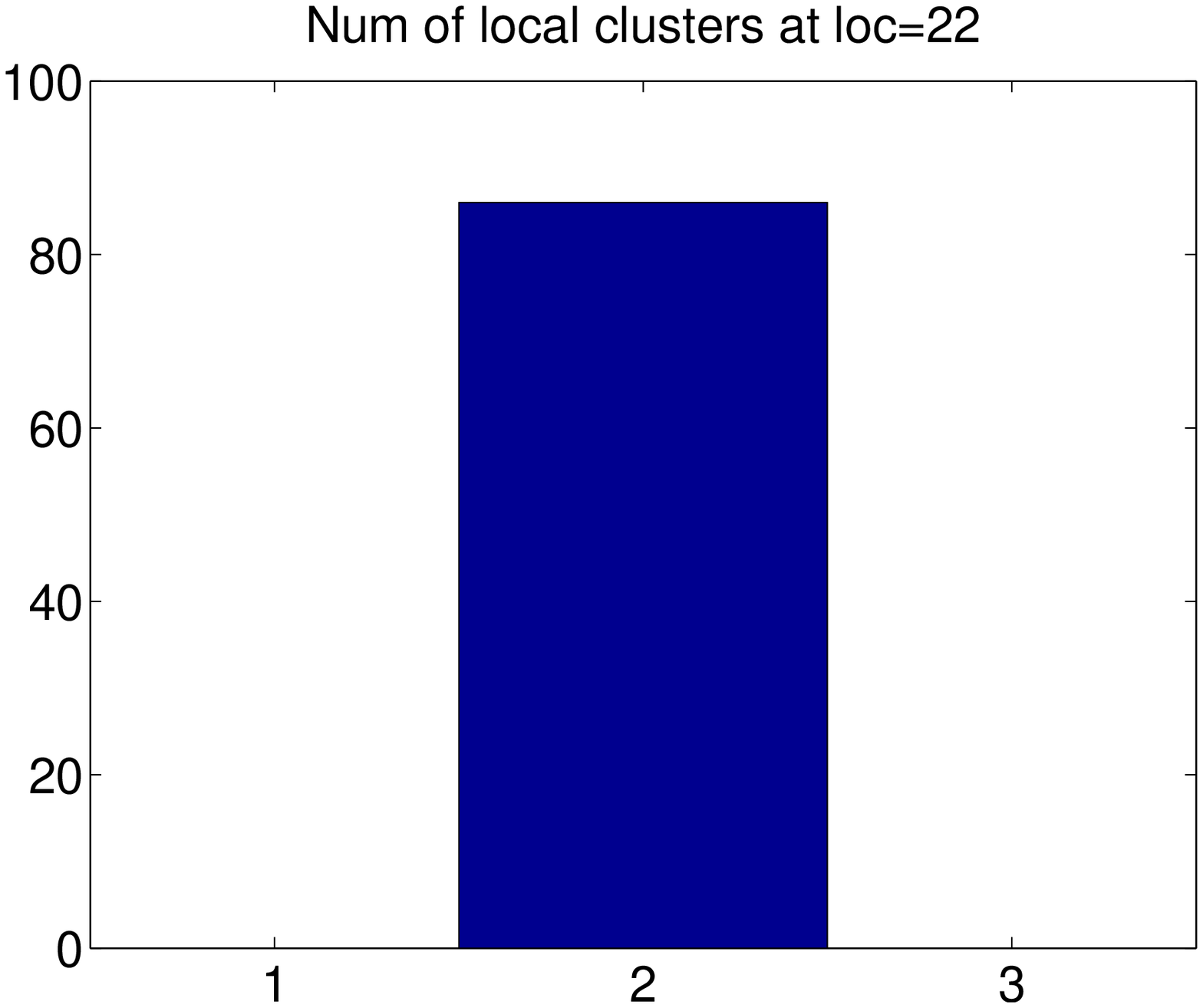} &
\includegraphics[keepaspectratio,width = 0.18\textwidth]{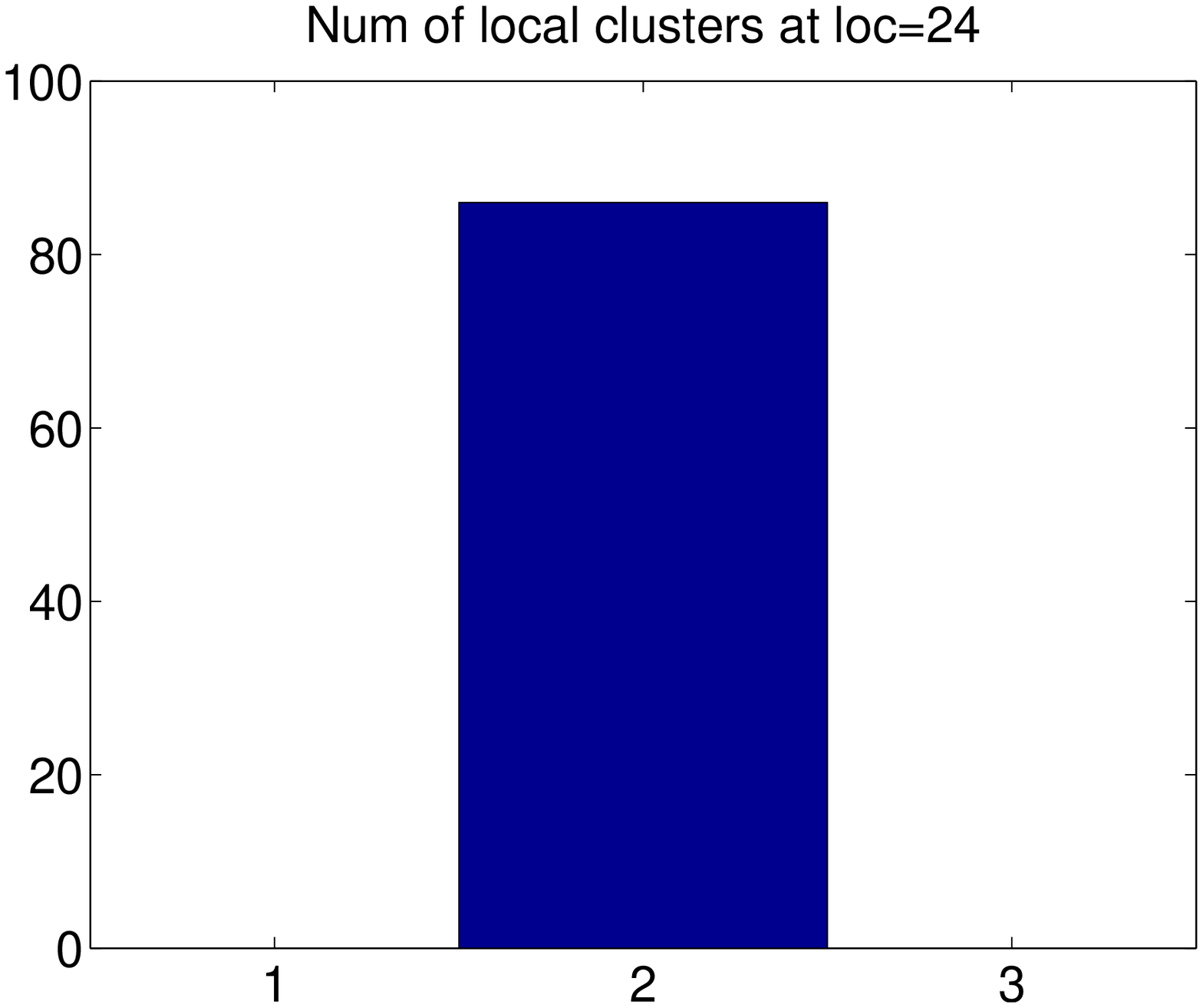}
\end{tabular}
\end{center}
\caption{The leftmost panel shows the posterior distribution of the number of global clusters,
while remaining panels show the the number of local clusters associating with
group index $u$.}
\label{Fig-local-PGD}
\end{figure}

\begin{figure}[t]
\begin{center}
\begin{tabular}{cc}
\includegraphics[keepaspectratio,width = 0.45\textwidth]{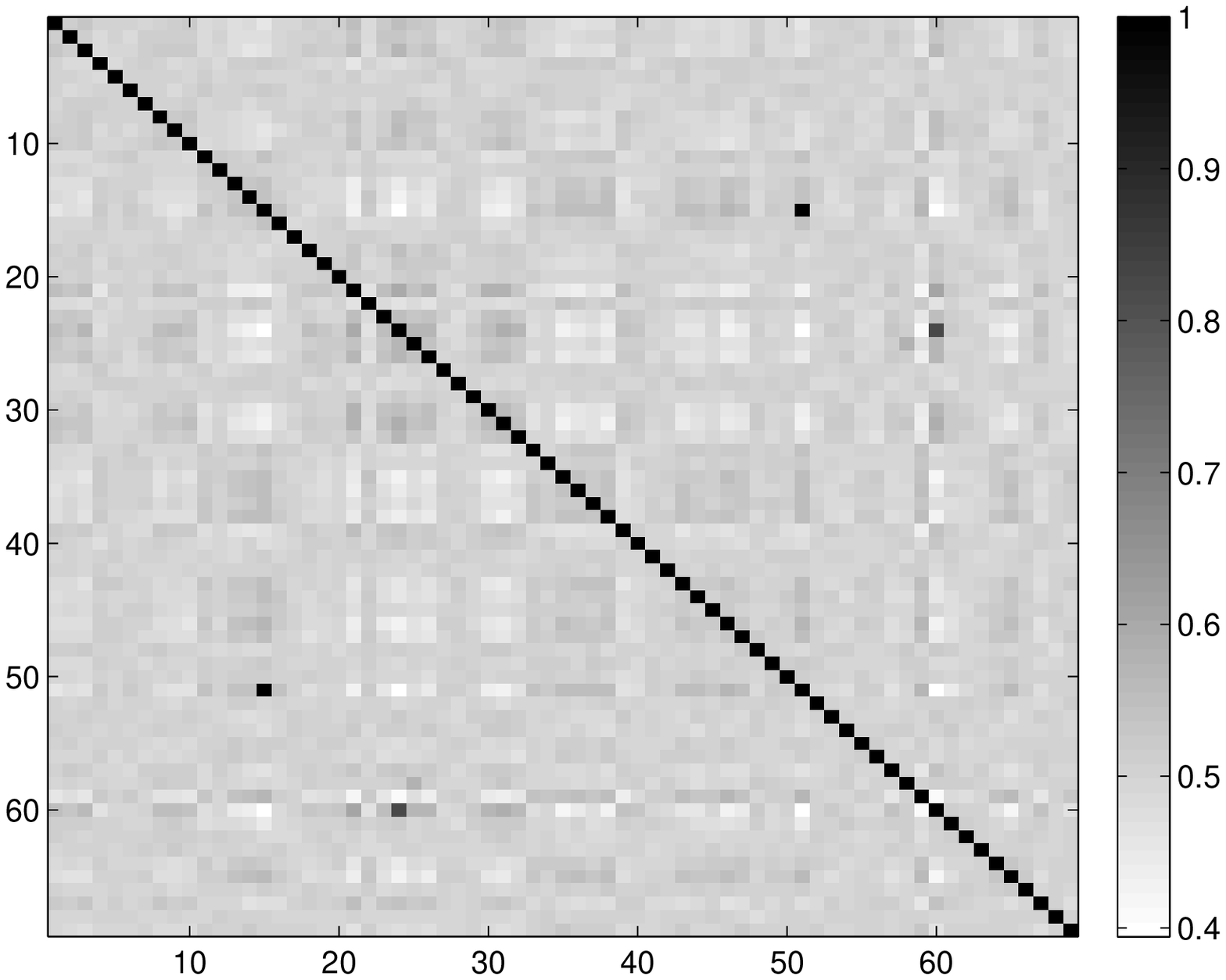}  &
\includegraphics[keepaspectratio,width = 0.45\textwidth]{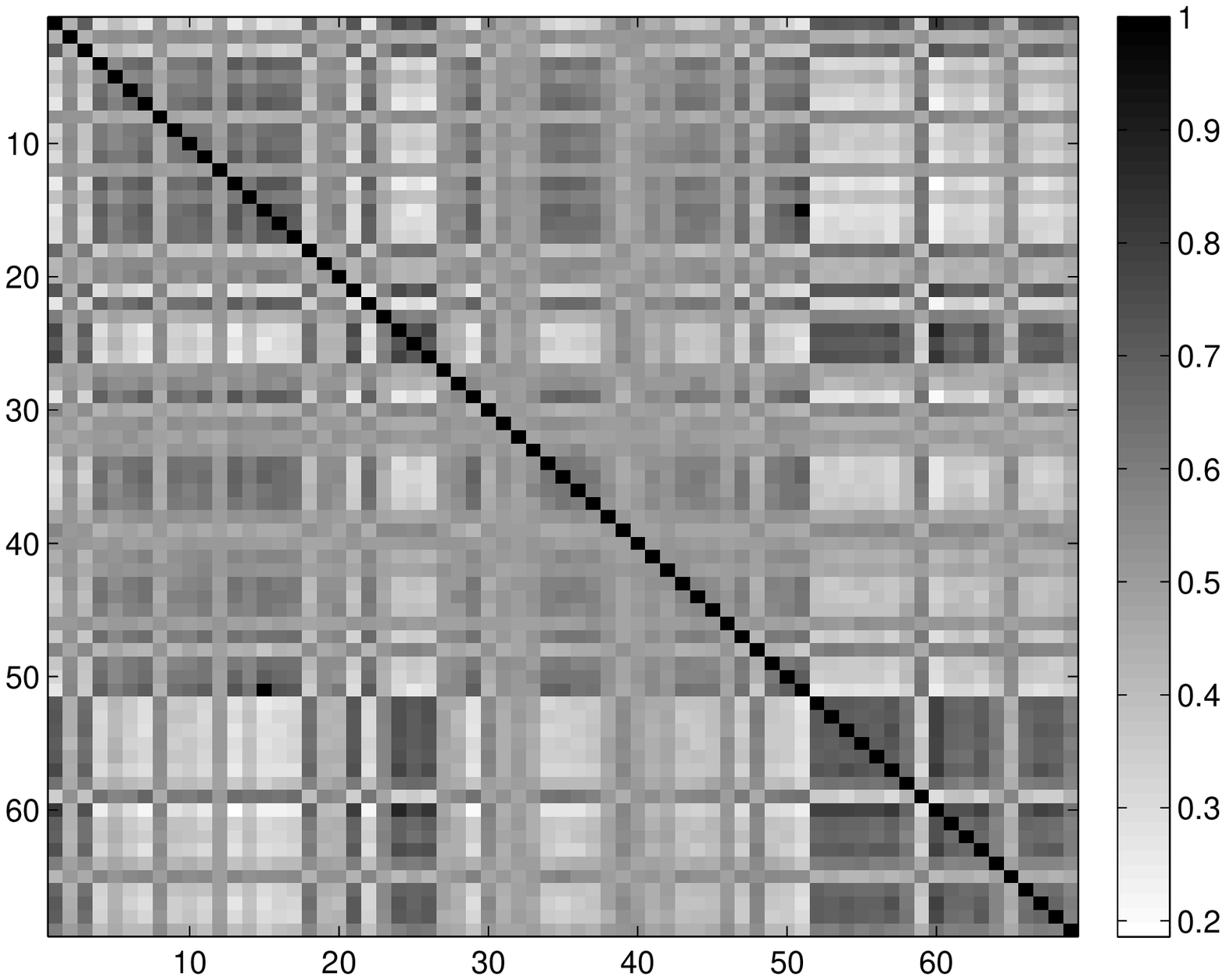} 
\end{tabular}
\end{center}
\caption{Pairwise comparison of individual hormone curves using the
hybrid-DP~\citep{Petrone-etal-09}. Each entry
in the heatmap depicts the posterior probability that the two curves
share the same \emph{local} clusters, averaged over a fixed interval
([1,20] in the left, and [21,24] in the right figure) in the
menstrual cycle.}
\label{Fig-Heat-dlp}
\end{figure}

\section{Illustrations}
\label{sec-examples}

\paragraph{Simulation studies.}
We generate two data sets of spatially varying 
clustered populations (see Fig.~\ref{Fig-Data-AB} for illustrations). 
In both data sets, we set $V = \{1,\ldots,15\}$.
For data set A, $K=5$ global factors $\vec{\phi_1},\ldots,\vec{\phi_5}$ 
are generated from a Gaussian process (GP). 
These global factors provide support for
15 spatially varying mixtures of normal distributions, each of which has
5 mixture components. The likelihood $F(\theta_{ui})$ is given by
$N(\theta_{ui},\sigma_\epsilon^2), \sigma_\epsilon = 0.1$.
For each $u$ we generated independently 100 samples from the corresponding
mixture (20 samples from each mixture components). 
Note that each circle in the figures denote a data sample.
This kind of data can be encountered in tracking problems, where
the samples associating with each covariate $u$ can be viewed as a 
snapshot of the locations of moving particles at time point $u$. The particles move 
in clusters. They may switch clusters at any time, but the identification
of each particle is \emph{not} known as they move from one time step to the next.
The clusters themselves move in relatively smoother paths. Moreover, the number
of clusters is not known. It is of interest to 
estimate the cluster centers, as well as their moving paths.
\footnote{Particle-specific tracking is possible if the identity of
the specific particle is maintained across snapshots.}
For data set B, to illustrate the variation in the number
of local clusters at different locations, we generate a number of
global factors that simulate the bifurcation behavior in a 
collection of longitudinal trajectories. Here a trajectory corresponds to a 
global factor. Specifically, we set $V=\{1,\dots,15\}$. Starting at $u=1$ there 
is one global factor, which is a random draw from a relatively 
smooth GP with mean function $\mu(u) = \beta_\mu u$,
where $\beta_\mu \sim \textrm{Unif}(-0.2,0.2)$
and the exponential covariance function parameterised by $\sigma = 1$, 
$\omega = 0.05$. At $u = 5$, the 
global factor splits into two, with the second one also an 
independent draw from the same GP, which is re-centered 
so that its value at $u=4$ is the same as the value of the previous global 
factor at $u=4$. At $u=10$, the second global factor splits
once more in the same manner. These three global factors 
provide support for the local clusters at each $u\in V$. The likelihood
$F(\cdot|\theta_{ui})$ is given by a normal distribution with 
$\sigma_\epsilon = 0.2$. At each $u$ we generated 30 independent
observations. 

Although it is possible to perform clustering analysis for data at
each location $u$, it is not clear how to link these clusters across
the locations, especially given that the number of clusters might
be different for different $u$'s. The nHDP mixture model provides
a natural solution to this problem. It is fit for both data sets 
using essentially the same prior specifications. The concentration
parameters are given by $\gamma \sim \textrm{Gamma}(5,.1)$
and $\alpha \sim \textrm{Gamma}(20,20)$. $H$ is taken to be a mean-0 GP 
using $(\sigma,\omega) = (1,0.01)$ for data set A, and $(1,0.05)$
for data set B. The variance $\sigma_\epsilon^2$ is endowed with 
prior $\textrm{InvGamma}(5,1)$. 
The results of posterior inference (via MCMC sampling) for both
data sets are illustrated by Fig.~\ref{Fig-global-A} and Fig.~\ref{Fig-global-B}.
With both data sets, the number global clusters are estimated
almost exactly (5 and 3, respectively, with probability $>90\%$). 
The evolution of the posterior distributions on the number of local clusters
for data set B is given in Fig.~\ref{Fig-local-B}. 
%
In both data sets, the local factors are accurately estimated
(see Figs.~\ref{Fig-global-A} and~\ref{Fig-global-B}). For data set B, due to the 
varying number of local clusters, there are regions for $u$,
specifically the interval $[5,10]$ where multiple global factors alternate
the role of supporting local clusters, resulting in wider credible
bands. 


In Section~\ref{sec-properties} we discussed the implications of
prior specifications of the base measure $H$ for the identifiability
of global factors. We have performed a sensitivity analysis for data
set A, 
and found that the inference for global factors is robust when
$\omega$ is set to be in $[.01,.1]$. For $\omega = 0.5$,
for instance, which implies that $\phi_u$ are weakly dependent across $u$'s,
we are not able to identify the desired global factors
(see Fig.~\ref{Fig-identify}), despite the fact that local factors
are still estimated reasonably well. 

The effects of prior specification for $\sigma_\epsilon$ on the inference
of global factors are somewhat similar to the hybrid DP model: a smaller
$\sigma_\epsilon$ encourages higher numbers of and less smooth global curves
to expand the coverage of the function space (see Sec. 7.3 of
\cite{Nguyen-Gelfand-09}). Within our context, the prior for $\sigma_\epsilon$
is relatively more robust than that of $\omega$ as discussed above.
The prior for concentration parameter
$\gamma$ is extremely robust while the priors for $\alpha_u$'s
are somewhat less. We believe the reason for this robustness is due
to the modeling of the global factors in the second stage of the
nested hierarchy of DPs, and the inference about these factors has the effect of
pooling data from across the groups in the first stage. In practice,
we take all $\alpha_u$'s to be equal to increase the robustness
of the associated prior.

%
\paragraph{Progesterone hormone clustering.}
We turn to a clustering analysis of Progesterone hormone data.
This data set records the natural logarithm of the progesterone metabolite, 
measured by urinary hormone assay, during a monthly cycle for 51 female subjects. 
Each cycle ranges from -8 to 15 (8 days pre-ovulation to 15 days post-ovulation).
We are interested in clustering the hormone levels per day, and assessing
the evolution over time. We are also interested in global clusters, i.e.,
identifying global hormone pattern for the entire monthly cycle and
analyzing the effects on contraception on the clustering patterns.
See Fig.~\ref{Fig-Data-PGD} for the illustration and~\cite{Brumback-Rice98}
for more details on the data set.

For prior specifications, we set $\gamma\sim \textrm{Gamma}(5,0.1)$,
and $\alpha_u = 1$ for all $u$. Let $\sigma_\epsilon \sim \textrm{InvGamma}(2,1)$.
For $H$, we set $\mu = 0$, $\sigma= 1$ and $\omega = 0.05$.  
It is found that the
there are 2 global clusters with probability close to 1. 
In addition, the mean estimate of global clusters match very well with 
the sample means from the two groups of women, a group of those using
contraceptives and a group that do not (see Fig.~\ref{Fig-Data-PGD-dlp}).
Examining the variations of local clusters, there is a significant 
probability of having only one local cluster during the first 20 days.
Between day 21 and 24  the number of local clusters is 2 with probability close to 1. 

To elaborate the effects of contraception on the hormone behavior
(the last 17 female subjects are known to use contraception),
a pairwise comparison analysis is performed.
For every two hormone curves, we estimate the posterior probability that they share
the same local cluster on a given day, which is then averaged over
days in a given interval. It is found that the hormone levels
among these women are almost indistinguishable in the first 20 days
(with the clustering-sharing probabilities in the range of $75\%$), but in the last 4 days, 
they are sharply separated into two distinct regimes (with the clustering-
sharing probability between the two groups are dropped to $30\%$). 

We compare our approach to the hybrid Dirichlet process (hybrid-DP) approach 
~\citep{Petrone-etal-09,Nguyen-Gelfand-09}, perhaps the only existing 
approach in the literature for joint modeling of global and local clusters. 
The data are given to the hybrid-DP as the replicates of a random 
functional curve, whereas in our approach, such functional identity information
is not used. In other words, for us only a collection of hormone levels
across different time points are given
(i.e., the subject ID of hormone levels are neither revealed nor matched
with one another across time points).
For a sensible comparison, the same prior specification for base measure $H$
of the global clusters were used for both approaches. The inference results are
illustrated in Fig.~\ref{Fig-Data-PGD-dlp}.
A close look reveals that the global clusters obtained by the 
hybrid-DP approach is less faithful to the contraceptive/no
contraceptive grouping than ours. This can be explained by the fact that 
hybrid-DP is a more complex model that directly specifies the local cluster 
switching behavior for functional curves. It is observed in this example that
an individual hormone curve tends to over-switch the local cluster assignments
for $u \geq 20$,
resulting in significantly less contrasts between the two group of women 
(see Fig.~\ref{Fig-Heat} and~\ref{Fig-Heat-dlp}).
This is probably due the complexity of the hybrid-DP, which can only
be overcome with more data (see Propositions 7 and 8 of~\cite{Nguyen-Gelfand-09} for a theoretical
analysis of this model's complexity and posterior consistency).
Finally, it is also worth noting that the hybrid-DP approach practically
requires the number of 
clusters to be specified a priori (as in the so-called $k$-hybrid-DP 
in~\cite{Petrone-etal-09}), 
while such information is directly infered from data using the nHDP mixture.

\section{Discussions}
\label{sec-discussions}
We have described a nonparametric approach to the inference of global
clusters from locally distributed data. We proposed a nonparametric 
Bayesian solution to
this problem, by introducing the nested Hierarchical
Dirichlet process mixture model. This model has
the virtue of simultaneous modeling of both local clusters and
global clusters present in the data. The global clusters are
supported by a Dirichlet process, using a stochastic process
as its base measure (centering distribution).
The local clusters are supported by the global clusters. Moreover,
the local clusters are randomly selected using another hierarchy
of Dirichlet processes. As a result, we obtain a collection of
local clusters which are spatially varying, whose spatial dependency
is regulated by an underlying spatial or a graphical model.
\comment{
We provided an analysis of the model properties, including
a stick-breaking and a P\'olya-urn scheme characterization, which
are inherited from the properties of the canonical Dirichlet processes.
The graphical and spatial dependency were investigated, along with a
discussion of model identifiability. We presented two MCMC
sampling methods, and discussed the computational implications
of using a graphical model distribution and a spatial distribution
as the base measure in the nHDP model. 
}
The canonical aspects of the nHDP (because of its use of the Dirichlet
processes) suggest straightforward extensions to accomodate richer behaviors
using Poisson-Dirichlet processes (also known as the Pittman-Yor processes),
where they have been found to be particularly suitable for certain applications, and
where our analysis and inference methods can be easily adapted. It would also be
interesting to consider a multivariate version of the nHDP model.
Finally, the manner in which global and local clusters are combined
in the nHDP mixture model is suggestive of ways of
direct and simultaneous global and local clustering for  
various structured data types.

\section{Appendix}

\subsection{Marginal approach to sampling} 
\label{sec-marginal}
The P\'olya-urn characterization 
suggests a Gibbs sampling algorithm to obtain posterior
distributions of the local
factors $\theta_{ui}$'s and the global factors $\vec{\psi}_t$'s,
by integrating out random measures $Q$ and $G_u$'s.
%
%
Rather than dealing with the $\theta_{ui}$'s and $\vec{\psi}_{t}$ directly,
we shall sample index variables $t_{ui}$ and $k_t$ instead,
because $\theta_{ui}$'s and $\vec{\psi}_t$'s
can be reconstructed from the index variables and the $\vec{\phi}_k$'s.
This representation is generally thought to make the MCMC sampling more 
efficient.
Thus, we construct a Markov chain on the space of $\{\vec{t}, \vec{k}\}$. 
Although the number of variables is in principle unbounded, only 
finitely many are actually associated to data and represented explicitly.

A quantity that plays an important role in the computation of
conditional probabilities in this approach is the conditional
density of a selected collection of data items, given the remaining data.
For a single observation $i$-th at location $u$, define the conditional 
probability of $y_{ui}$ under a mixture component $\phi_{uk}$, 
given $\vec{t},\vec{k}$ and all data items except $y_{ui}$:
\begin{equation}
\label{Eqn-integrate-phi}
f_{uk}^{-y_{ui}}(y_{ui}) = \frac{\int F(y_{ui}|\phi_{uk}) \prod_{u'i'\neq ui;z_{u'i'} = k}
F(y_{u'i'}|\phi_{u'k})H(\vec{\phi}_k) d\vec{\phi_k}}
{\int \prod_{u'i'\neq ui;z_{u'i'} = k}
F(y_{u'i'}|\phi_{u'k})H(\vec{\phi}_k) d\vec{\phi}_k}.
\end{equation}
Similary, for a collection of observations of all data $y_{ui}$ such that $t_{ui} = t$
for a chosen $t$, which we denote by vector $\vec{y}_t$, let
$f_{k}^{-\vec{y}_{t}}(\vec{y}_{t})$ be the
conditional probability of $\vec{y}_t$ under the mixture component
$\vec{\phi}_{k}$, given $\vec{t},\vec{k}$ and all data items 
except $\vec{y}_t$.
\comment{
\begin{equation}
\label{Eqn-integrate-phi-2}
f_{k}^{-\vec{y}_{t}}(\vec{y}_{t}) = \frac{\int \prod_{ui: t_{ui} = t} F(y_{ui}|\phi_{uk}) 
\prod_{u'i': t_{u'i'}\neq t;z_{u'i'} = k}F(y_{u'i'}|\phi_{u'k}) H(\vec{\phi}_{k}) d\vec{\phi}_{k}}
{\int \prod_{u'i': t_{u'i'}\neq t;z_{u'i'} = k}F(y_{u'i'}|\phi_{u'k}) H(\vec{\phi}_{k}) d\vec{\phi}_{k}}.
\end{equation}
}

\myparagraph{Sampling $\vec{t}$.} Exploiting the exchangeability of the $t_{ui}$'s
within the group of observations indexed by $u$, we treat $t_{ui}$ as 
the last variable being sampled in the group. To obtain the conditional
posterior for $t_{ui}$, we combine the conditional prior distribution
for $t_{ui}$ 
with the likelihood of generating data $y_{ui}$. 
Specifically, 
the prior probability that $t_{ui}$ takes on a 
particular previously used value $t$ is proportional to $n_{ut}^{-ui}$, 
while the probability that it takes on a new value $t^{\textrm{new}} = m_{u} + 1$
is proportional to $\alpha_u$. The likelihood due to $y_{ui}$ given 
$t_{ui} = t$ for some previously used $t$ is
$f_{uk}^{-y_{ui}}(y_{ui})$. Here, $k = k_t$. 
The likelihood for $t_{ui} = t^{\textrm{new}}$ is calculated by integrating 
out the possible values of $k_{t^{\textrm{new}}}$:
%
\begin{equation}
\label{Eqn-llh-y}
p(y_{ui}|\vec{t}^{-ui}, t_{ui} = t^{\textrm{new}}, \vec{k}, \textrm{Data}) 
= \sum_{k=1}^{K}\frac{q_k}{q_{\cdot} + \gamma} f_{uk}^{-y_{ui}}(y_{ui}) +
\frac{\gamma}{q_{\cdot} + \gamma} f_{uk^{\textrm{new}}}^{-y_{ui}}(y_{ui}),
\end{equation}
where $f_{uk^{\textrm{new}}}^{-y_{ui}}(y_{ui}) = \int F(y_{ui}|\phi_{u})H_u(\phi_u) 
d\phi_u$ is the prior density of $y_{ui}$. As a result, the conditional
distribution of $t_{ui}$ takes the form
\begin{equation}
\label{Eqn-sample-t}
p(t_{ui} = t| \vec{t}^{-{ui}}, \vec{k}, \textrm{Data}) \propto
\begin{cases}
n_{ut}^{-ui}f_{uk_t}^{-y_{ui}}(y_{ui}) & \;\mbox{if}\; t \;\mbox{previously used} \\
\alpha_u p(y_{ui}|\vec{t}^{-ui}, t_{ui} = t^{\textrm{new}}, \vec{k}) & \;\mbox{if} 
\; t = t^{\textrm{new}}.
\end{cases}
\end{equation}
If the sampled value of $t_{ui}$ is $t^{\textrm{new}}$, we need to obtain
a sample of $k_{t^{\textrm{new}}}$ by sampling from Eq.~\eqref{Eqn-llh-y}:
\begin{equation}
p(k_{t^{\textrm{new}}} = k | \vec{t},\vec{k}^{-t^{\textrm{new}}}, \textrm{Data}) \propto
\begin{cases}
q_k f_{uk}^{-y_{ui}}(y_{ui}) & \; \mbox{if} \; k \; \mbox{previously used}, \\
\gamma f_{uk^{\textrm{new}}}^{-y_{ui}}(y_{ui}) & \; \mbox{if} \; k = k^{\textrm{new}}.
\end{cases}
\end{equation}
\comment{
If as a result of updating $t_{ui}$ some index $t$ becomes unoccupied, 
i.e., $n_{ut} = 0$, then the probability that this index will be reoccupied
in the future will be zero, since this is always proportional to $n_{ut}$.
As a result, we may delete the corresponding $k_t$ from the data structure.
If as a result of deleting $k_{ut}$ some mixture component $\vec{\phi}_k$
become unallocated, we delete this mixture component as well.
}

\myparagraph{Sampling $\vec{k}$.} As with the local factors within each
group, the global factors $\vec{\psi}_t$'s are also exchangeable. Thus
we can treat $\vec{\psi}_t$ for a chosen $t$ as the last variable sampled
in the collection of global factors. Note that 
changing index variable $k_t$ actually changes
the mixture component membership for relevant data items (across all
groups $u$) that are associated with $\vec{\psi}_t$, the likelihood 
obtained by setting $k_t = k$ is given by 
$f_{k}^{-\vec{y}_{t}}(\vec{y}_{t})$, where $\vec{y}_{t}$
denotes the vector of all data $y_{ui}$ such that $t_{ui} = t$. So,
the conditional probability for $k_t$ is:
\begin{equation}
\label{Eqn-sample-k}
p(k_{t} = k | \vec{t},\vec{k}^{-t},\textrm{Data}) \propto
\begin{cases}
q_k f_{k}^{-\vec{y}_{t}}(\vec{y}_{t})  & \; \mbox{if} \; k \; \mbox{previously used}, \\
\gamma f_{k^{\textrm{new}}}^{-\vec{y}_{t}}(\vec{y}_{t})  & \; \mbox{if} \; k = k^{\textrm{new}},
\end{cases}
\end{equation}
where $f_{k^{\textrm{new}}}^{-\vec{y}_{t}}(\vec{y}_{t}) = \int \prod_{ui: t_{ui} = t}
F(y_{ui}|\phi_{u}) H(\vec{\phi}) d\vec{\phi}$.

\myparagraph{Sampling of $\gamma$ and $\alpha$.} We follow the
method of auxiliary variables developed by~\cite{Escobar-West}
and~\cite{Teh-etal}.
Endow $\gamma$ with a $\textrm{Gamma}(a_\gamma,b_\gamma)$ prior.
At each sampling step, we draw $\eta \sim \textrm{Beta}(\gamma+1,q_{\cdot})$.
Then the posterior of $\gamma$ is can be obtained as a gamma
mixture, which can be expressed as 
$\pi_\gamma \textrm{Gamma}(a_\gamma + K, b_\gamma - \log(\eta))
+ (1-\pi_\gamma) \textrm{Gamma}(a_\gamma + K - 1, b_\gamma - \log(\eta))$,
where $\pi_\gamma = (a_\gamma + K -1)/(a_\gamma+K-1 + q_{\cdot}(b_\gamma - 
\log(\eta)))$. The procedure is the same for each $\alpha_u$,
with $n_u$ and $m_{u}$ playing the role of $q_{\cdot}$ and $K$,
respectively. Alternatively, one can force all $\alpha_u$ to
be equal and endow it with a gamma prior, as in~\cite{Teh-etal}.

\bibliographystyle{asa}
{\small
\bibliography{NPB}
}

\end{document}